\documentclass[structabstract]{aa}  
\usepackage{graphicx}
\usepackage{txfonts}
\usepackage{xspace}
\begin{document}

  \title{Integral field spectroscopy of nitrogen overabundant blue compact dwarf
  galaxies}
\titlerunning{A PMAS study of high N/O BCDs}

  \author{E. P\'erez-Montero \inst{1}
  \and J. M. V\'\i lchez \inst{1}
  \and B. Cedr\'es \inst{1}
  \and G. F. H\"agele \inst{2,3} \thanks{CONICET, Argentina}
  \and M. Moll\'a \inst{4}
  \and C. Kehrig  \inst{5} \\
   \and A.I. D\'\i az \inst{2}
    \and R. Garc\'\i a-Benito \inst{6}
     \and D. Mart\'\i n-Gord\'on \inst{1}}

 \offprints{E. P\'erez-Montero}

  \institute{
Instituto de Astrof\' \i sica de Andaluc\' \i a - CSIC. Apdo. 3004, 18008, Granada, Spain \\
            \email{epm@iaa.es, jvm@iaa.es, bce@iaa.es, dmg@iaa.es}
\and
Departamento de F\'\i sica Te\'orica, 
         C-XI, Universidad Aut\'onoma de Madrid,
              28049, Cantoblanco, Madrid, Spain \\
 \email{guille.hagele@uam.es, angeles.diaz@uam.es}
  	  \and
 	  Facultad de Cs. Astron\'omicas y Geof\'\i sicas., Universidad Nacional de La Plata, Paseo del Bosque s/n, 1900, La Plata, Argentina 
 	   \and
       Departamento de Investigaci\'on B\'asica, CIEMAT, Avda. Complutense 22, 28040 Madrid, Spain\\
       \email{mercedes.molla@ciemat.es}  
       \and
    	Leibniz-Institut f\"ur Astrophysik Postdam, innoFSPEC Postdam, An der Sternwarte 16, 14482 Postdam, Germany \\
    	\email{ckehrig@aip.de}
    	\and
		Kavli Institute of Astronomy and Astrophysics, Peking University, 100871, Beijing, China \\            	
          \email{luwen@pku.edu.cn}   	
}

   \date{}


%

   \abstract
{The oxygen  and nitrogen-to-oxygen abundances relation is  characterized by a plateau around log(N/O) = -1.6 
at metallicities lower than 12+log(O/H) $\sim$ 8.25.
However, there is a subset of Blue Compact Dwarf galaxies (BCDs) with unexpectedly high N/O values.}
{We want to study the spatial distribution of the physical properties and of oxygen and nitrogen abundances in three BCDs 
(HS 0128+2832, HS 0837+4717 and Mrk 930) with a reported excess of N/O in order to investigate  the nature of this excess and,
particularly, if it is associted with Wolf-Rayet (WR) stars}
{We have observed  these BCDs by using PMAS integral field spectroscopy in the optical spectral range (3700 - 6900 {\AA}),
mapping the brightest emission lines and, hence, their interstellar medium (ISM) physical-chemical
properties (reddening, excitation, electron density and temperature, and O and N chemical abundances),
using both the direct method and appropriate strong-line methods. 
We make a statistical analysis of the resulting distributions and we compare them with the integrated properties of the galaxies.}
{Outer parts of the three galaxies are placed on the "AGN-zone" of the 
[NII]/Hα vs. [OIII]/Hβ diagnostic diagram most likely due to a high N/O
combined with the excitation structure in these regions. 
From the statistical analysis, it is assumed that a certain property can 
be considered as spatially homogeneous (or uniform) if a normal 
gaussian function fits its distribution in several regions of the 
galaxy. Moreover, a disagreement between the integrated 
properties and the mean values of the distribution usually appears when 
a  gaussian does not fit the corresponding distribution.
We find that: 1) for HS 0128+2832, only the N/O ratio as derived 
from the direct method is uniform; 2) for HS 0837+4717, all
properties are uniform; and 3) for Mrk 930, the uniformity is found for 
all parameters, except for electron density and reddening. 
The rotation curve together with the Ha map and UV
images, reveal a perturbed morphology and possible interacting processes
in Mrk930.
Finally, if the 
N/O is constant at spatial scales of the order of several kpc, as our 
analysis indicates,  we conclude that the number of WR stars estimated 
from spectroscopy, is not sufficient to pollute the ISM and to produce 
the observed N/O excess in these objects.
 }
{} 
\maketitle

%

\section{Introduction}

The understanding of the formation and evolution of starburst galaxies
has been based mainly so far on the study of different integrated properties of 
their star-forming knots, which depend on their content of
stars, gas, and dust. In the optical spectrum of these knots, 
the blue light emitted by the massive young star clusters and
the emission lines coming from the ionized gas stand out. These lines provide
information about the physical properties and the chemical abundances
of the gas and the stars. Between the different types of starburst galaxies, 
Blue Compact Dwarf galaxies (BCDs),
also known as H{\sc ii} galaxies, have been characterized very
  accurately by analyzing their optical spectra in a wide sample of these objects
({\em e.g.} Terlevich et al., 1981: Kunth \& \"Ostlin, 2000).  Since the metallicity distribution of these objects
peaks at a low value (about 1/5 $\cdot$ Z$_{\odot}$), the
cooling rate in the ionized gas is much less efficient than in metal-rich giant H{\sc ii} regions in
galaxy-disks and consequently their electron temperatures are much higher. This makes the determination
of chemical abundances using the so-called {\em direct method} ({\em i.e.} using 
collisionally excited lines and electron temperatures) much more accurate than
in other metallicity regimes ({\em e.g.} P\'erez-Montero \& D\'\i az, 2003, H\"agele et al., 2006).
This feature of BCDs, together with their compact aspect and gas richness,
make them ideal scenarios to study the star formation processes in the local universe
and in low metallicity environments resembling those in a younger Universe.
Particularly, it is interesting to look into the ionic abundance ratios between species with an
assumed different stellar origin to probe the chemical evolution of these objects. 

This is the case of the nitrogen-to-oxygen ratio (N/O), since their stellar yields
are quite different and mostly proceeding from low- and intermediate-mass and
massive stars, respectively, and, therefore, it gives important clues about the star formation rate and history of star-forming galaxies (Moll\'a et al., 2006).
This ratio has been well studied since some decades ago. Edmunds \& Pagel
(1978) and Alloin et al. (1979) concluded that the
constant value of N/O [log(N/O) $\sim$ -1.6]   in the low metallicity
regime ({\em i.e.} for 12+log(O/H) $<$ 8) is consistent with a primary origin
of the nitrogen, while the correlation between N/O and O/H at higher oxygen
abundances corresponds to a secondary production of N. These findings
were later confirmed by McCall, Rybsky \& Shields (1985) and Vila-Costas \& Edmunds (1993), among others.

Traditionally, both the {\sl plateau} and the dispersion found in the N/O {\sl
  vs.} O/H plot at the low metallicity regime have been explained in terms of
the delay between the ejections of O
and N, which appears if  O is produced by massive stars and N by low- and intermediate-mass stars. 
In that case, different star formation histories included in chemical
evolution models may change the exact value of O/H for which the secondary N
would appear and, therefore, the observed dispersion is easily explained, such
as Henry et al. (2000), Prantzos (2003) and Moll\'a et al (2006) found. At the same time the plateau appears as a consequence of a very low and continous star formation rate and of metallicity-dependent stellar yields which,  as expected, give a high proportion of primary N when O/H is low, and a higher secondary N production when O/H increases. 

Besides, the relative production of N is sensitive to the assumed theoretical stellar yields. 
Thus, Renzini \& Voli (1981) predicted a certain amount of primary N
produced by the low- and intermediate-mass stars. Other works also give a proportion of primary N for this stellar mass range, such as
Van der H\"oek \& Groenewegen (1997) and, more recently, Gavil\'an et al. (2006). These stellar yields, computed for a wide range of metallicities and masses, and by simplified synthesis stellar models have been also corroborated by precise calculations from the stellar evolution field (Marigo, 2001; Ventura et al., 2002;, Dray, 2003).
Additionally, Meynet \& Maeder (2002) and
Chiappini et al. (2006) predict an enhancement of the N production in the low
metallicity regime assuming rapid rotation of massive and intermediate mass stars, fitting the observations coming from the halo stars of our Galaxy. 

Moreover, in any case, by assuming a recursive star formation history with several bursts
is also possible to find some extra-enhancements of the N/O ratio in the starburst phases: O
increases when massive stars die, decreasing simultaneously N/O, and then, when intermediate mass stars die, N increases again. 
Hence, some peaks should appear in the O/H vs. N/O diagram (Garnett, 1990; Pilyugin, 1992).
Several authors have appealed to hydro-dynamical effects, such as outflows of enriched gas by super-massive galactic winds (van Zee et al., 1998) or inflows of metal-poor gas (K\"oppen \& Hensler, 2005), which makes the metallicity of the galaxy to be lower, keeping the N/O ratio at a high value, to explain the observations.

In this context, there is a subsample of BCDs which present high
values of N in comparison to the expected value for their metal content,
even taking into account the observational dispersion (P\'erez-Montero \& Contini, 2009).
Two examples are Mrk 996 [Thuan et al., 1996, 12+log(O/H)=7.96,
log(N/O)=-0.80] and NGC 5253 [Walsh \& Roy, 1987, 1989;
Kobulnicky et al., 1997, 12+log(O/H)=8.12,
    log(N/O)=-0.83 in the region HII-1].  
More recently,  "Green pea" galaxies,
which resemble in many aspects BCDs, have been found to be also depicted by
low metallicities and high N/O ratios (Amor\'\i n, P\'erez-Montero \& V\'\i lchez, 2010).
    In a number of these metal-poor
BCDs with an  enhanced N/O ratio has been detected broad prominent emissions
at $\sim$ 4650 {\AA} and $\sim$ 5808 {\AA}. 
These broad features are emitted by Wolf-Rayet (WR) stars, being characterized
by the ejection of material via strong stellar winds which pollute the
interstellar medium (ISM) with the products of the H and He-burning in the outer layers of these stars. 
This suggests that WR stars could be also the responsible for the high N abundances in these galaxies.

Among BCDs showing high N/O and where WR stars were detected, are UM420,
UM448, the merger compact group Mrk 1089 (all these in Guseva et al., 2000), HS
0837+4717 (Pustilnik et al., 2004) and NGC 5253 (L\'opez-S\'anchez et al.,
2007; Monreal-Ibero et al., 2010). In the study of galaxies with WR signatures from the
Sloan Digital Sky Survey (SDSS) carried out by Brinchmann et al. (2008), it is also
found that WR galaxies show an elevated N/O relative to non-WR galaxies.
Additionally, a  N/O excess is reported in other six BCDs 
with log(N/O) $>$ -1.3 and a prominent WR blue bump in their optical
integrated spectra (H\"agele et al., 2006, 2008, P\'erez-Montero et al.,
2010). On the other hand, the high N/O ratio found in metal-poor halo stars indicates that another
mechanism polluting the ISM with N,  other
than WR star winds, might exist.


Nevertheless, most of the contributions which establish the
relation between the local N pollution and the presence of WR stars are based on long-slit
integrated observations and, therefore, almost no information
about the relative spatial position of massive stars and possible chemical inhomogeneities exist in this
sample of BCDs. 
 One of the few examples is described in Kehrig et
al. (2008) who ﬁnd an excess of N/O simultaneously with the detection of
WR stellar population across the BCD IIZw70 by means of integral ﬁeld
spectroscopy (IFS).
These are also the cases of NGC5253 (Monreal-Ibero et al., 2010)
and IC10  (L\'opez-S\'anchez et al., 2010).
IFS constitutes a powerful tool to study in
detail the spatial distribution  of the different physical properties and chemical
abundances across the objects. Besides, an appropriate statistical
analysis of these distributions, helps to distinguish between the spatial uniformity of
some properties and to detect local inhomogeneity or pollution of chemical elements in the
ISM of galaxies, and possibly to
link them with the positions and properties of the star-forming knots.

In this work, we use IFS to analyze three BCDs (HS 0128+2832, HS 0837+4717, and Mrk 930) selected because of their
overabundance of N reported in the literature.
In HS 0837+4717 and Mrk 930, the presence of WR stars has been confirmed 
by long-slit spectroscopy. IFS lead to a 2D spatial characterization 
of both the stellar population and the ISM physical-chemical properties allowing us
to check whether the N/O excess is spatially correlated with the WR stars or if
it is due to a more global process affecting the whole galaxy.

The paper is organized as follows. In Section 2, we describe our sample of
objects as well as the observations and data reduction. In Sections 3 and 4,
we present and discuss our results. Finally, Section 5 summarizes the main
conclusions derived from this work.


\begin{table*}
\centering
\caption[]{List of the observed objects, with some additional basic information taken from
Izotov \& Thuan. (2004; HS 0128+2832), Pustilnik et al. (2004; HS 0837+4717) and Izotov \& Thuan (1998; Mrk 930).
The table includes for each object coordinate position, adopted distance,
redshift, logarithm of H$\alpha$ flux in erg/s/cm$^2$/\AA, and oxygen and nitrogen chemical abundances. These abundances have been recalculated using the
procedure described in the text and taking the emission-line data provided in the same sources.}
\label{objects}
\begin{tabular} {l c c c c c c c c }
\hline
\hline
 \multicolumn{1}{c}{Object  ID}  &  Other names  & R.A. (2000) &  $\delta$ (2000) & D(Mpc) & redshift & log F(H$\alpha$) & 12+log(O/H) & log(N/O) \\

\hline
HS 0128+2832 &  & 01h31m21.3s & +28d48m12s &  63.1 & 0.016130  & -13.23 & 8.12 $\pm$ 0.05 & -1.23 $\pm$ 0.14\\
HS 0837+4717 & SHOC 220 & 08h40m29.9s & +47d07m10s &   166.0  & 0.041950 & -13.06 & 7.62 $\pm$ 0.06 & -0.75 $\pm$ 0.20\\
Mrk 930 &   &  23h31m58.3s  &  +28d56m50s  &  83.2 & 0.018296  &   -12.71 & 8.07 $\pm$ 0.03 & -1.42 $\pm$ 0.10 \\
\hline
\end{tabular}
\end{table*}

\section{Data}

\subsection{The sample}

\begin{figure}
\centering
     \includegraphics[width=8cm,clip=]{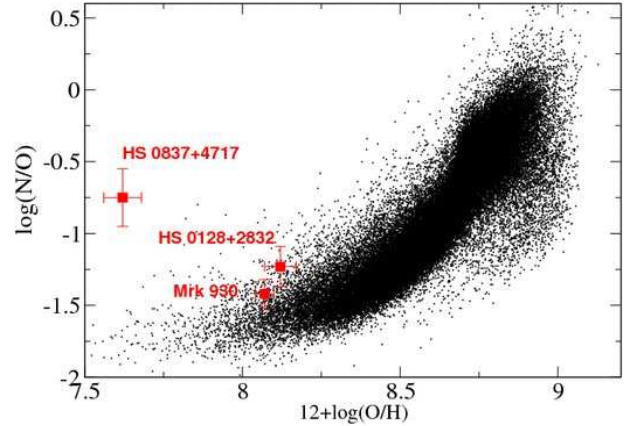}
   
   \caption{Relation between O abundance and N/O ratio as derived from emission-line spectra of
   star-forming objects of the SDSS. The red squares represent the objects of the sample described in the text, with
   their values recalculated using the emission lines reported in the literature
    }
    \label{ohno}
    \end{figure}

Three BCDs with different degrees of N overabundance
as derived using integrated long-slit spectroscopy
available in the literature and visible from the Northern hemisphere
at the epoch of observations were selected. 
The main properties of these objects as taken from the literature
 are listed in Table \ref{objects}, including names, position, adopted distance, redshift and
 H$\alpha$ flux in the slit without aperture correction.
The available data comprise the required emission lines to derive at least 
one electron temperature allowing
the direct and precise determination of the O and N  chemical abundances.
These were recalculated using the methodology described in the section 3.6, and they
are also listed in Table \ref{objects}. In Fig. \ref{ohno}, we show the derived O abundance and
N/O ratio for these three objects compared to the values from the star-forming
galaxies of the SDSS (Amor\'\i n et al., 2010). As we see, all of them present values of N/O higher
than the average N/O for their same metallicity regime.

\begin{figure}
\centering
     \includegraphics[width=9cm,clip=]{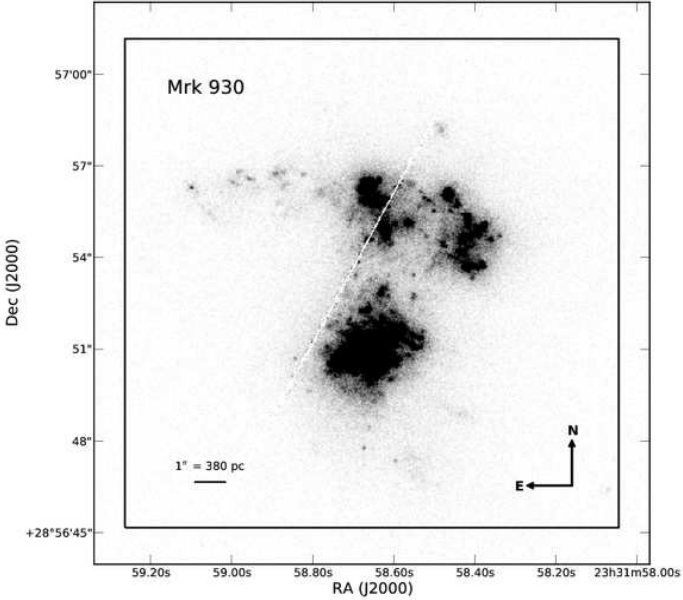}
   
   \caption{HST-ACS image of Mrk 930 taken with the F140-LP filter. The square shows the 
   field of view of the PMAS instrument encompassing the whole galaxy. 
   The figure also shows the complex structure of the ionizing stellar clusters in this galaxy. 
   North points towards up and east to the left.
    }
    \label{acs}
    \end{figure}

HS 0128+2832 is a compact galaxy studied by Izotov \& Thuan (2004), who
derived a low O abundance [12+log(O/H) = 8.12] combined with an elevated abundance of
N  [log(N/O) = -1.23]. 
No trace of WR stars was reported in the integrated spectrum of this object.
HS 0837+4717 is apparently another compact galaxy which has been extensively studied by Pustilnik et al. (2004), 
who found very low O abundance [12+log(O/H) = 7.65] and a very high value of the relative abundance of N
[log(N/O) = -0.75]. These authors
detected the WR blue bump which was used to estimate approximately 1000 WR stars in the starburst.
Although the stellar origin of the He{\sc ii} line shown by Pustilnik is not clear, this
galaxy also appears in the catalog of WR objects in the SDSS (WR101) as a class 2
object ({\em i.e.} convincing WR feature but not obvious after continuum subtraction).
Our third object, Mrk 930,  presents a quite disturbed morphology, characterized by an ellongated shape and whose
main burst of star formation is in the south, but other several fainter knots are also well visible in
the north part.  Fig. \ref{acs} shows an ACS-HST\footnote {This image has been taken from the
Multimission Archive at STScI (MAST) in the webpage http://archive.stsci.edu/} image taken with the F140LP filter
centered in the ultraviolet at a wavelength $\sim$ 1400 {\AA},
 where it is remarkable the presence of these super star clusters along the
galaxy in a complex structure.
Guseva et al. (2000) analyzed both the blue and red WR bumps 
detected by Izotov \& Thuan (1998) for this galaxy and they estimated
around 2000 WR stars, although this is very uncertain due to the noise of the spectrum.
A very noisy hint of this WR feature is also pointed out by Adamo et al. (2011)
using long-slit spectroscopy of the brightest knot of this galaxy.
The chemical analysis carried out by Izotov \& Thuan (1998) for this galaxy revealed a low metallicity
  [12+log(O/H) = 8.07] and a
value of the N/O ratio slightly enhanced [log(N/O) = -1.42].

The study of these three galaxies by means of IFS constitutes an ideal
test to find out to what extent the relative enrichment of N found in
some BCDs is related to the presence of WR stars in
their knots of star formation.


\subsection{Observations}

\label{spectra}


Observations were carried out using the integral field unit (IFU) Postdam
Multi-Aperture Spectrophotometer (PMAS),
developed at the Astrophysikalisches Institut Potsdam (Roth et al. 2005). 
PMAS is attached to the 3.5 m Telescope in the Calar
Alto Observatory (Almer\'\i a, Spain). 
The data were acquired on 2008, October, 29th,  under
good conditions of transparency and with a typical seeing close to 2$^{\prime\prime}$. 
To avoid major differential atmospheric reffraction effects, all
expositions were taken at an air mass lower than 1.2.

One single pointing was taken for each galaxy, covering in all cases
the most intense burst of star formation and its surroundings.

The log of observations is given in Table~\ref{tab2}.  The V600
grating, with a dispersion of 1.6~\AA/pixel in the 2x2 binning mode
to reduce the reading time and readout noise, was used in two spectral ranges,
covering both the blue (3700-5200 {\AA}) and red (5350-6850 {\AA}) sides.
The PMAS spectrograph is equipped with 256 fibers coupled to a $16\times16$
lens array. Each fiber has a spatial sampling of
$1\hbox{$^{\prime\prime}$}\times1\hbox{$^{\prime\prime}$}$ on the
sky, resulting in a field of view of $16\hbox{$^{\prime\prime}$}\times16\hbox{$^{\prime\prime}$}$. 
This field is shown in relation to the size of Mrk 930 in Fig. \ref{acs}. 
At the reported distances of the three objects (see Table 1), each spaxel size of
$1\hbox{$^{\prime\prime}$}$ corresponds to 330 pc in HS 0128+2832,
840 pc in HS 0837+4717, and 380 pc in Mrk 930.
Calibration images were taken following the science
exposures and consisted of emission line lamp spectra (HgNe), and
spectra of a continuum lamp needed to locate the 256 individual
spectra on the CCD. Observations of the spectrophotometric standard
stars BD +28$^\circ$4211 and Hz44 were taken during the observing
night for flux calibration.


\begin{table}
\caption{Log of observations.}
\label{tab2}
\centering
\begin{tabular}{lcc}

\hline\hline

Object 	&  B exptime &  R exptime \\
  	&   (s)  &   (s) \\
 \hline
HS 0128+2832  	& 2x1200 + 3x900 	& 2x1200 + 3x900 \\
HS 0837+4717  	& 2x1200 + 1x900 	& 2x1200 + 1x900 \\
Mrk 930   	&  2x1200 	&  2x1200 \\
\hline
\end{tabular}
\end{table}

\subsection{Data reduction}

We reduced the data using the software R3D (S\'anchez 2006). Different
exposures taken at the same pointing were combined using
IRAF\footnote{IRAF is distributed by the National Optical Astronomy
Observatories.} tasks.  The expected locations of the spectra were
traced on a continuum-lamp images taken before each target
exposure. After bias subtraction, we extracted the target spectra by
adding the signal from the 5 pixels around the central traced pixel
(which is the total object spectrum width).  With exposures of Hg and
Ne lamps taken immediately after the science exposures, the spectra
were wavelength calibrated. We checked the accuracy of the wavelength
calibration using sky emission lines, and find typical deviations of
1 \AA~ and 0.5 \AA~ for the blue and red ranges, respectively.
The effective spectral resolution, derived
by measuring the width of the arc lines is 3.5 {\AA} FWHM (corresponding to an
instrumental dispersion about 90 km/s at 5007 {\AA} and 70 km/s at H$\alpha$
wavelength).
The continuum-lamp exposure was also used to determine the response of
the instrument for each fiber and wavelength (the so-called
flat-spectra). We used these flat-spectra in order to homogenize the
response of all the fibers. 
For the standard star observations we co-added the spectra of
the central fibers and compared the one-dimensional standard star
spectrum with table values to create a sensitivity function. The
spectra were flux calibrated using IRAF.

The reduced spectra were contained in a data cube for each object
and spectral range and they were sky-subtracted and corrected for the effect
of DAR using the R3D package
(S\'anchez 2006). 


\begin{figure*}
\centering
     \includegraphics[width=6cm,clip=]{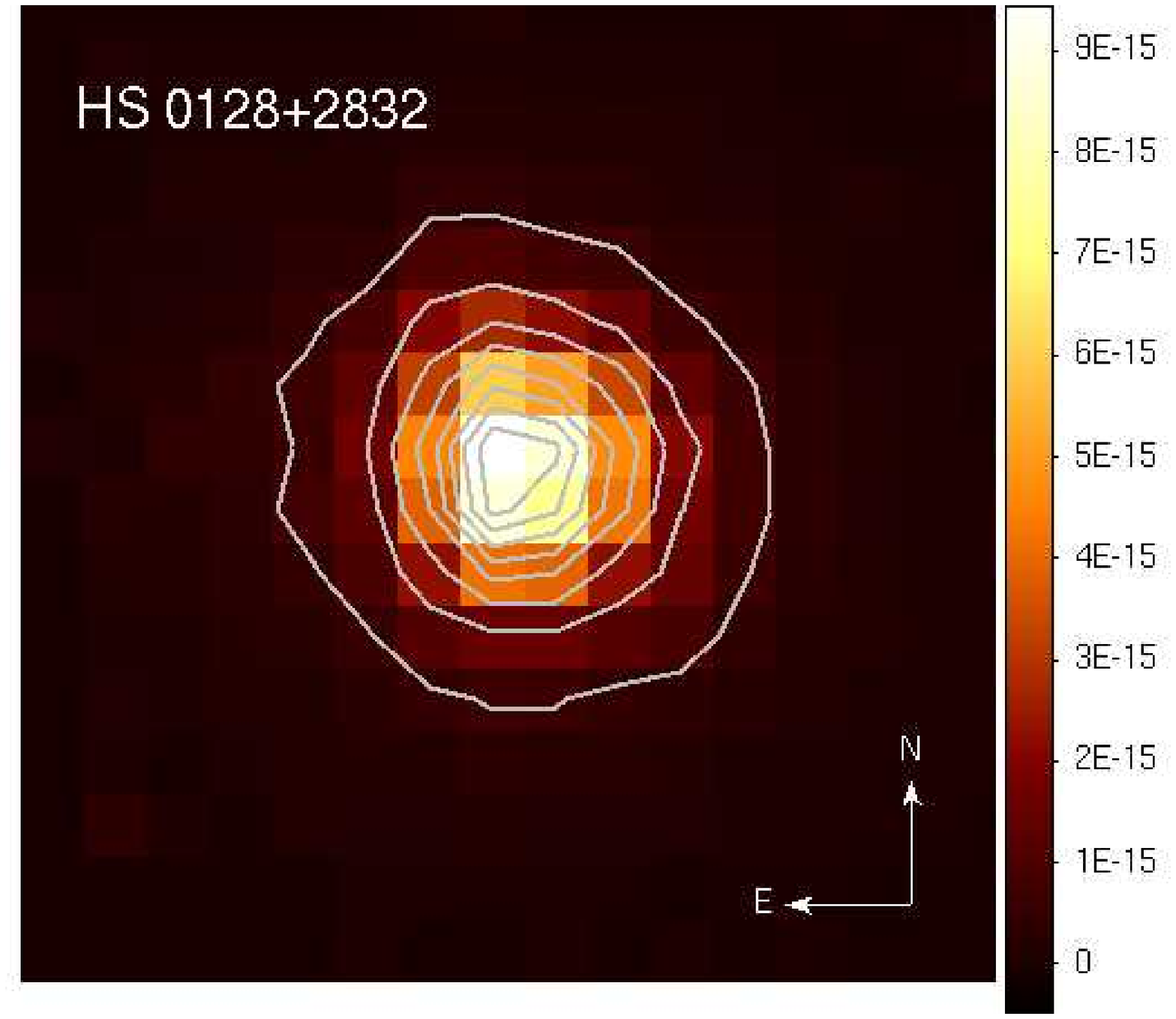}
     \includegraphics[width=6cm,clip=]{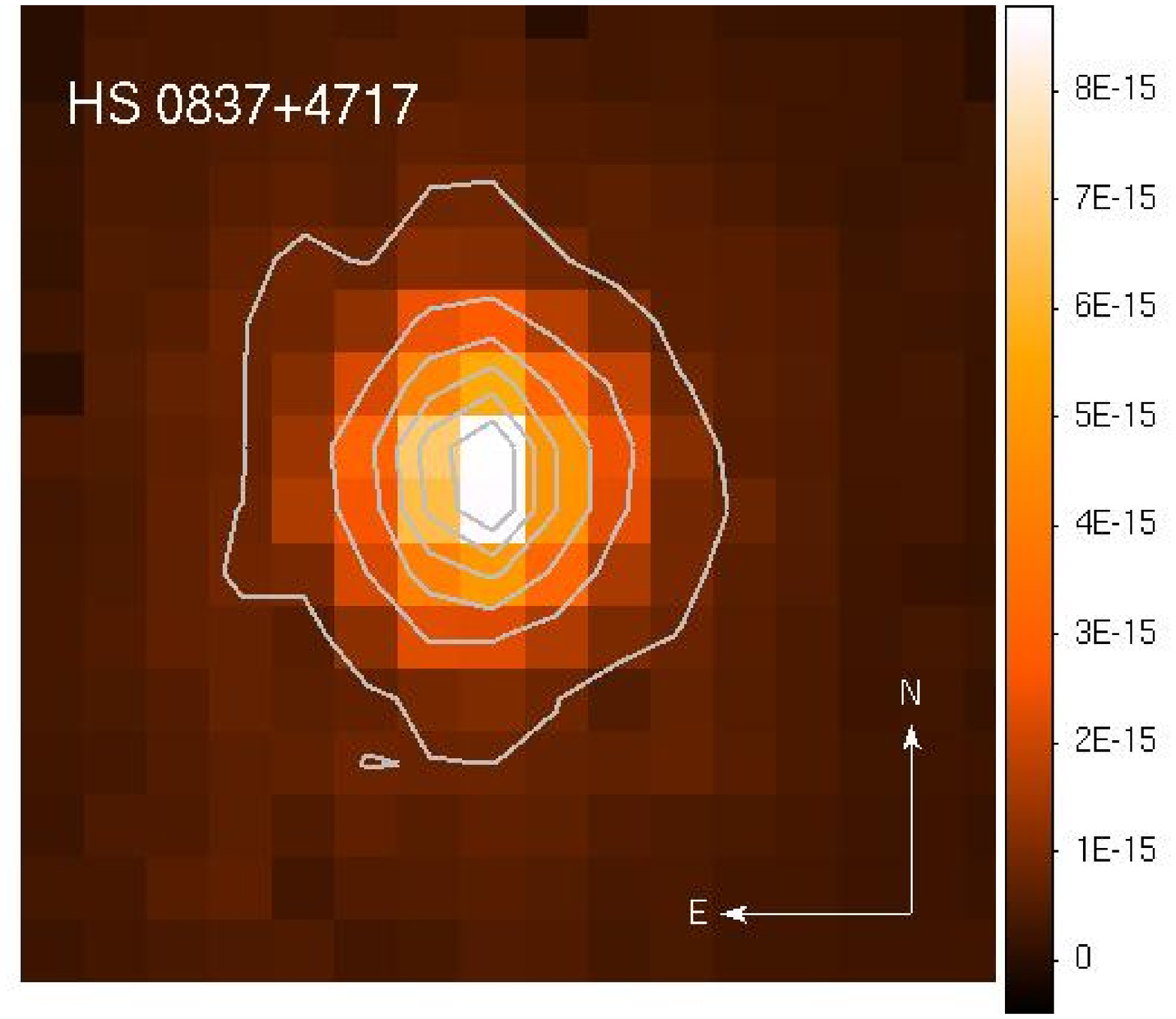}
       \includegraphics[width=6cm,clip=]{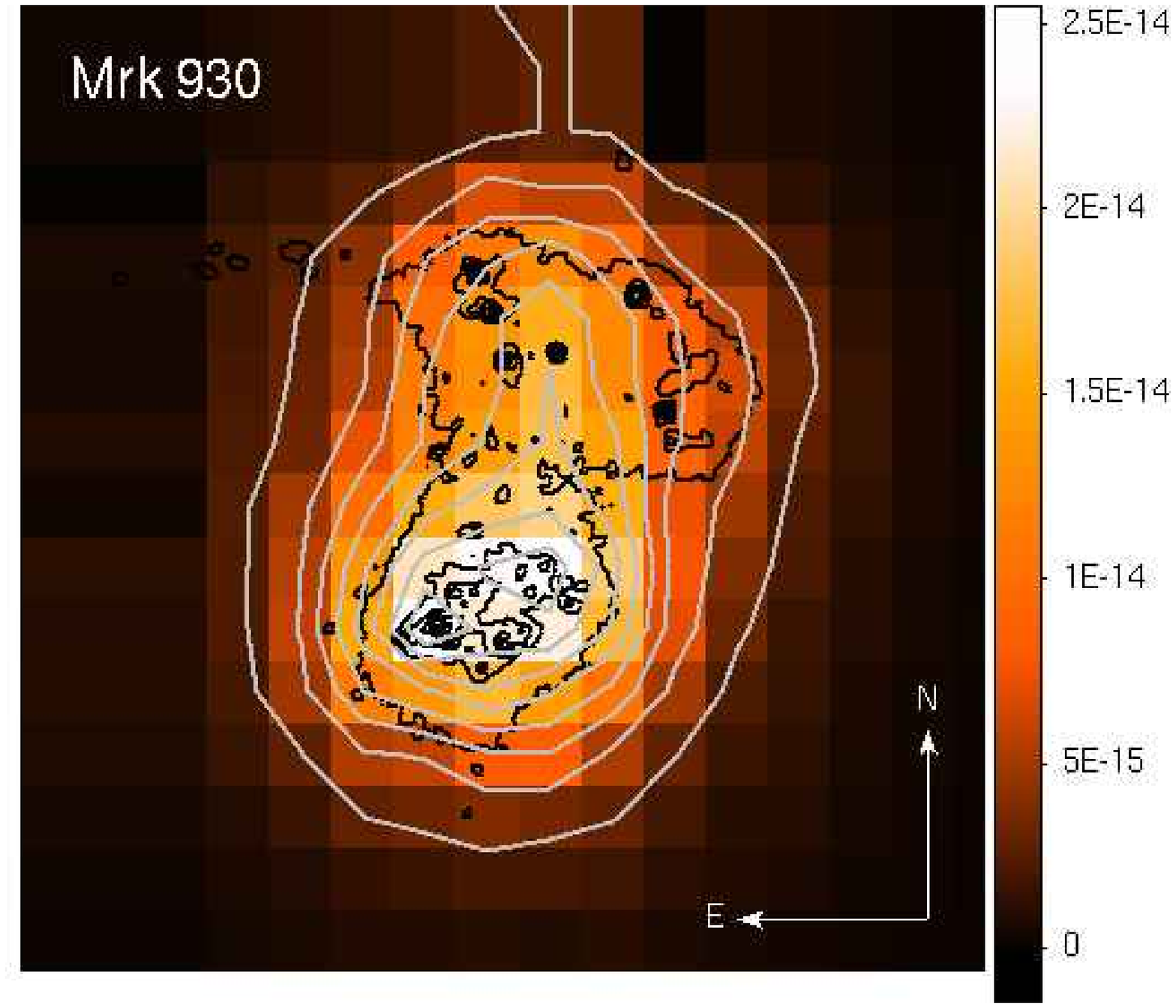}
   
   \caption{Extinction corrected H$\alpha$ maps of HS 0128+2832, HS 0837+4717, and Mrk 930 from left to right respectively. The grey solid line
   contours show the isophotes in units of 12.5 \% of the maximum of H$\alpha$.  In the Mrk 930 image, black
   solid lines represent the contours extracted from the ACS-HST UV image shown in Fig. \ref{acs}, which represent
   the position of the brightest super star clusters. 
   In all images, each spaxel has 1'' of
   resolution, north points to up and east to left. Units are in erg / s / cm$^ 2$ / {\AA}. }
    \label{Halfa}
    \end{figure*}



\begin{table*}
\centering
\caption[]{Results of the Lilliefors test and the gaussian fits to the histograms of the derived physical conditions and chemical abundances
in the three studied objects for different subsets of spaxels: Region 1: spaxels with a relative F(H$\alpha$) larger than 25 \% 
of the maximum F(H$\alpha$) in the galaxy,  Region 2: larger than 12.5\% of the maximum flux, 
Region 3: all spaxels in the field with enough signal-to-noise for the corresponding involved emission lines.
For each region the number of spaxels considered for the statistics is done.
Sign. stands
for the significance level of the null hypothesis in the Lilliefors test. If the null hypothesis cannot be rejected with
more than a 10 \% of confidence, then the gaussian fit is assumed and the corresponding mean and dispersion
values are taken (in bold font). Otherwise, the mean and dispersion are those of the not fitted distribution.}
\label{gaussian}
\begin{tabular} {l l | c c c | c c c | c c c}
\hline
\hline
Object &  &   \multicolumn{3}{c}{HS 0128+2832}  &   \multicolumn{3}{|c|}{HS 0837+4717} &  \multicolumn{3}{c}{Mrk 930}  \\
Region &  &   1   &  2   &  3  &  1   &  2  & 3  &   1   &   2    &    3  \\
\hline
log F(H$\alpha$) (erg/s/cm$^2$/\AA) &  & -13.00  &    -12.83   &   -12.57   &   -12.84  &   -12.69   &    -12.42   &   - 12.06   &  -11.95   &   -11.87  \\
\hline
c(H$\beta$) &  Spaxels &  13 &   30   & 55  &  18  &  40   &  77  & 42  &  73  & 157 \\
  & Mean  &  0.37  &  0.40  &  0.39  &  {\bf 0.70} & {\bf 0.79} & {\bf 0.87 } & 0.48 & 0.46 & 0.51 \\
 & Dispersion               &  0.25  &  0.28  &  0.29  &  {\bf 0.34} & {\bf 0.24} & {\bf 0.23}  & 0.24 & 0.24 & 0.31 \\
 & Sign. (\%)               &  $<$ 10        & $<$ 10         & $<$ 10        & {\bf 71} & {\bf 80} & {\bf 49} &  $<$ 10   &  $<$ 10 & $<$ 10 \\
\hline
log ([O{\sc ii}]/[O{\sc iii}]) &  Spaxels  &  14  &  29  &  55  &   17  &  28  & 37  &  47  &  66  &  143 \\
  & Mean  &  -0.97  & -0.96 &  -0.80 &  -0.94 & {\bf -0.88} & {\bf -0.83} & {\bf -0.31} & {\bf -0.33} &  -0.22 \\
 & Dispersion                   &   0.10  &  0.20 & 0.41 &  0.21 & {\bf 0.21} & {\bf 0.26}  &  {\bf 0.18} & {\bf 0.18} &  0.21 \\
 & Sign. (\%)                  &  $<$ 10  &  $<$ 10  &  $<$ 10 &  $<$ 10 & {\bf 13} &  {\bf 59} & {\bf 38} & {\bf 64} & $<$ 10\\
\hline
n([S{\sc ii}])  (cm$^{-3}$) &  Spaxels & -- & -- & -- & -- & -- & -- & 38 &  67 & 107 \\
 & Mean &  --  &  --  &  -- &  --  & -- &  -- & 101  & 213 & 271 \\
 & Dispersion & -- & -- & -- & -- & -- & -- & 135 & 275 & 296  \\
  & Sign. (\%) & -- & -- & -- & -- & -- & -- & $<$ 10 & $<$ 10 & $<$ 10 \\
  \hline
 t([O{\sc iii}]) (10$^4$K) &  Spaxels & 13 & 22  & 22  & 10 & 12 & 13 &  39 &  52 & 52 \\
 &  Mean  &  1.35 & 1.37  & 1.37 & {\bf 1.64} &  {\bf 1.64}  &  {\bf 1.62} & {\bf 1.26} & {\bf 1.29} & {\bf 1.29} \\
 & Dispersion                           & 0.10 &   0.19 & 0.19 & {\bf 0.25} & {\bf 0.24} & {\bf 0.24}  & {\bf 0.18}  &{\bf 0.21} & {\bf 0.21} \\
 & Sign. (\%)                           & $<$ 10  &  $<$ 10      & $<$ 10      & {\bf 22}    & {\bf 37}   & {\bf 43} & {\bf 11} & {\bf 15} & {\bf 15} \\
\hline 
 12+log(O/H)    & Spaxels &   9 & 21  &  21  &   10  & 12  & 13  &  35 & 47  & 47 \\
 (direct method) &Mean         & 8.00 &   7.99 &  7.99 & {\bf 7.83} & {\bf 7.82}  & {\bf 7.85} & {\bf 8.18} & {\bf 8.15} & {\bf 8.15} \\
                           & Dispersion  & 0.09 &  0.10 &  0.10  & {\bf 0.15} & {\bf 0.14} & {\bf 0.18} & {\bf 0.16} & {\bf 0.18}  & {\bf 0.18} \\
                          & Sign. (\%)   & $<$ 10 & $<$ 10         &   $<$ 10     & {\bf 41}    &  {\bf 15}   & {\bf 11}   & {\bf 55}   & {\bf 32} & {\bf 32} \\
\hline 
 12+log(O/H)  & Spaxels &  14  &  29  & 50  &  16 &  27 & 32 & 41 &  79  & 79 \\
 (R23-P method) &  Mean          &  8.01 & {\bf 7.99}  & 7.96 & {\bf 7.88} & {\bf 7.90}  & {\bf 7.92} & {\bf 8.18} &  8.14 &  8.14 \\
                         & Dispersion   &  0.09 &  {\bf 0.11} & 0.28  & {\bf 0.18} & {\bf 0.19} & {\bf 0.23} & {\bf 0.16} &  0.19  &  0.19 \\
                       & Sign. (\%)   & $<$ 10      &  {\bf 25}  &   $<$ 10     & {\bf 98}    &  {\bf 17}   & {\bf 18}   & {\bf 13}  & $<$ 10 & $<$ 10 \\                
\hline
log(N/O)           &  Spaxels & 13 & 17  & 17  & 9 & 11 & 12 & 36 & 49 & 51 \\
(direct method) & Mean            &  {\bf -1.19} & {\bf -1.17}  & {\bf -1.17} & {\bf -0.82}  & {\bf -0.81}  & {\bf -0.82} & {\bf -1.43} & {\bf -1.44} & {\bf -1.43} \\
                          & Dispersion   &  {\bf 0.11} &  {\bf 0.16} & {\bf 0.16}  & {\bf 0.16} & {\bf 0.15} & {\bf 0.14} & {\bf 0.13} &  {\bf 0.19}  & {\bf 0.18} \\
                          & Sign. (\%)     & {\bf 72}      &  {\bf 67}  &   {\bf 67}  & {\bf 82}  &  {\bf 83}   & {\bf 57}   & {\bf 22}  & {\bf 12} & {\bf 38} \\  
 \hline                        
log(N/O)    &  Spaxels &  15 &  25  & 25  & 17  &  28   & 34 &   47 & 84 & 124 \\
(N2O2 method) &  Mean          &  -1.18 & -1.12  & -1.12 &  -0.85 & {\bf -0.80}  &  {\bf -0.79} & {\bf -1.36} & -1.43 &  -1.41 \\
                           & Dispersion   &  0.25 &   0.27 & 0.27  &  0.26 & {\bf 0.28} &  {\bf 0.29} & 0.24 &  0.17  & {\bf 0.11} \\
                          & Sign. (\%)   & $<$ 10      &  $<$ 10  &   $<$ 10  & $<$ 10  &  {\bf 18}   & {\bf 27}  & {\bf 28}   & $<$ 10 &  $<$ 10 \\   
\hline 
 
\end{tabular}
\end{table*}



\begin{table*}
\centering
\begin{minipage}{160mm}
\caption[]{Dereddened emission line intensities, relative to I(H$\beta$) = 1000, and physical properties and oxygen and
nitrogen chemical abundances with their corresponding propagated errors, 
as derived from the integrated spectra of the studied galaxies. Mrk 930 has been
divided between the brightest southern knot, and those situated in the northern region.  }
\label{int_spectra}
\begin{tabular} {l c c c c c}
\hline
\hline
 &  HS 0128+2832  &  HS 0837+4717 &  Mrk 930-S & Mrk 930-N \\
\hline
$\lambda$ (\AA) & I($\lambda$) & I($\lambda$) & I($\lambda$) & I($\lambda$)  \\
\hline
3727 [O{\sc ii}]       &   788 $\pm$ 116        &   1045 $\pm$ 87  &   2672 $\pm$ 43   & 2532 $\pm$ 58          \\
4102 H$\delta$      &   270 $\pm$ 9        &   271 $\pm$ 22  &  277 $\pm$ 10    &  292 $\pm$ 18    \\
4340 H$\gamma$ &   455 $\pm$ 11        &   470 $\pm $ 24 & 484 $\pm$ 11    &  474 $\pm$ 18   \\
4363 [O{\sc iii}]       &   100 $\pm$ 8        &    149 $\pm$ 12   &   54 $\pm$ 6   &  40 $\pm$ 5    \\
4861 H$\beta$		&   1000 $\pm$ 27         &  1000 $\pm$ 45 &   1000 $\pm$ 15    &  1000 $\pm$ 20   \\
4959 [O{\sc iii}]       &   2269 $\pm$ 48         &   2021 $\pm$ 77 &  1593 $\pm$ 22     &  1452 $\pm$ 21    \\
5007 [O{\sc iii}]        &   6962 $\pm$ 139         &    6064 $\pm$ 195 & 4817 $\pm$ 59     &  4441 $\pm$ 65   \\
6563 H$\alpha$       &   2811 $\pm$ 66         &    2774 $\pm$ 102   &  2830 $\pm$ 47     &  2849 $\pm$ 53   \\
6584 [N{\sc ii}]         &    78 $\pm$ 8       &    152 $\pm$ 10  &   128 $\pm$ 10      &  136 $\pm$ 20     \\
6717 [S{\sc ii}]         &    231 $\pm$ 20        &        --    & 257 $\pm$ 10    &  309 $\pm$ 30        \\
6731 [S{\sc ii}]        &    176 $\pm$ 20        &        --   &  198 $\pm$ 10  &  215 $\pm$ 20         \\
\hline 
 c(H$\beta$)         &    0.14 $\pm$ 0.05      &    0.80 $\pm$ 0.10 &   0.43 $\pm$ 0.07 & 0.32 $\pm$ 0.08       \\
 log([O{\sc ii}]/[O{\sc iii}]  &   -1.07 $\pm$ 0.09    &  -0.89 $\pm$ 0.07 &   -0.37 $\pm$ 0.02      &    -0.38 $\pm$ 0.01        \\
 \hline
 n([S{\sc ii}])  (cm$^{-3}$ )   &   200$^{+260}_{-170}$     &           --            &    $<$ 190       &    32$^{+20}_{-19}$     \\
 t([O{\sc iii}])  (10$^4$ K)   &   1.32 $\pm$ 0.05           &     1.69 $\pm$ 0.10 &    1.20 $\pm$ 0.05     &    1.11 $\pm$ 0.05        \\
 t([O{\sc ii}])   (10$^4$ K)\footnote{Derived from t([O{\sc iii}]) using the models from
 P\'erez-Montero \& D\'\i az (2003) and P\'erez-Montero \& Contini (2009) for t([O{\sc ii}]) and t([N{\sc ii}]), respectively.}  &    1.12 $\pm$ 0.08           &     1.45 $\pm$ 0.20 &  1.16 $\pm$ 0.05  &  1.11 $\pm$ 0.07          \\
  t([N{\sc ii}])   (10$^4$ K)$^a$   &   1.25 $\pm$ 0.03          &     1.41 $\pm$ 0.04 &   1.19 $\pm$ 0.03     &     1.14 $\pm$ 0.03          \\
\hline
12+log(O$^+$/H$^+$)    &  7.09 $\pm$ 0.24           &     6.90 $\pm$ 0.21  &    7.69 $\pm$ 0.10       &      7.89 $\pm$ 0.13           \\
12+log(O$^{2+}$/H$^+$)    &  7.99 $\pm$ 0.06           &     7.67 $\pm$ 0.07  &  7.95 $\pm$ 0.06         &    8.01 $\pm$ 0.07     \\
12+log(O/H)                          &    8.04 $\pm$ 0.08          &      7.73 $\pm$  0.10  &    8.14 $\pm$ 0.05      &  8.24 $\pm$ 0.12      \\
12+log(O/H)  (R23-P)          &     7.95        &      7.90     &   8.19     &    8.10       \\
\hline
12+log(N$^+$/H$^+$)    &    5.95 $\pm$ 0.10       &     6.14 $\pm$ 0.05  &   6.14 $\pm$ 0.03        &     6.22 $\pm$ 0.03          \\
log(N/O)                          &  -1.14 $\pm$ 0.25            &      -0.76 $\pm$  0.23  &   -1.47 $\pm$ 0.10      &   -1.59 $\pm$ 0.15    \\
log(N/O)  (N2O2)                     &     -1.14       &      -0.98  &   -1.41      &  -1.50      \\

\end{tabular}
\end{minipage}
\end{table*}



\begin{figure*}[t]
\centering
    \includegraphics[width=6cm,height=5cm,clip=]{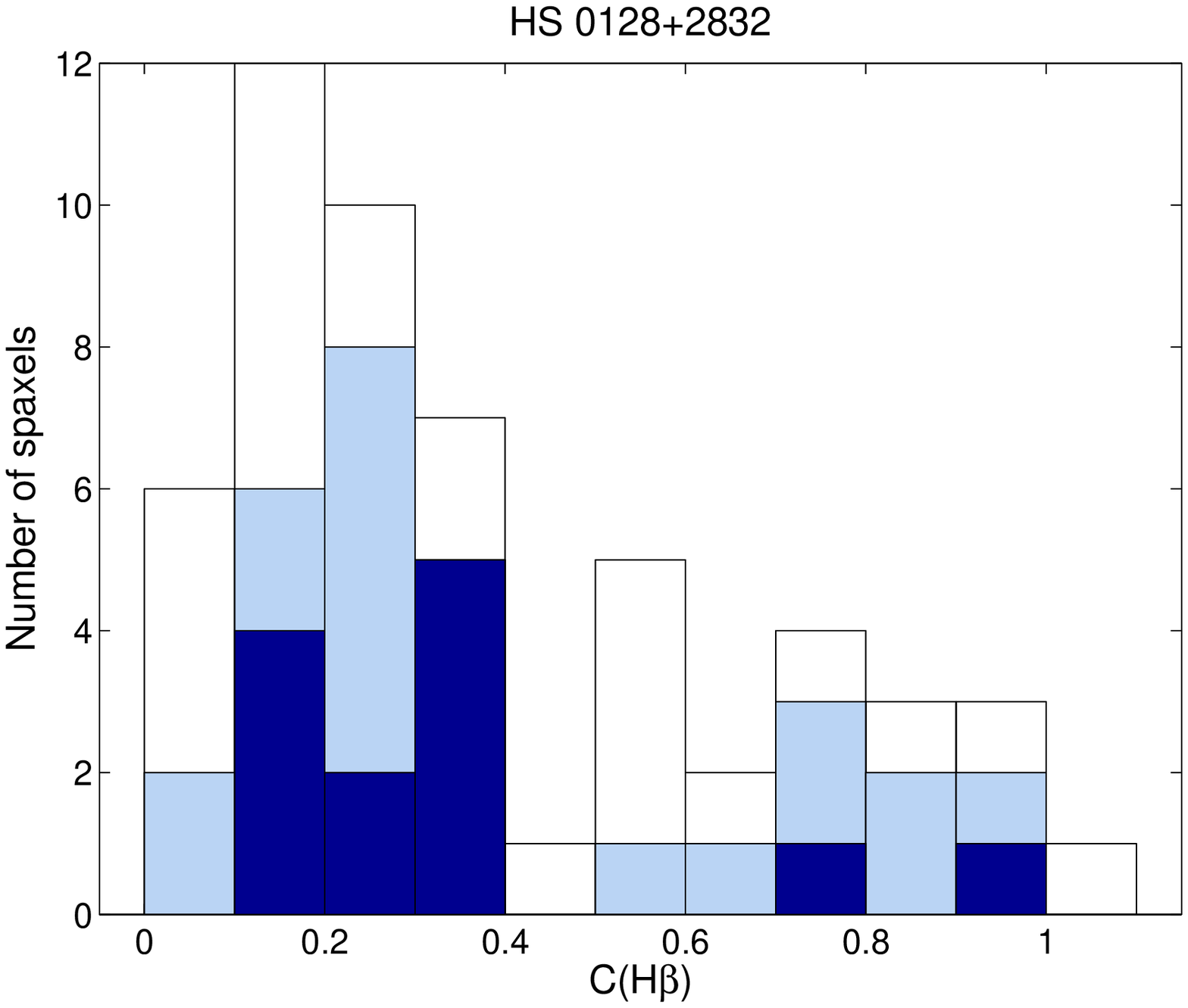}
     \includegraphics[width=6cm,height=5cm,clip=]{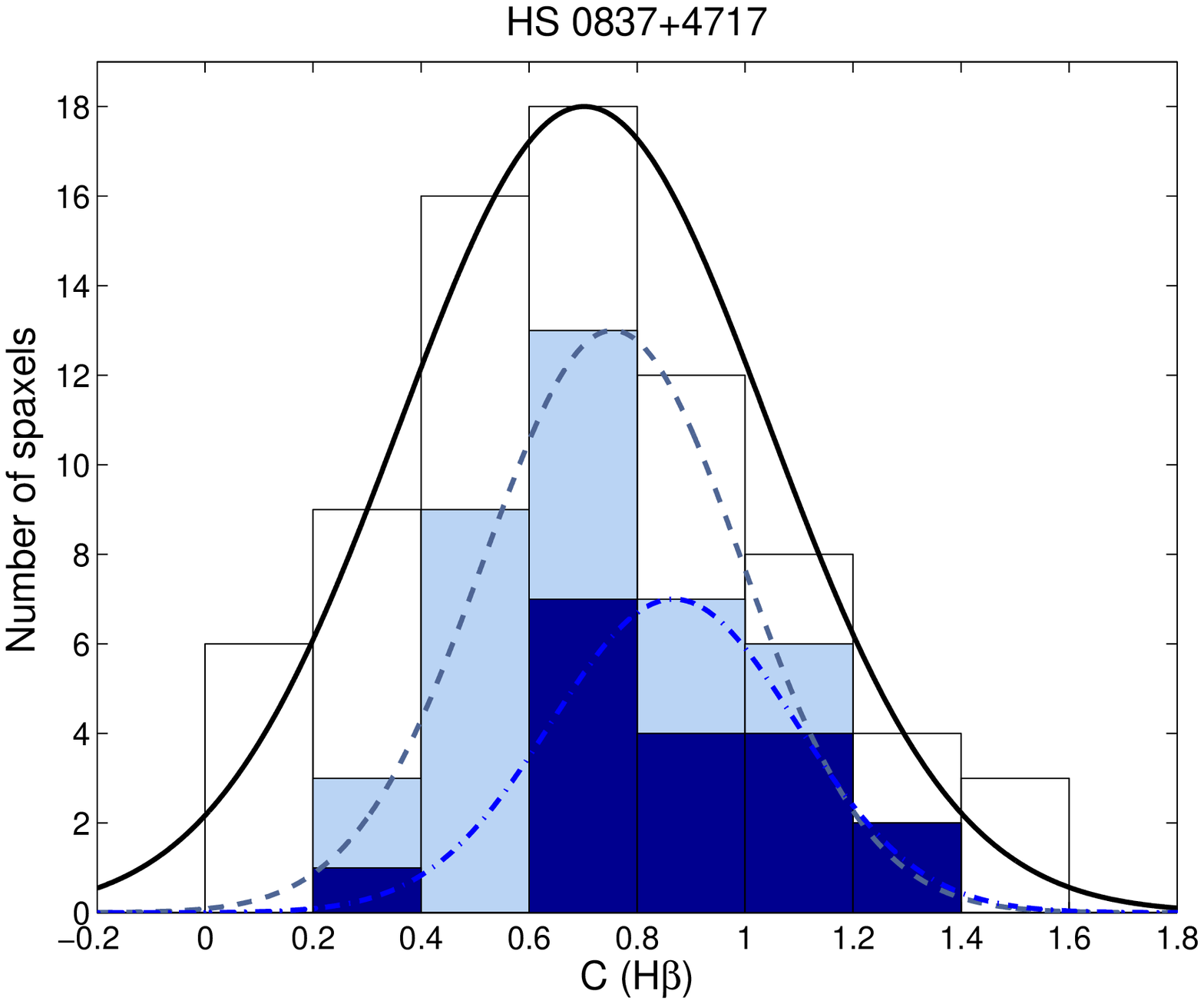}
       \includegraphics[width=6cm,height=5cm,clip=]{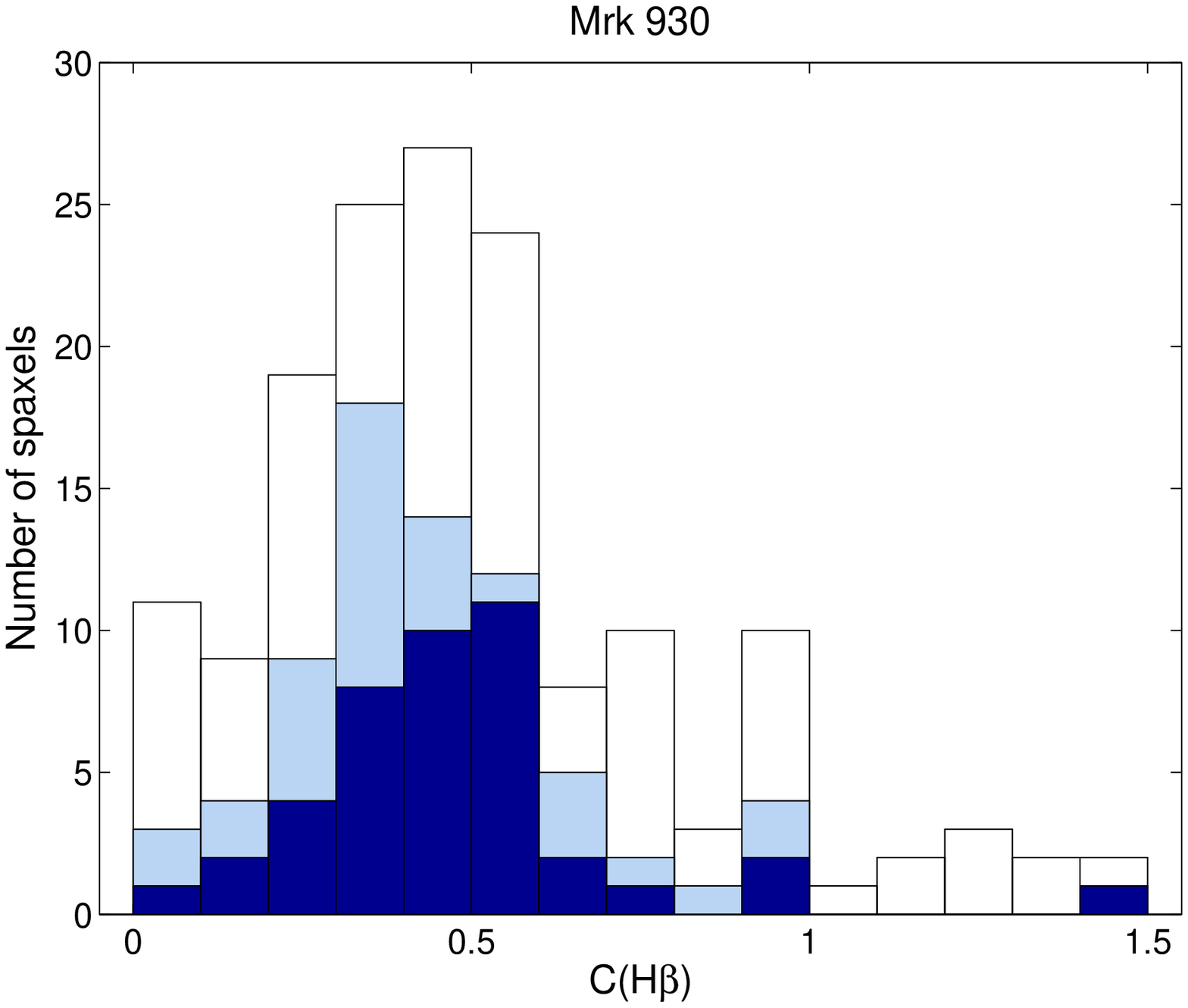}
     \includegraphics[width=6cm,clip=]{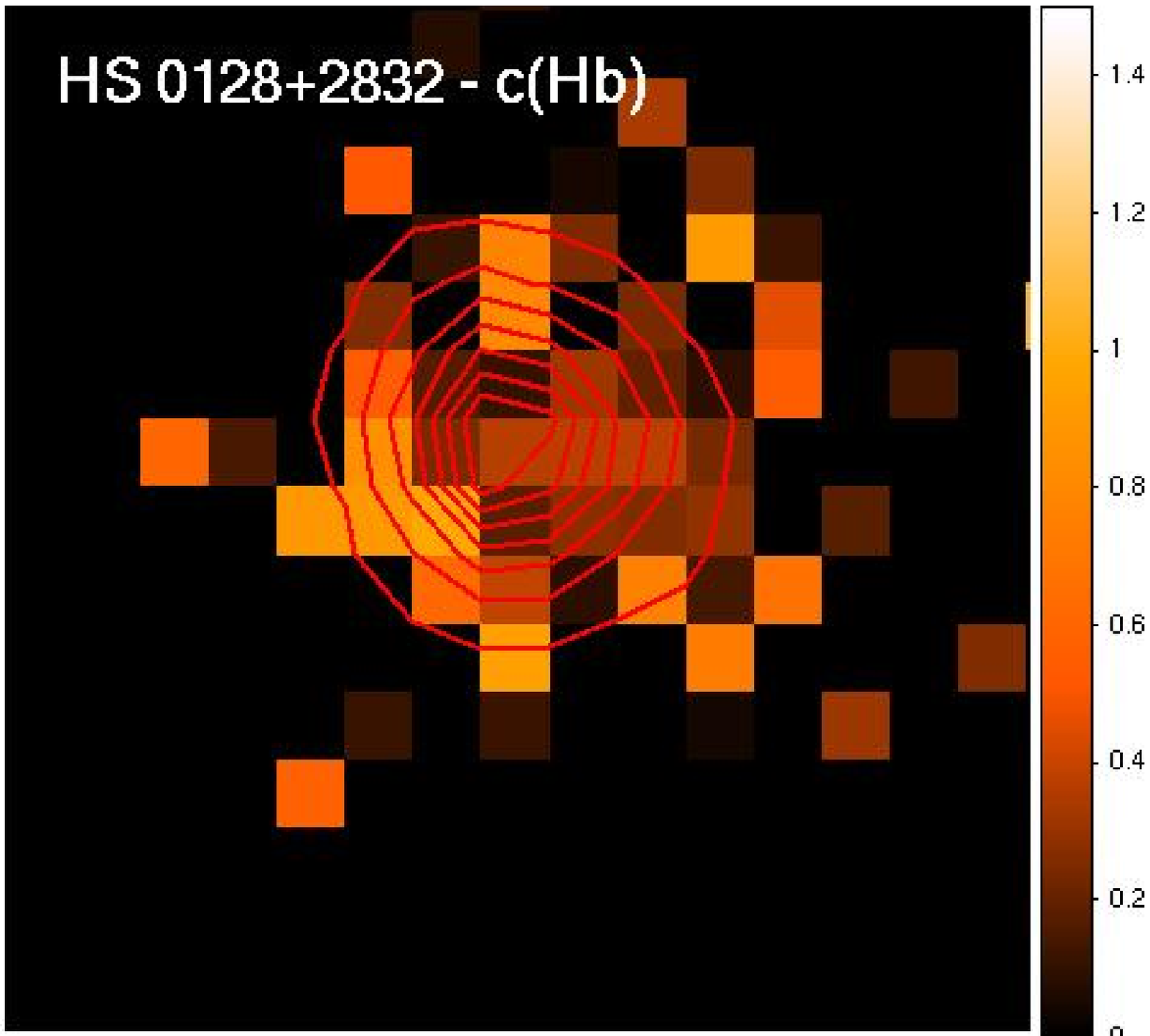}
     \includegraphics[width=6cm,clip=]{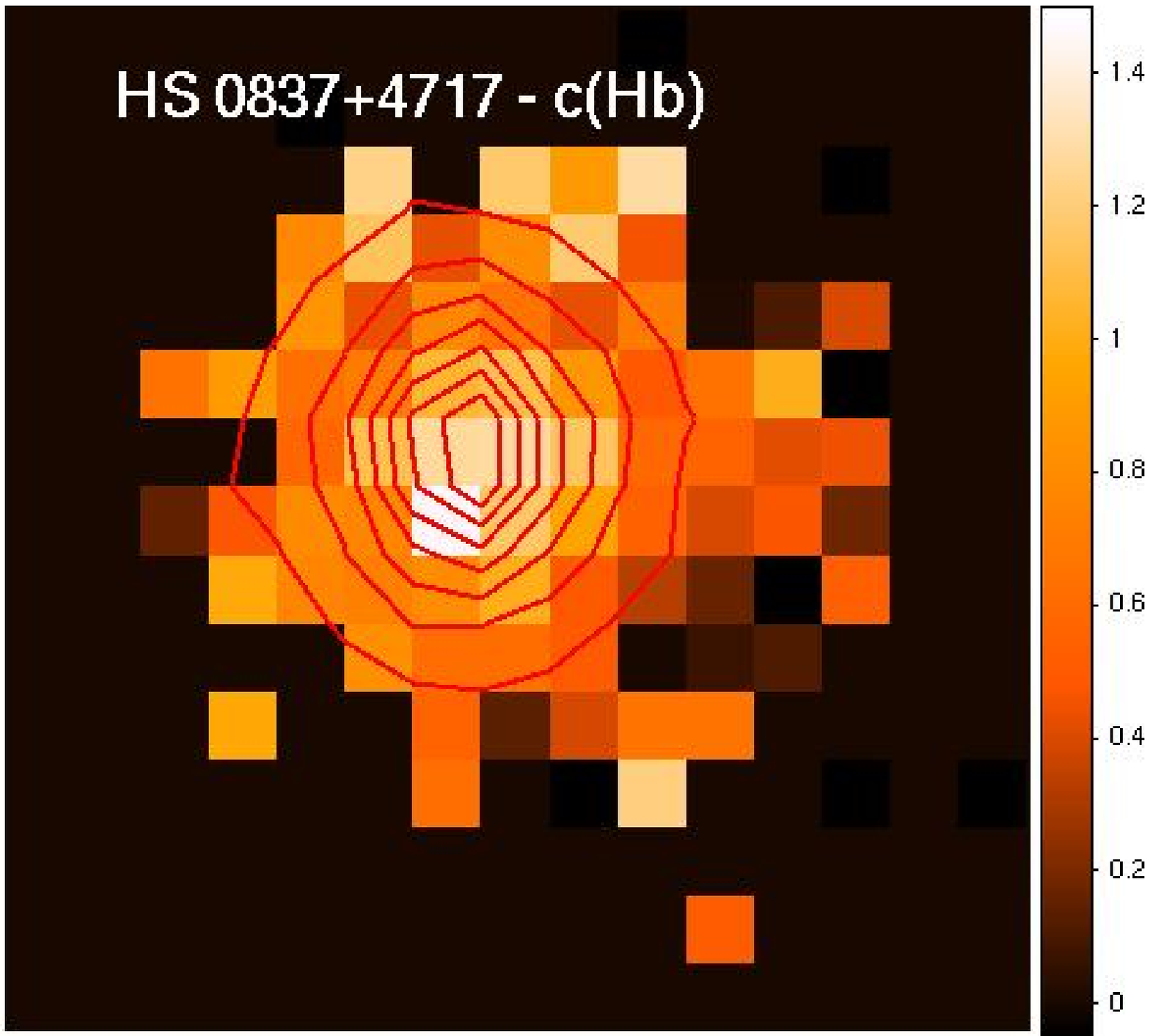}
     \vspace{0.3cm}
       \includegraphics[width=6cm,clip=]{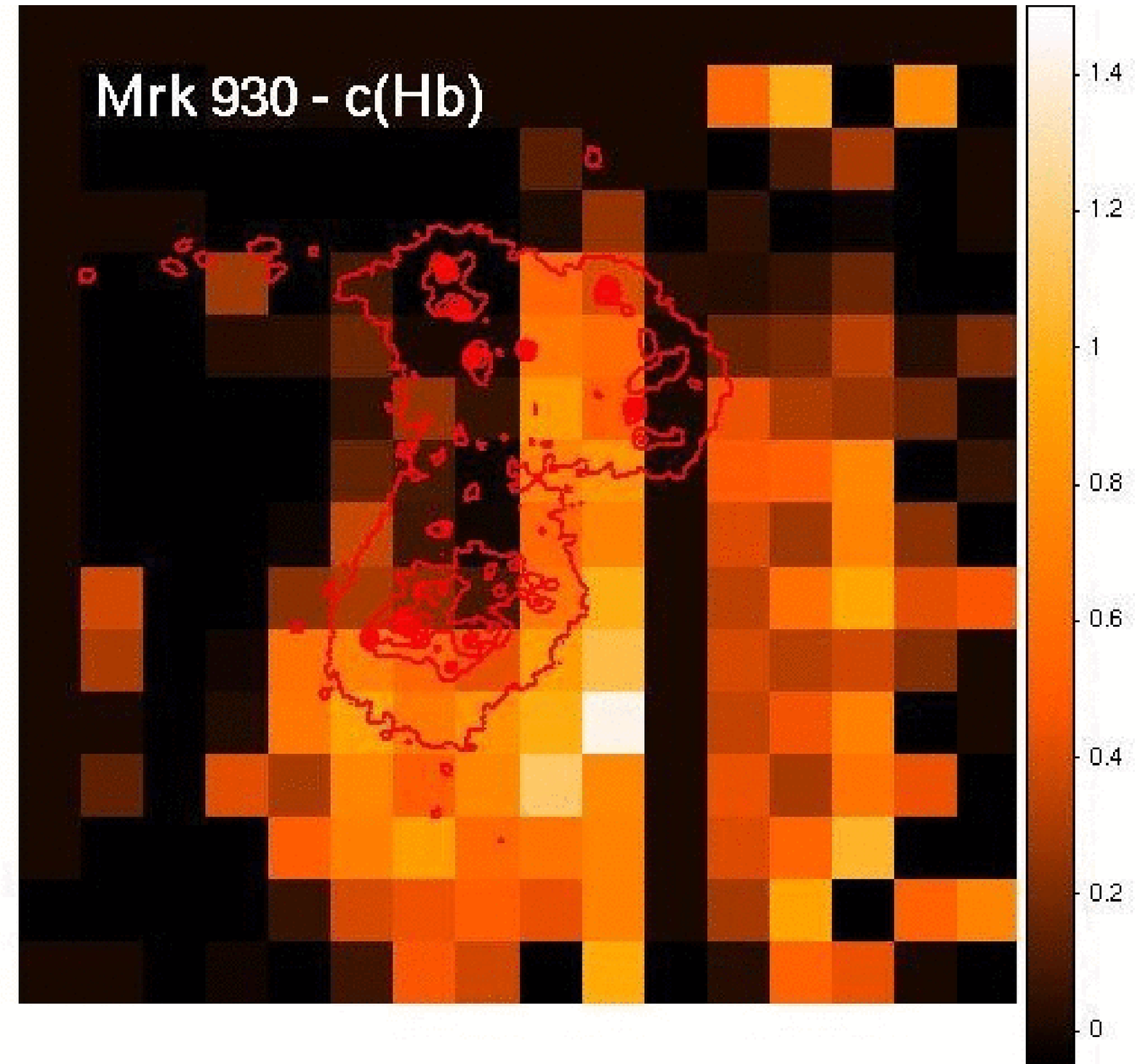}
   
   \caption{Reddening constant maps of HS 0128+2832, HS 0837+4717, and Mrk 930 from left to right, respectively. 
   The solid line contours are those described in Fig. 3. The histogram plots above represent the distribution of the extinction in different regions:
Dark blue bars represent the spaxels in Region 1 (spaxels with a F(H$\alpha$) larger than a 25 \% of the maximum), 
light blue bars, in Region 2 (spaxels with F(H$\alpha$) larger than 12.5 \% of the maximum), and white bars, Region 3 (all the
spaxels in the observed field with enough S/N).} 
    \label{CHb}
    \end{figure*}


\section{Results}

\subsection{Line measurement and H$\alpha$ maps}

We fitted each emission-line profile on the extracted one-dimensional spectra of each spaxel using 
a  gaussian function over the local position of the continuum. 

We used an automatic routine based on the IRAF task {\tt splot} to measure the
flux of most of the emission-lines. For the low S/N emission-lines ({\em e.g.}
[O{\sc iii}] 4363 {\AA}) and those lying close to the border of the CCD ({\em e.g.} 
[O{\sc ii}] 3727 {\AA} for HS 0128+2832 and Mrk 930), the fitting procedure
was repeated using a manual measurement.

The measured flux of the Balmer emission lines is sometimes underestimated
because of the the presence of an underlying stellar continuum ({\em e.g.}
Diaz, 1988). For objects showing high EW(H$\beta$) values (223 \AA\ in HS
0128+2832, 230 \AA\ in HS 0837+4717, and 93 \AA~ in Mrk 930), the effect of the
stellar absorption is negligible due to the low typical values of 1-2 \AA~ for
the hydrogen line EWs in absoprtion (see e.g. McCall et al. 1985).
However, to be careful, we carried out
an eye-inspection, fiber-by-fiber, of the H$\beta$ and H$\gamma$
emission lines, and we did not find any apparent stellar absorption feature
(e.g. wings of absorption lines) underlying these emission lines.


We calculated the statistical error of the line fluxes, $\sigma_{l}$, using the
expression $\sigma_{l}$ =
$\sigma_{c}$N$^{1/2}$[1+EW/N$\Delta$]$^{1/2}$ (see P\'erez-Montero \& D\'\i az, 2003)  where
$\sigma_{c}$ represents a standard deviation in a box centered close to
the measured emission line, N is the number of pixels used in the
measurement of the line flux, EW is the equivalent width of the line,
and $\Delta$ is the wavelength dispersion in \AA/pixel. This
expression takes into account the error in the continuum and the
photon counts statistics of the emission line. The error
measurements were performed on the extracted one-dimensional spectra.
These associated errors are quite different depending on the observed object. Hence,
the average relative error of H$\beta$ is 10\% in HS 0128+2832 and HS 0837+4717, but
only of 5\% in Mrk 930. Besides, to minimize errors in the ratios between a
certain emission line and H$\beta$,
we always take first its ratio in relation to the closest hydrogen emission line 
({\em i.e.} H$\alpha$ in the case of [N{\sc ii}] and [S{\sc ii}])
and then we renormalize it using the corresponding theoretical Balmer ratio ({\em i.e.}
at the electron temperature reported in the long-slit observations for each
object). We have checked that the variation of this temperature across the
field of view of the instrument does not introduce errors in the theoretical
ratio higher than those associated with the flux of the emission lines.\begin{scriptsize}\begin{footnotesize}\begin{small}\end{small}\end{footnotesize}\end{scriptsize}

H$\alpha$ emission line maps (continuum subtracted and extinction corrected) are shown in Fig.~\ref{Halfa}. 
In the same figure, the contours in grey solid line show the isophotes in units of 12.5 \% times the
emission of the maximum of H$\alpha$ in each galaxy. In the panel of Mrk 930, we also show the contours extracted from
the HST-ACS image shown in Fig. 2. Although the combined low spatial resolution of the
instrument (1''/spaxel) and poor seeing during the observing night ($\sim$ 2'') do not allow a precise
analysis of the morphology of the galaxies, it is seen that in HS 0128+2832 and
HS 0837+4717, the central compact burst dominates the emission of the galaxy, while
Mrk 930 displays a more extended distribution of the star formation in agreement
with what is observed in the UV image (see Fig. \ref{acs}).

\subsection{Statistical distributions and integrated spectra}

Line intensities with their corresponding errors measured for each spaxel have been
used to obtain the maps of several physical properties and chemical
abundances.  
Only those emission lines with a signal-to-noise ratio (S/N) $>$ 5 have been considered. 
In order to extract statistical information about the behaviour
of the various derived physical properties and chemical abundances,
we also provide their distribution histograms, taking into
account different regions around the H$\alpha$ peak intensity,
  F(H$\alpha$)$_{peak}$. Hence, hereafter in all histograms,  dark blue
  bars represent the spaxels for which F(H$\alpha$) $>$ 25\% $\cdot$ F(H$\alpha$)$_{peak}$ and
light blue bars, those where
F(H$\alpha) >$ 12.5\% $\cdot$ F(H$\alpha$)$_{peak}$.
Finally, white bars represent the distribution
of the corresponding physical-chemical property in all spaxels with enough S/N over the all IFU area.
In Table \ref{gaussian}, we show the total H$\alpha$ extinction-corrected fluxes
in logarithm units,
as measured in each of these regions for the three studied galaxies.
The ratios between the H$\alpha$ fluxes measured in the whole field of view of
the IFU and the fluxes measured in long-slit observations, listed in Table 1, 
can be considered as aperture correction factors, giving an
estimate for the H$\alpha$ flux lost when using long-slit observations.
In our cases, this gives ratios 
of 4.6 in HS 0128+2832, 4.4 in HS 0837+4717, and 6.9 in Mrk 930.
In the case of Mrk 930, this factor is almost three times the aperture correction factor estimated by
Guseva et al. (2000). 

In order to ease the comparison between the studied distributions and
the position of each defined region, we have overplotted
in all the maps the corresponding H$\alpha$ galaxy contours, with the exception
of Mrk 930, for which we have used the contours from the ACS image, with 
better spatial resolution, easing the 
 identification of the individual knots.

In order to ascertain if a physical-chemical property can be considered
as homogeneous across the IFU area, we have followed a
statistical criterion.  We have plotted the spaxel distribution of a given
property in each region of each galaxy and we have studied if that
distribution can be fitted by a normal function. 
As a first-order approximation,
it can be assumed that the measured variations around the mean value of
the distribution have mainly a statistical origin
and, hence, a
normal distribution of measurements/derivations can be considered.  With this
aim, a Lilliefors test (Lilliefors, 1967) was carried out for each of the distributions of the
derived properties and abundances in each of the regions defined above.  For
this test, the null hypothesis is that the data come from an unspecified
normal distribution. We then assumed that if the null hypothesis cannot be
rejected at the 10\% of significance level, the distribution is considered as
normal. In that case, the fitted gaussian is plotted in the corresponding
histogram. The number of spaxels used to do this test in each region of the
IFU, with the mean value, the dispersion of the fitted gaussian and the
significance level are listed in Table \ref{gaussian} and the corresponding
values are noted in bold font. Otherwise ({\em i.e.} if the significance level
is lower than 10\%), the mean and dispersion values are those of the
not fitted distribution.

A comparative analysis between the distribution of
the derived properties in the spaxels and their values in the integrated spectra has
been done by means of the
co-addition of the emission of those spaxels located in the central region
({\em i.e.} presenting H$\alpha$ emission higher than a 25 \% of the intensity
in the peak of each galaxy). Contrary to the statistical distribution method 
described above, this procedure gives a flux-weighted average of the
properties of the galaxy and it is more consistent to establish a comparison
with long-slit spectra.   In the case of
Mrk930, we have extracted two spectra: one around the southern brightest
knot  and a second one close to the northern knots.
The relative dereddened emission lines and the derived
properties for each of the integrated spectra, as described in the subsections
below are summarised in Table \ref{int_spectra}.

\subsection{Reddening correction and c(H$\beta$) maps}

For each fiber spectrum we derived its corresponding reddening
coefficient, c(H$\beta$), using the value of the Balmer decrement
derived from H$\alpha$/H$\beta$, H$\gamma$/H$\beta$, and H$\delta$/H$\beta$ flux emission-line ratios, 
as compared to the theoretical values
expected for recombination case B from Storey $\&$ Hummer (1995)
at the electron density and temperature reported in the literature 
using long-slit observations, and
applying the extinction law given by Cardelli et al. (1989) with R$_V$ = 3.1. 
Thus, the fluxes
of the emission lines for each fiber were corrected for extinction
using their corresponding c(H$\beta$) values. 


\begin{figure*}[t]
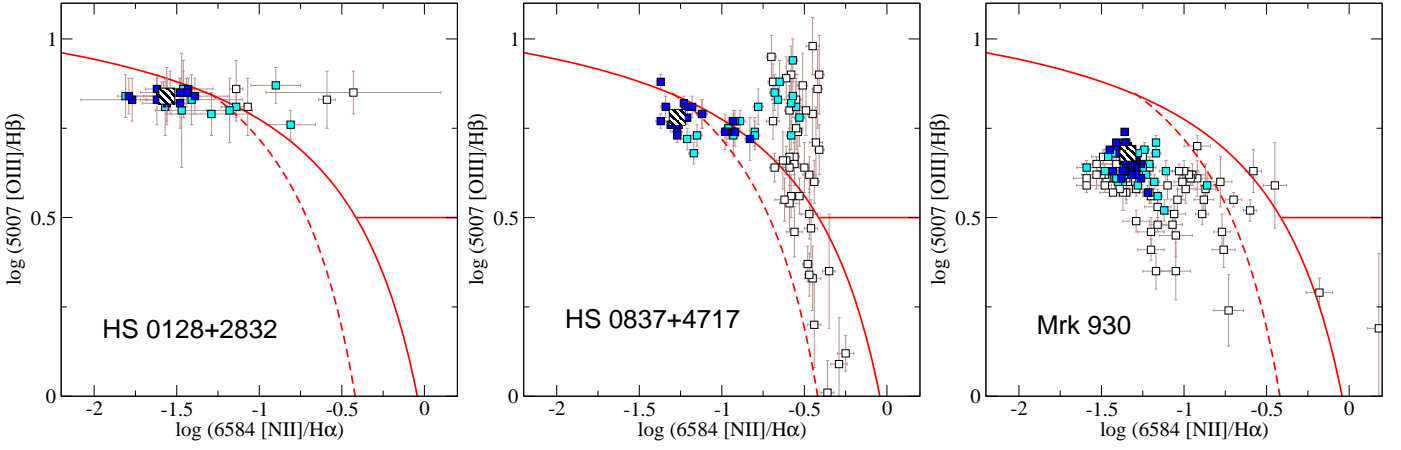

\centering
	    \includegraphics[width=6cm,clip=]{HS0128_N2_O3.eps}
     \includegraphics[width=6cm,clip=]{HS0837_N2_O3.eps}
       \includegraphics[width=6cm,clip=]{Mrk930_N2_O3.eps}
    
   \caption{Relation between log([N{\sc ii}]/H$\alpha$) and log([O{\sc iii}]/H$\beta$). one of the so-called BPT diagrams for the individual spaxels
   of the three studied galaxies. Dark blue squares represent the spaxels in Region 1, the light blue, Region 2, and the white all the rest.
   Large stripped squares represent the result of the integrated spectrum of Region 1.
   The red dashed line represents the empirical curve defined by Kauffman et al. (2003) to separate star-forming galaxies from active
   galactic nuclei. The solid line represents the theoretical separation defined by Kewley et al. (2001).}
    \label{n2o3}
    \end{figure*}



\begin{figure*}
\centering
	    \includegraphics[width=6cm,height=5cm,clip=]{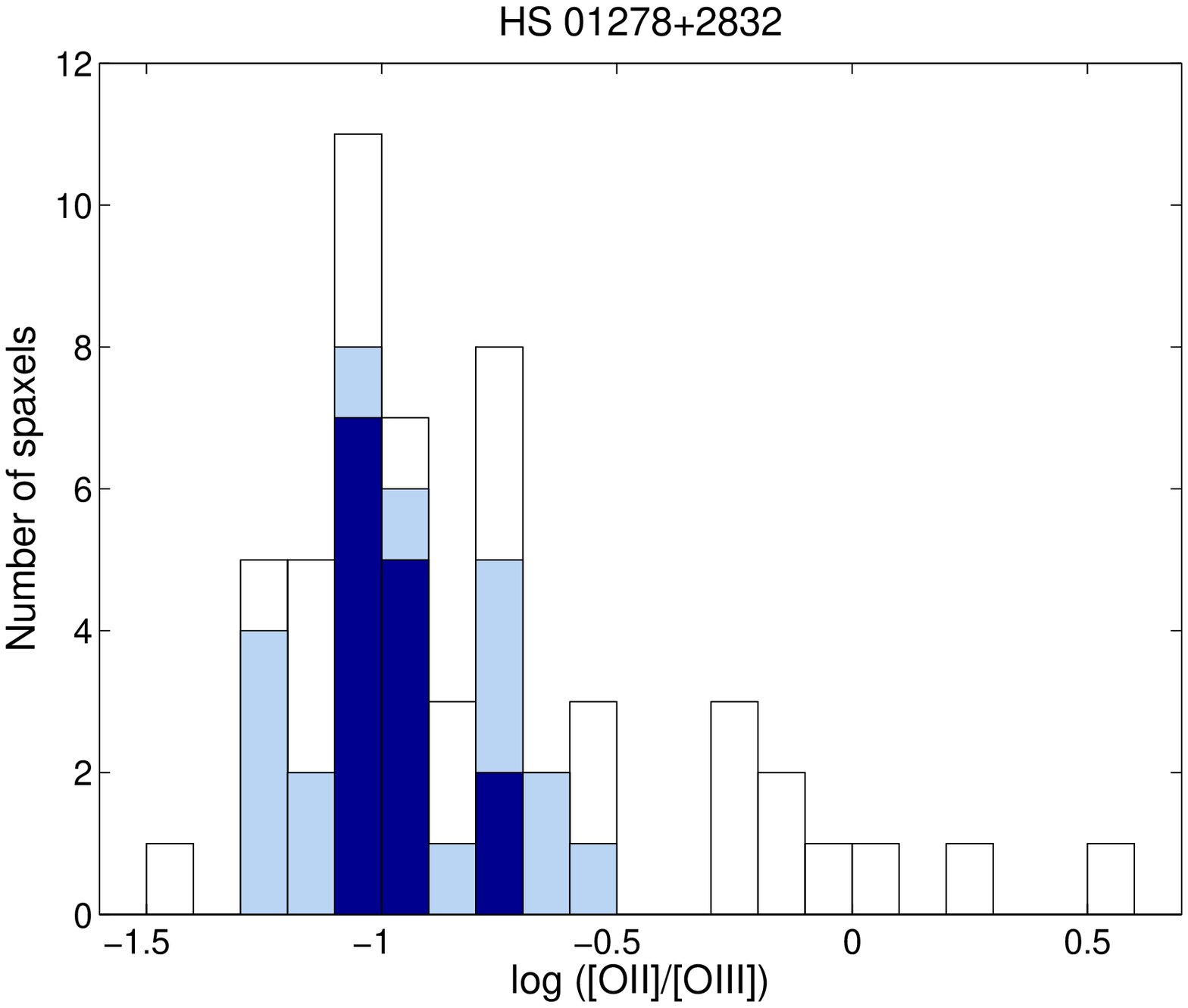}
     \includegraphics[width=6cm,height=5cm,clip=]{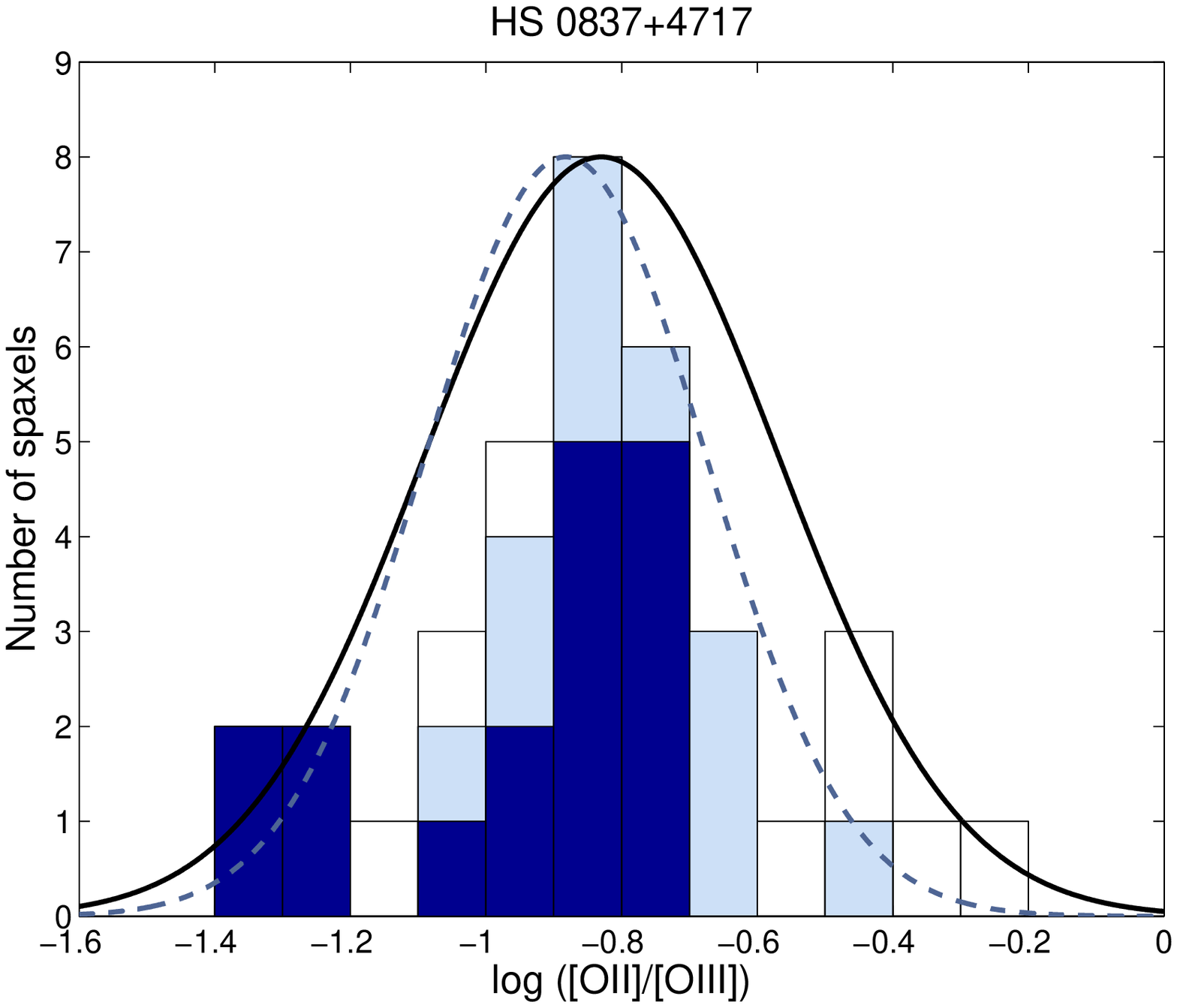}
       \includegraphics[width=6cm,height=5cm,clip=]{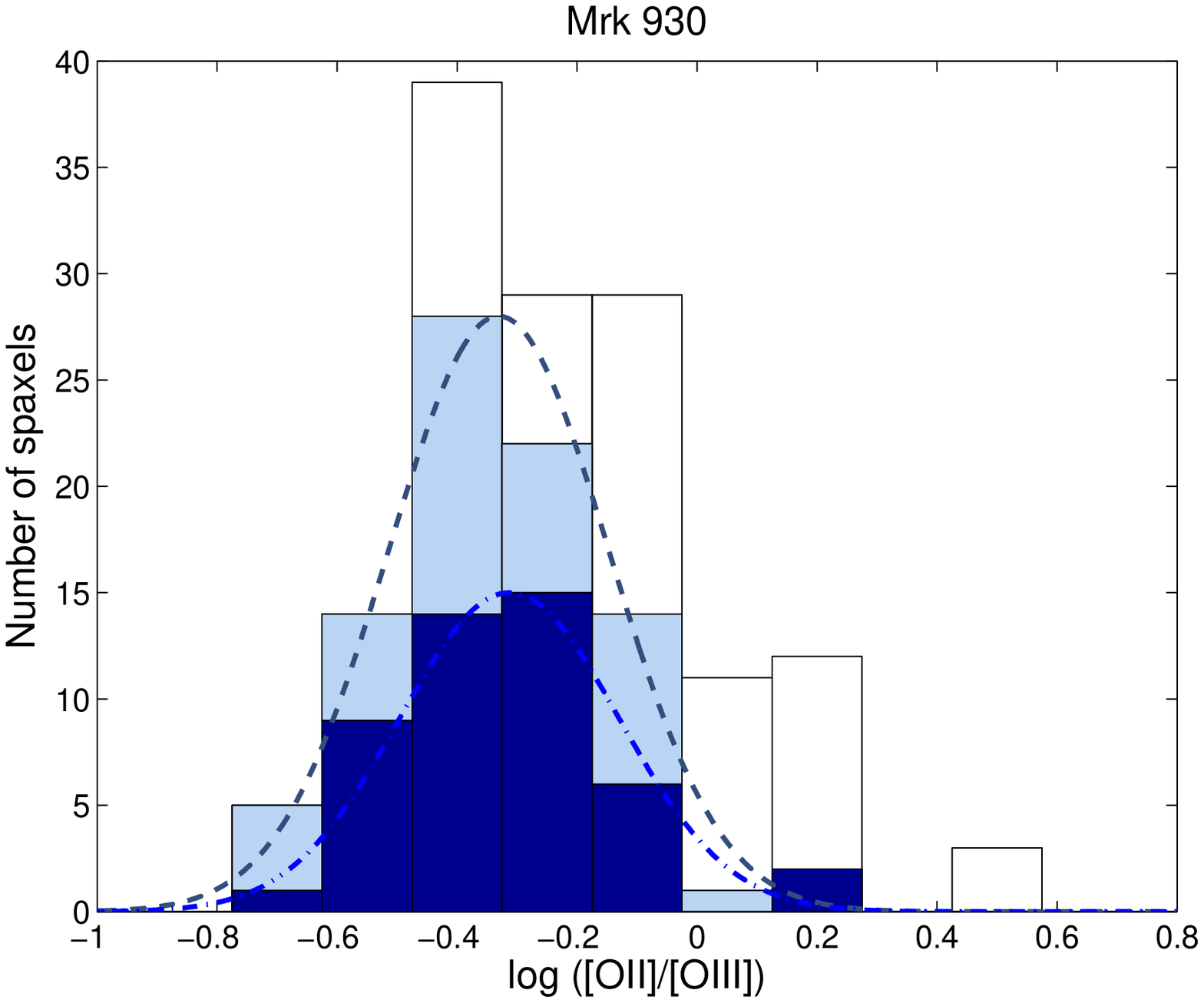}
     \includegraphics[width=6cm,clip=]{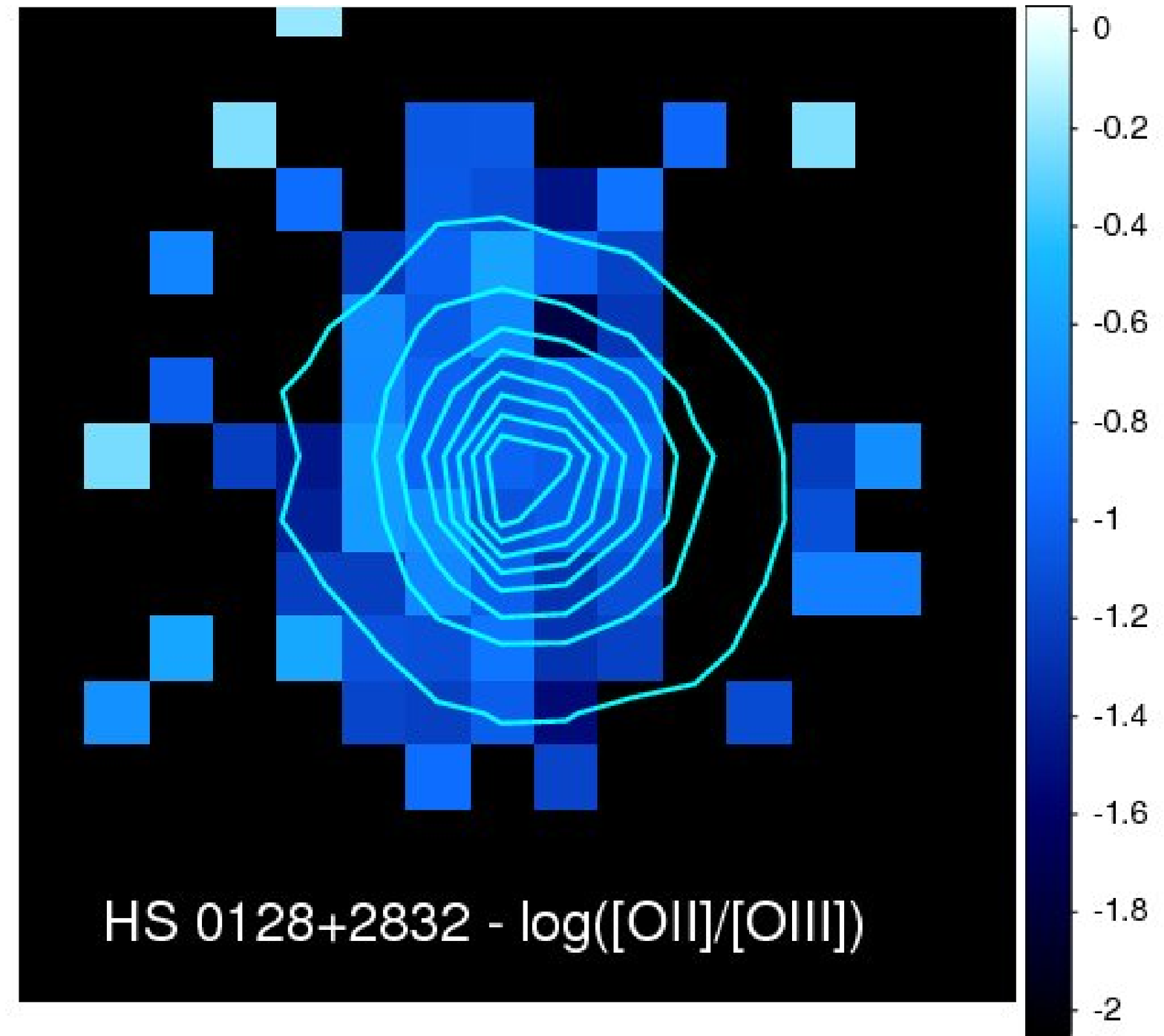}
     \includegraphics[width=6cm,clip=]{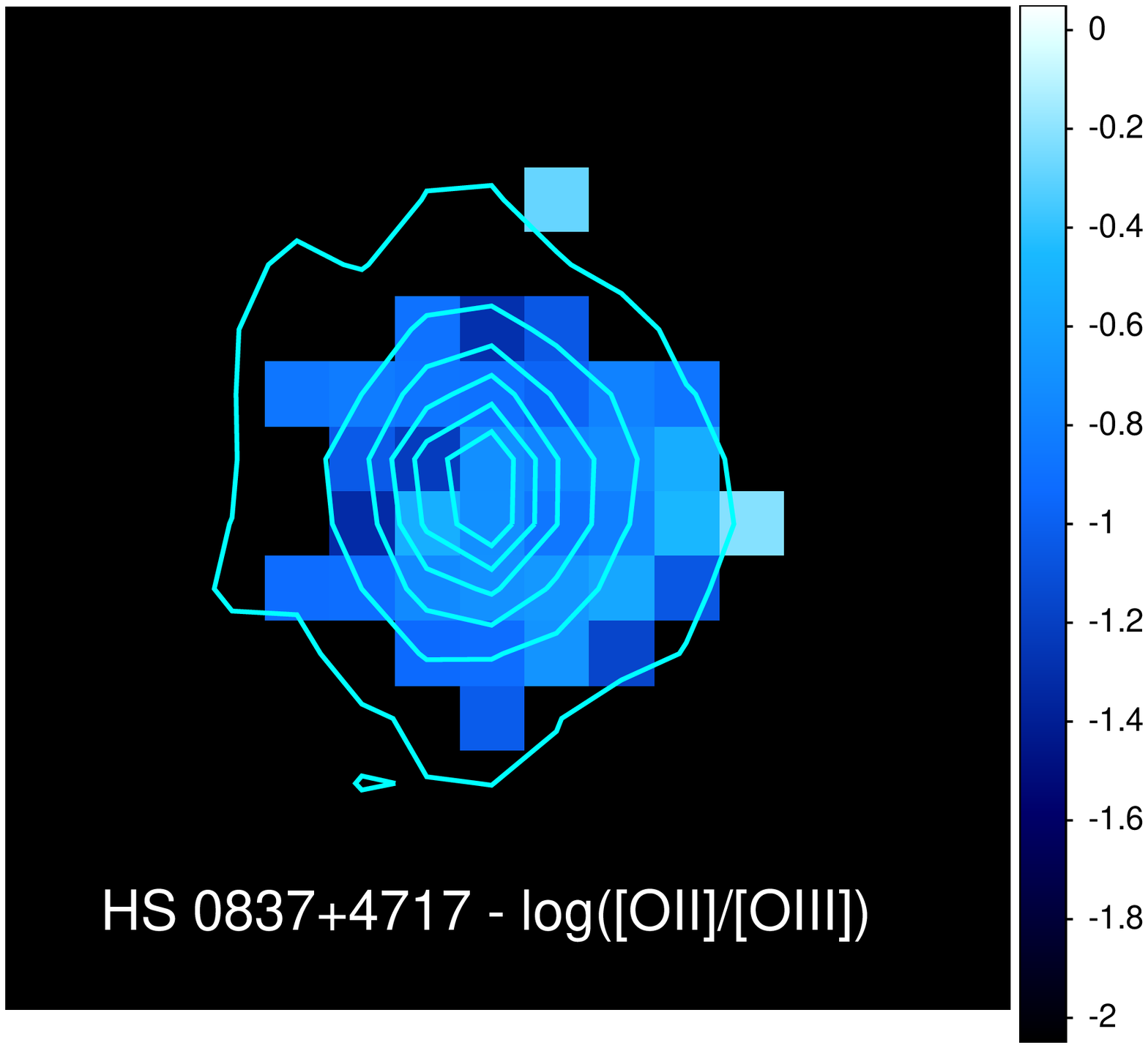}
       \includegraphics[width=6cm,clip=]{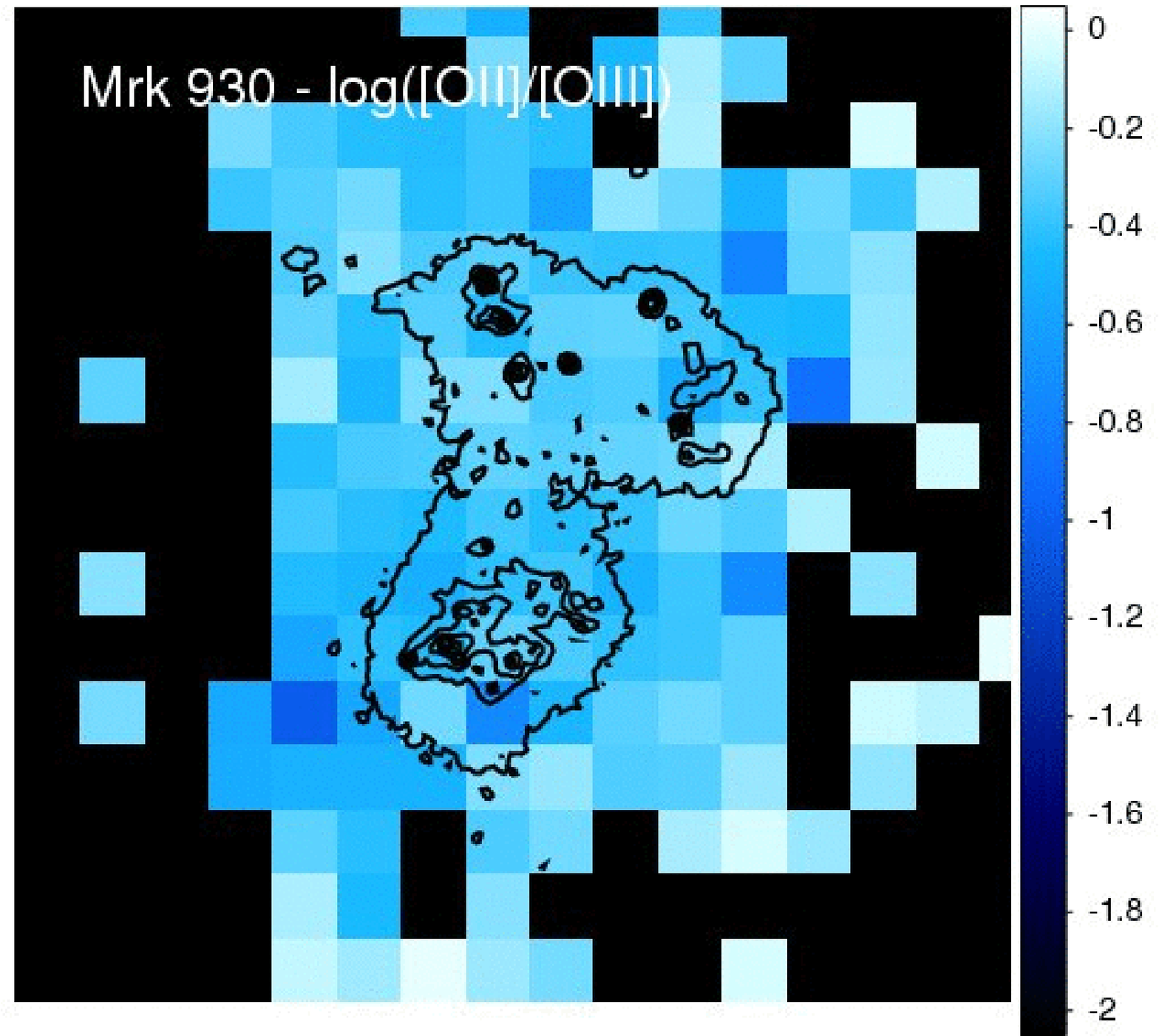}
   
   \caption{Distribution histograms (top) and maps (bottom) of the excitation maps in terms of log([O{\sc ii}]/[O{\sc iii}]) in HS 0128+2832, HS 0837+4717, and Mrk 930 
   from left to right respectively. The solid lines show the same contours as in Fig. 3. Units are in dex.
   The distribution regions are the same defined in Fig. \ref{CHb}.}
    \label{o2o3}
    \end{figure*}


The distribution functions and the maps of the reddening constants, c(H$\beta$), 
are shown in Fig. \ref{CHb} along with the same H$\alpha$ and UV (in the case of Mrk 930)
contours described above. The typical errors of the derived reddening 
constants are 0.2 in the three objects.  A gaussian fits the distribution of the three considered regions
of HS 0837+4717 with a mean value quite similar to that found in the
integrated spectrum [c(H$\beta$) = 0.80]. 
On the contrary, in HS 0128+2832 none of the three distributions are 
fitted by a gaussian function and the derived value for the integrated spectrum
of the inner-most region is much lower (0.14) than the mean of the
distribution for the same region (0.37), what could be indicative of a lower extinction
in the H$\alpha$ peak region. This behaviour also appears in the histogram of Region 1
for this galaxy, where a double-peak structure can be seen.
In Mrk 930 no gaussian fits any of the three distributions,
which are centeres at low values of the reddening, although
can be  seen some spaxels with reddening c(H$\beta$) $>$ 1.0 
breaking the normal distribution in all the regions.  By visual inspection of
the corresponding c(H$\beta$) map, these appear above all
in the south-west part of the galaxy, resembling somehow the ring-like structure
shown by Adamo et al. (2011).

\subsection{Ionization structure and excitation}

Diagnostic diagrams (BPT, Baldwin, Philips \& Terlevich, 1981) are frequently used to
distinguish between different ionization sources of the surrounding gas emitting the
detected bright emission lines.  Active galactic nuclei (AGNs), shocks or massive stars can be selected,
by comparing the ratios of low excitation-to-Balmer emission lines, like [N{\sc ii}] at
6584 {\AA}, [S{\sc ii}] at 6717 {\AA}, 6731 {\AA} or [O{\sc i}] at 6300 {\AA},
to H$\alpha$ in relation to a high excitation-to-Balmer emission line ratios, like [O{\sc iii}] at 5007 {\AA}
to H$\beta$.  In all studies about our three galaxy sample, their integrated spectra show 
a clear star-forming ionization origin in all these three diagrams. Nevertheless, the analysis of the
individual spaxels in our observations reveals unexpected behaviours when we
look at the [N{\sc ii}]/H$\alpha$ vs. [O{\sc iii}]/H$\beta$ diagram, shown in Fig. \ref{n2o3}.
In order to be compared with our observations, we have taken as a reference the theoretical curve
given by Kewley et al. (2001, in red solid line) and the empirical fitting described by Kauffmann et al. (2003),
in dashed red line). Following Kewley et al. (2006), the region between these two lines could correspond
to galaxies with a composite source of ionization, partially from star-formation and partially from X-rays, while 
the area under the curves could correspond to pure star-forming regions and the area over the curves to pure
AGNs. LINERs would lie in the low region at right. As can be seen, for 
all the three objects, most of the spaxels in Region 1 (dark blue squares) and
the integrated spectrum (represented by a large stripped square) lie in the SF region, although the
points are close to the Kewley's curve for HS 0837+4717. Nevertheless, 
a large fraction of the spaxels in Regions 2 and 3 ({\em i.e.} those in the outer regions of the
galaxies) lie both in the composite and AGN region in HS 0128+2832 and
Mrk 930. In the case of HS 0837+4717, most of these spaxels lie in the AGN region.

The excitation of the ionized gas is studied by using the 
ratio of emission-line fluxes of [O{\sc ii}] at 3727 {\AA} and [O{\sc iii}] at
4959, 5007 {\AA}.
Although this ratio has a certain dependence on the effective equivalent temperature
of the ionizing cluster and on the metallicity of the gas, it depends mainly on 
the ionization parameter ({\em i.e.} the ratio between the number of ionizing photons and particles,
D\'\i az, 1998). In Fig. \ref{o2o3} we see the maps of the logarithm of this
ratio and their corresponding distributions. 
The typical errors of this ratio range from 0.15 dex
in HS 0128+2832 and HS 0837+4717 to only 0.03 dex in Mrk 930.
This ratio tends to be
higher as a larger number of spaxels in the outer regions is considered, consistently
with a larger distance to the ionizing central stellar cluster. In the case
of Mrk 930 the gaussian fits are found in the inner regions. On the contrary, in HS 0837+4717,
the gaussian fits only the outer regions, although this is possibly due
to a lower number of spaxels ({\em i.e.}
for all the spaxels in the field of view with the required S/N).
For the three objects, the values in the integrated spectra are  lower than the averages found
in the histograms of the different regions, especially in HS 0837+4717.
In the case of Mrk 930, almost no difference has been found between the excitation of
the southern and northern knots which, besides, are in good
agreement with the values found for the gaussian fit of the inner-most region.

\begin{figure*}
\centering
     \includegraphics[width=6cm,height=5cm,clip=]{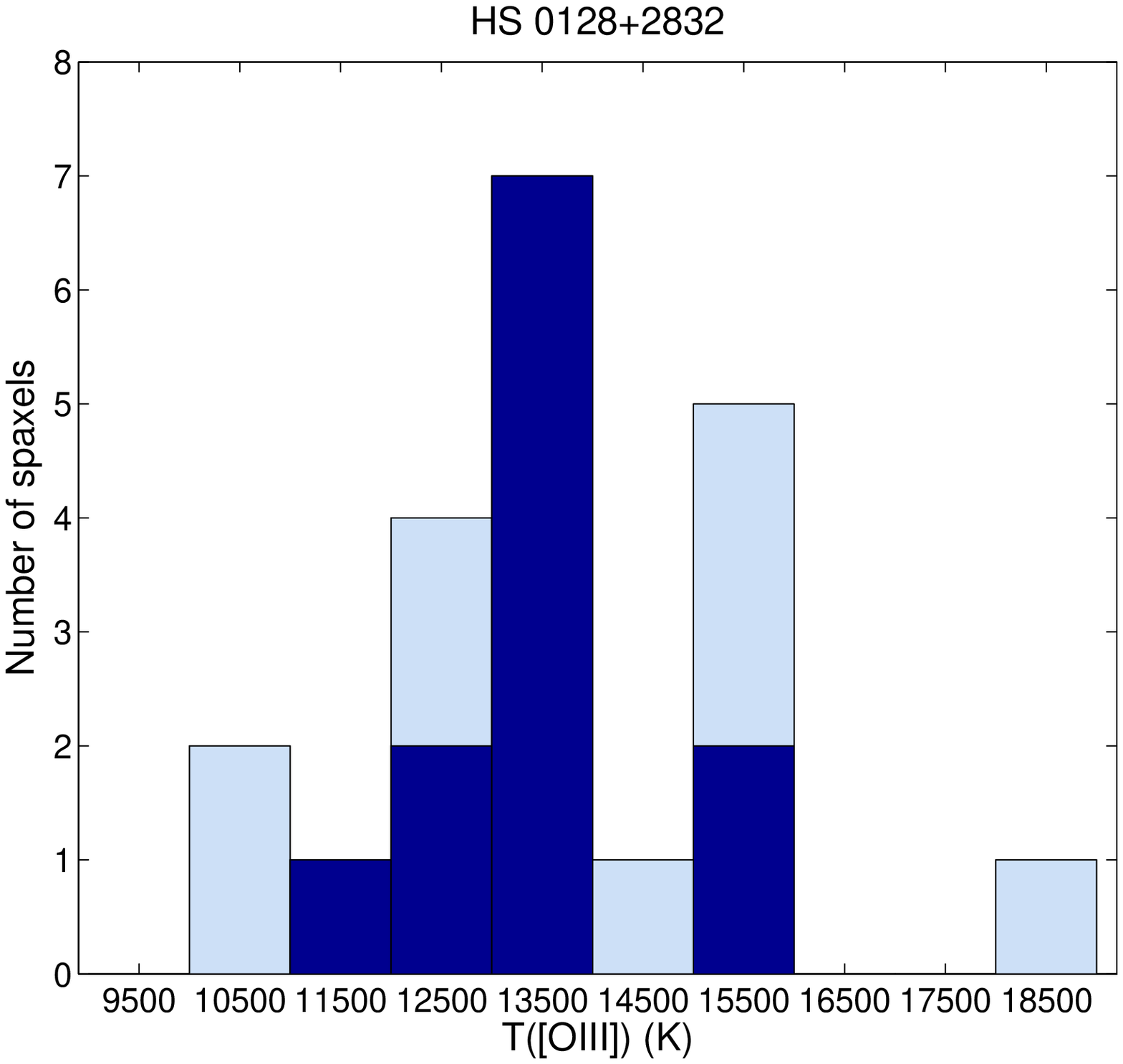}
     \includegraphics[width=6cm,height=5cm,clip=]{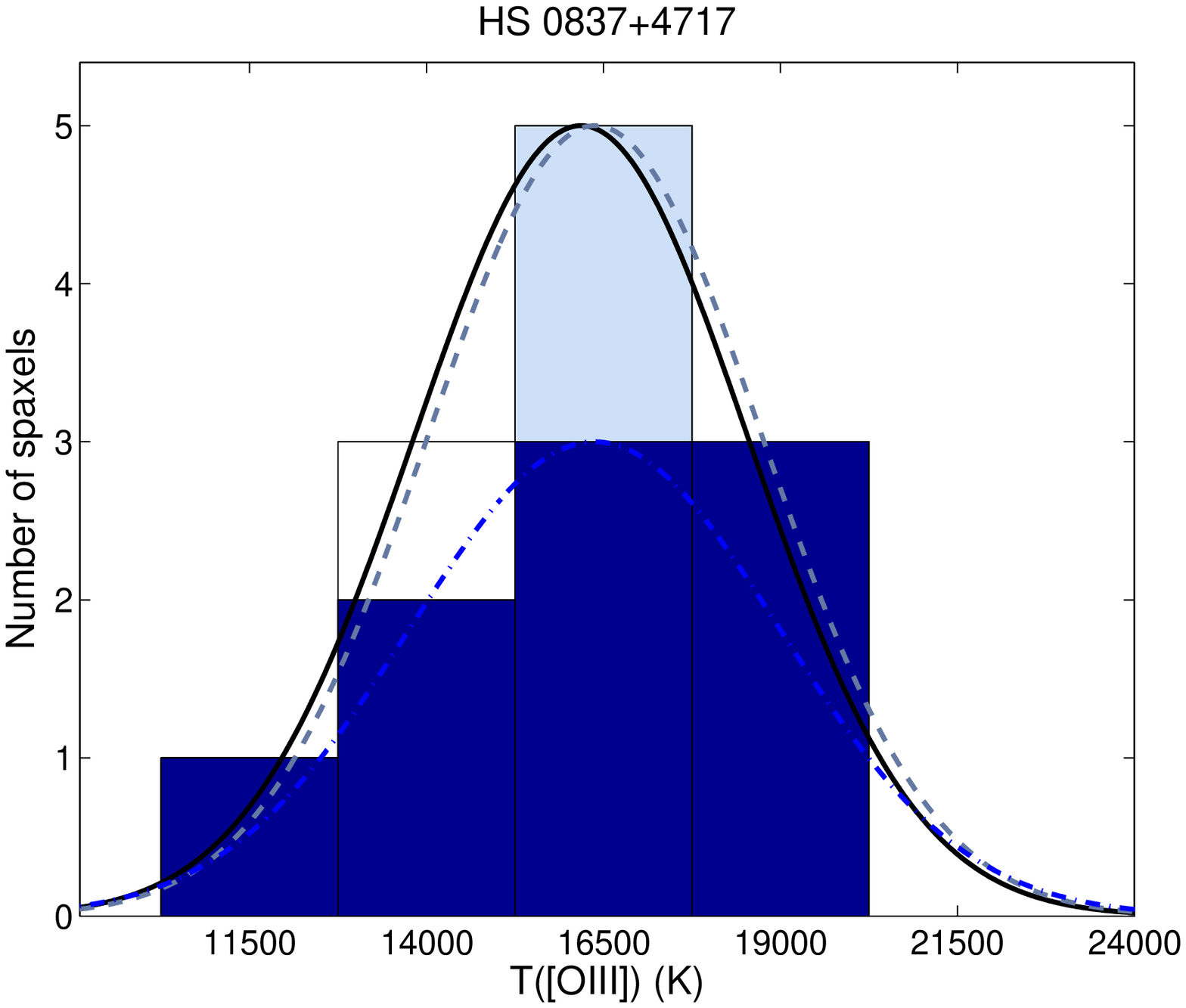}
     \includegraphics[width=6cm,height=5cm,clip=]{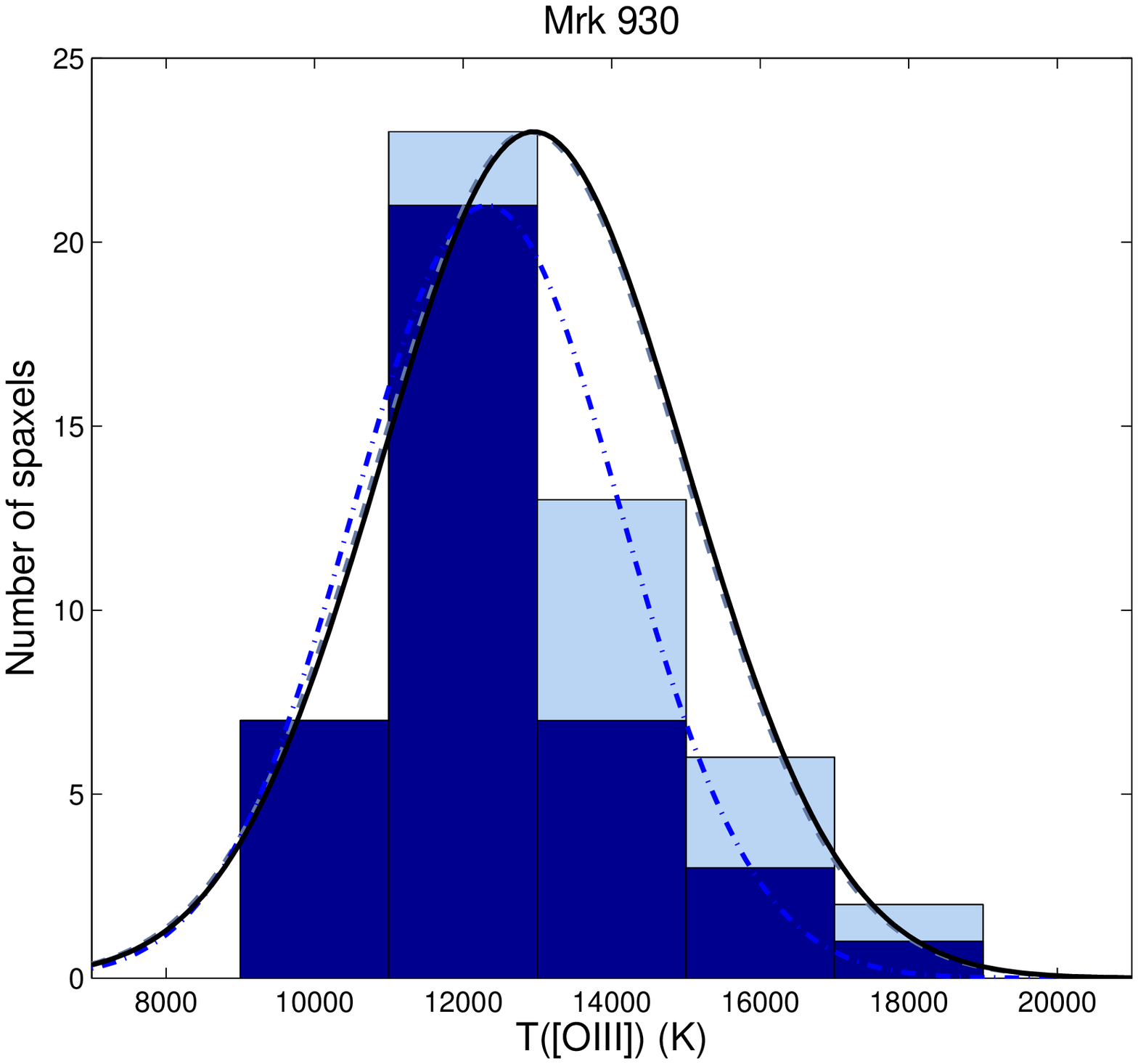}
     \includegraphics[width=6cm,clip=]{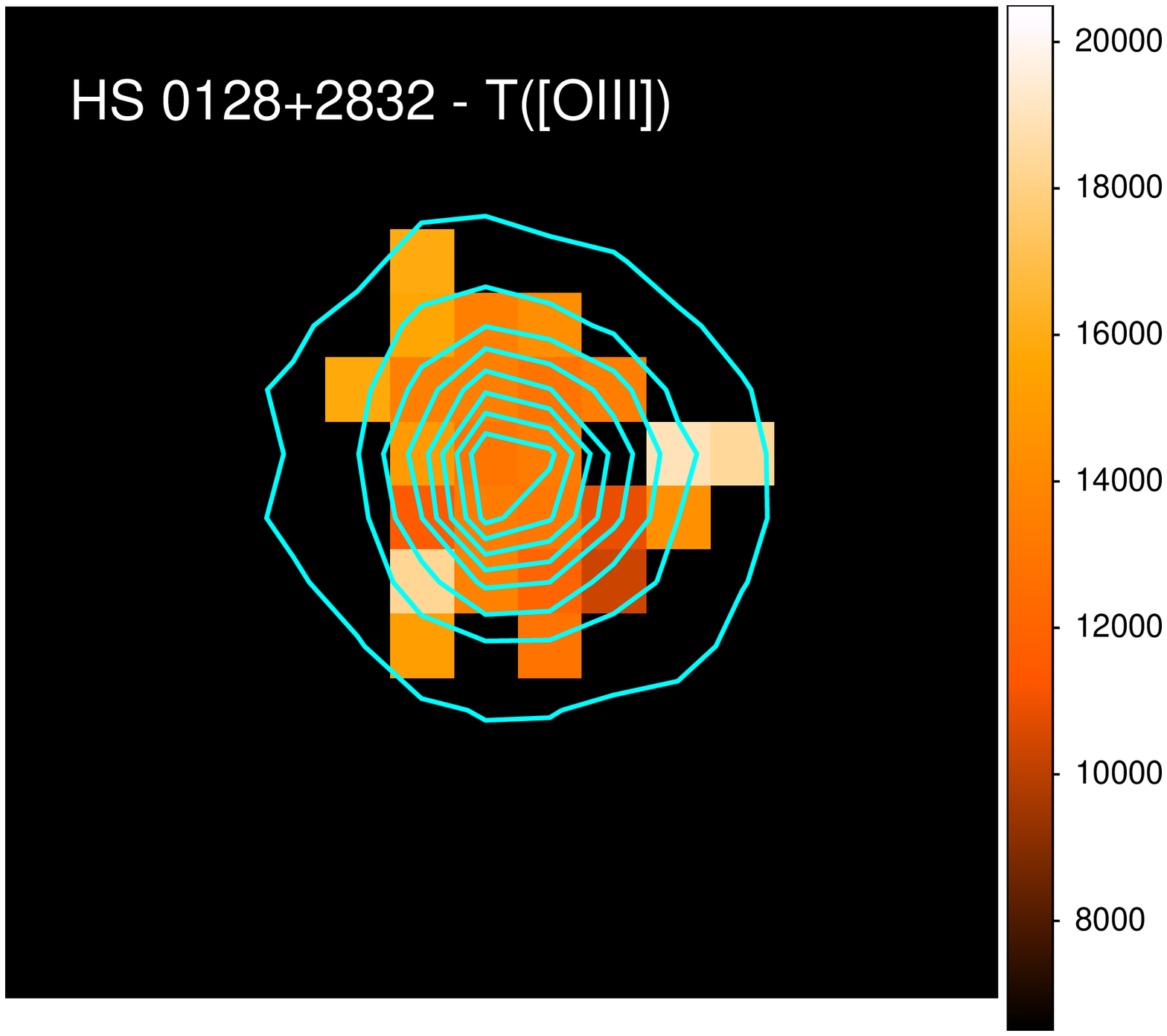}
     \includegraphics[width=6cm,clip=]{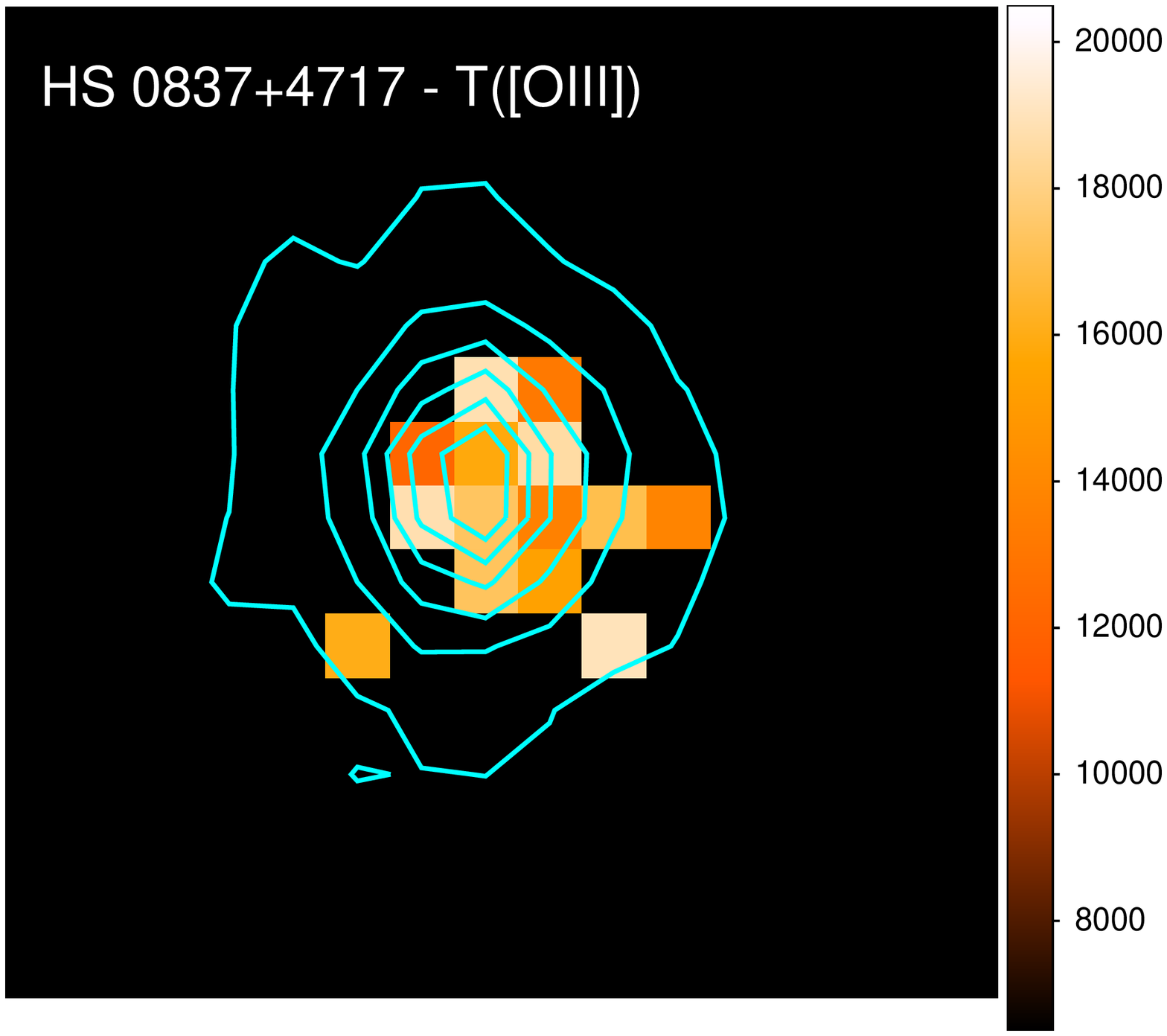}
     \includegraphics[width=6cm,clip=]{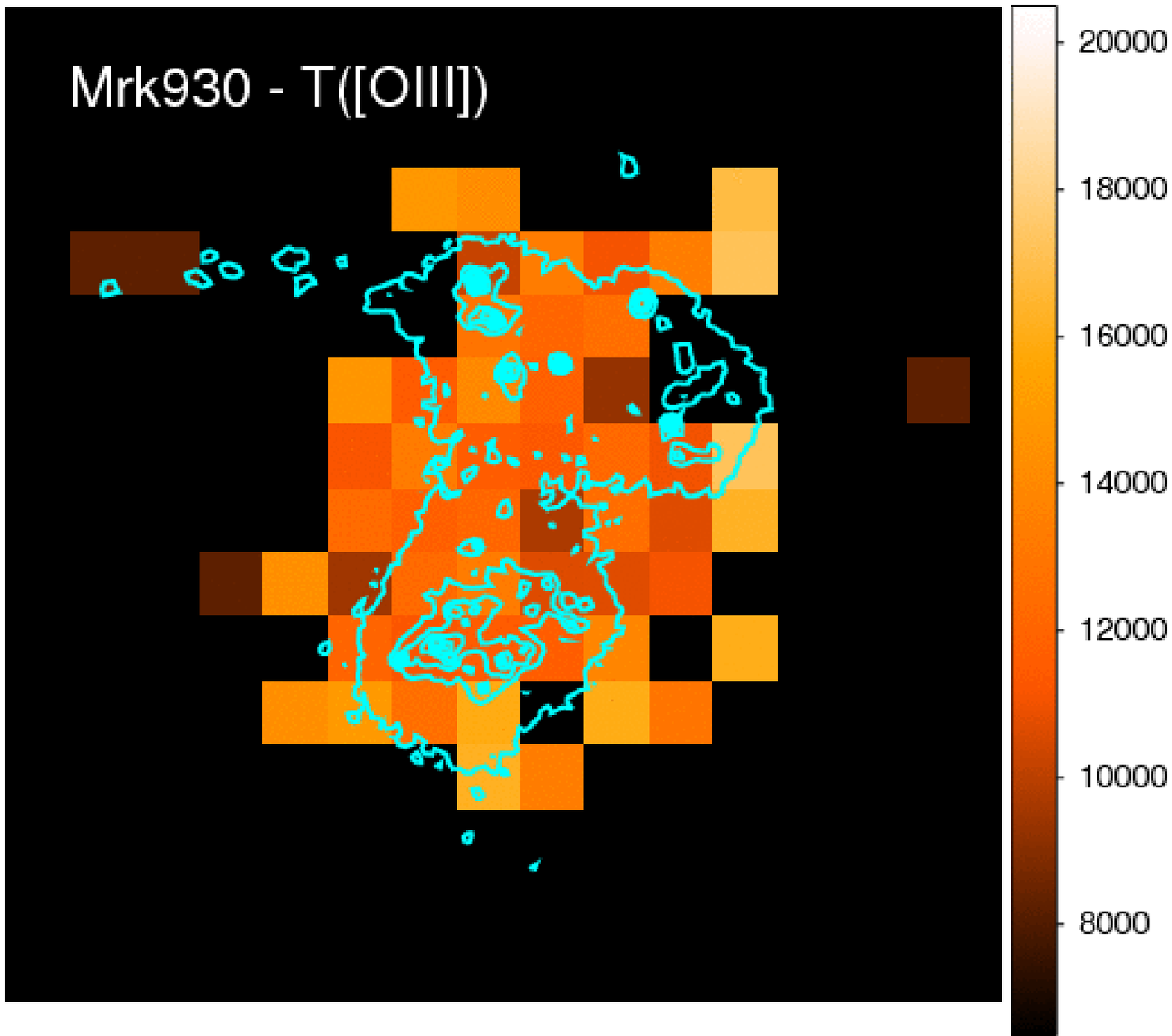}
   
   \caption{Distribution histograms (top) and maps (bottom) of the electron temperature of [O{\sc iii}] in HS 0128+2832, HS 0837+4717, and Mrk 930 from left to right respectively. The solid lines show the same contours as in Fig 3. Units are in K. The distribution regions are the same defined in Fig. \ref{CHb}. }
    \label{to3}
    \end{figure*}



\begin{figure}[h!]
\centering
       \includegraphics[width=7cm,clip=]{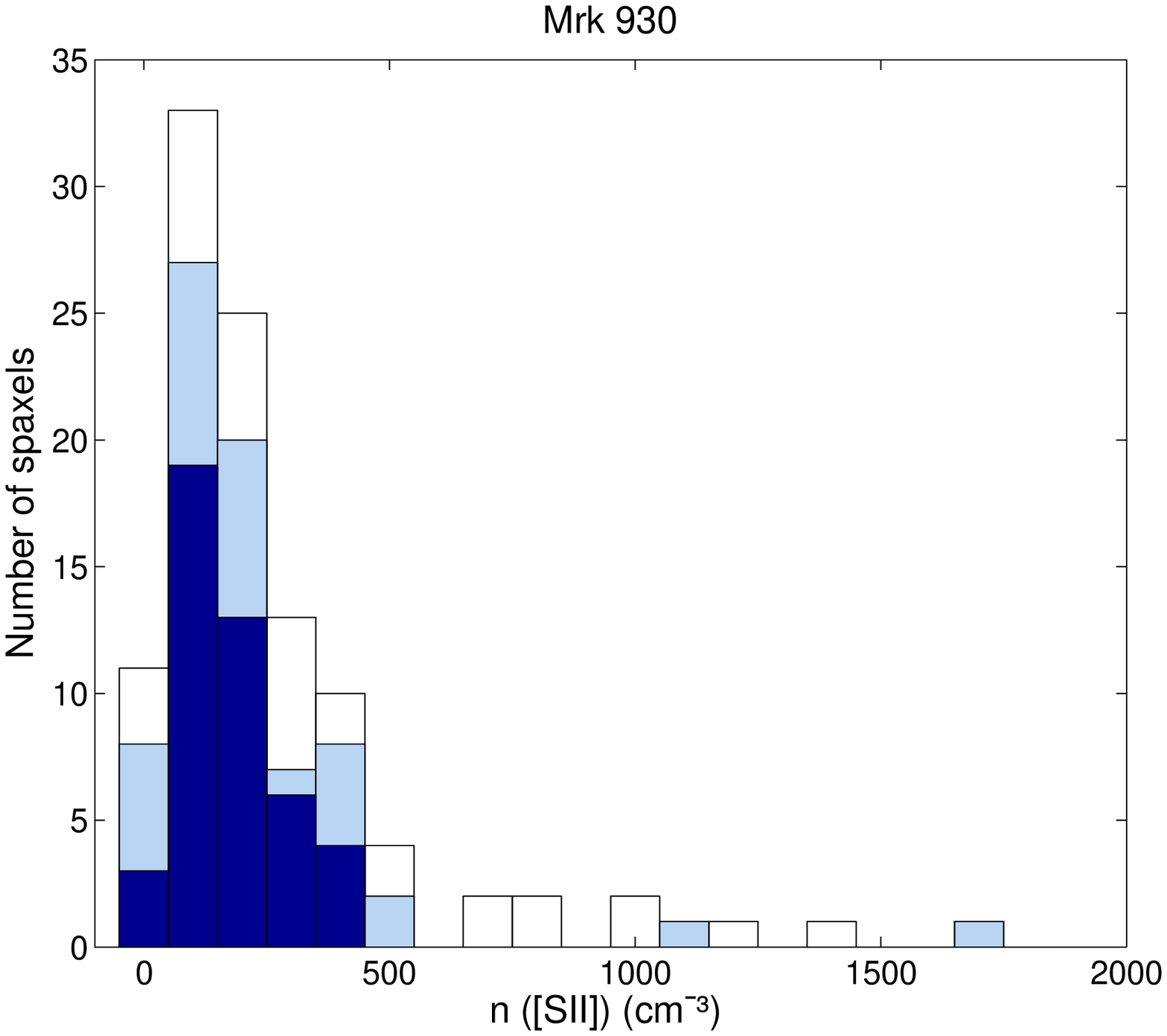}
       \includegraphics[width=6cm,clip=]{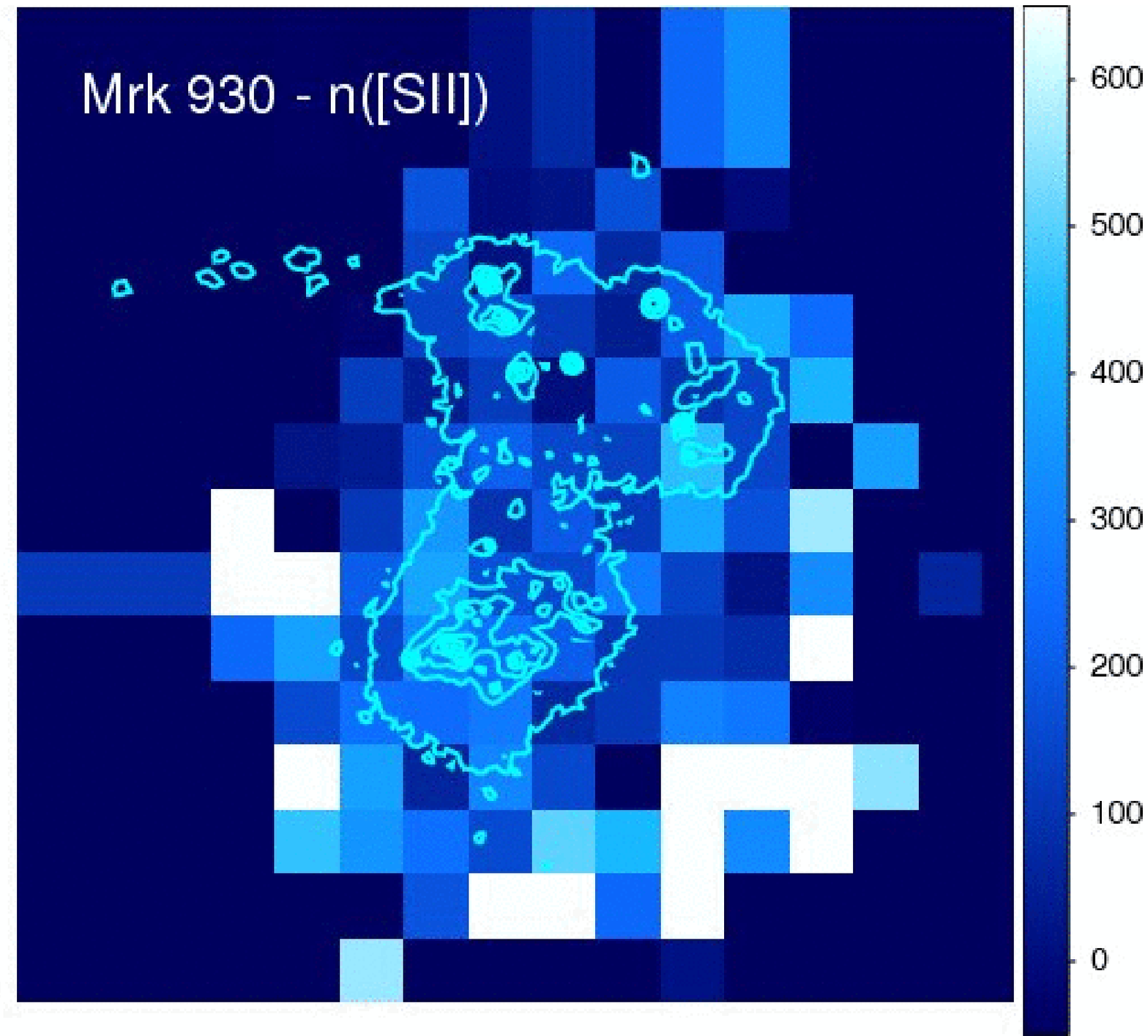}
   
   \caption{Distribution histograms (top) and map (bottom) of the electron density from [S{\sc ii}] in Mrk 930 
   The solid lines show the same contours as in Fig 3. Units are in cm$^{-3}$. The distribution regions are the same defined in Fig. \ref{CHb}}
    \label{nS2}
    \end{figure}


\subsection{Electron temperature and density}

Due to the low metallicity content of these galaxies, the cooling of the gas is not
very efficient and the electron temperature is high. 
Therefore, the electron temperature of [O{\sc iii}] has been estimated
in a large number of spaxels for our three galaxies by means of the
flux ratio between the strong nebular emission-lines [O{\sc iii}]
4959, 5007 {\AA} and the auroral weaker line at 4363 {\AA}. In order to
calculate the electron temperature t([O{\sc iii}]), we use the
expressions from H\"agele et al. (2006, 2008) that are based on the
TEMDEN task under the IRAF nebular package. The distribution functions and the maps of
this temperature in the three observed galaxies are shown in
Fig. \ref{to3}. The average error in the individual spaxels are
much higher for HS 0128+2832, where it reaches 2000 K. For HS 0837+4717 and Mrk 930 the uncertainty is $\sim$ 700 K.
A gaussian fit is found for the temperature values of all the regions in HS 0837+4717 
and in Regions 1 and 2 in Mrk 930, while no gaussians have
been found for any region in HS 0128+2818,
although for this galaxy the mean value of t([O{\sc iii}]) (13200 K) agrees
within the errors with the value measured in the integrated spectrum.  For
HS 0837+4717, an agreement is also found between the electron temperature of the integrated spectrum
(16900 K) and the mean values derived in the three studied regions. For Mrk 930 the electron temperatures derived in
the integrated spectra of both southern and northern knots are slightly lower
(12000 and 11000 K, respectively), but
still within the dispersion of the corresponding gaussian.

We derived the electron densities from the ratio between the [S{\sc ii}]
emission lines at 6717,6731 {\AA} using the same TEMDEN routine.
Only for Mrk 930 it has been
possible to measure these lines with enough confidence in a large
fraction of spaxels.  In the case of HS 0128+2832, only an upper limit
to the density can be provided, although a reliable estimate of the
electron density has been made in the analysis of the integrated
spectrum of the brightest region of this galaxy, but the errors are
too high to achieve any precise determination. In the case of HS
0837+4717, the redshift of this galaxy placed the [SII] lines outside of the spectral range covered in this work.
The distribution function and the map of the
electron density in Mrk 930 are shown in
Fig. \ref{nS2}. The average error in the spaxels is 50 cm$^{-3}$. 
A gaussian fit has not been found with enough confidence level for any of
the analyzed regions. In fact, over the outer southern part of the galaxy, the electron
density map displays high values, reaching up to 1700 cm$^{-3}$ that
is still below the critical density limit for collisional de-excitation. This makes the average density
of all distributions to be higher in the outer regions and the density measured
in the integrated spectrum of the southern knot to be slightly higher than in
the northern knots.


\begin{figure*}[t]
\centering
     \includegraphics[width=6cm,height=5cm,clip=]{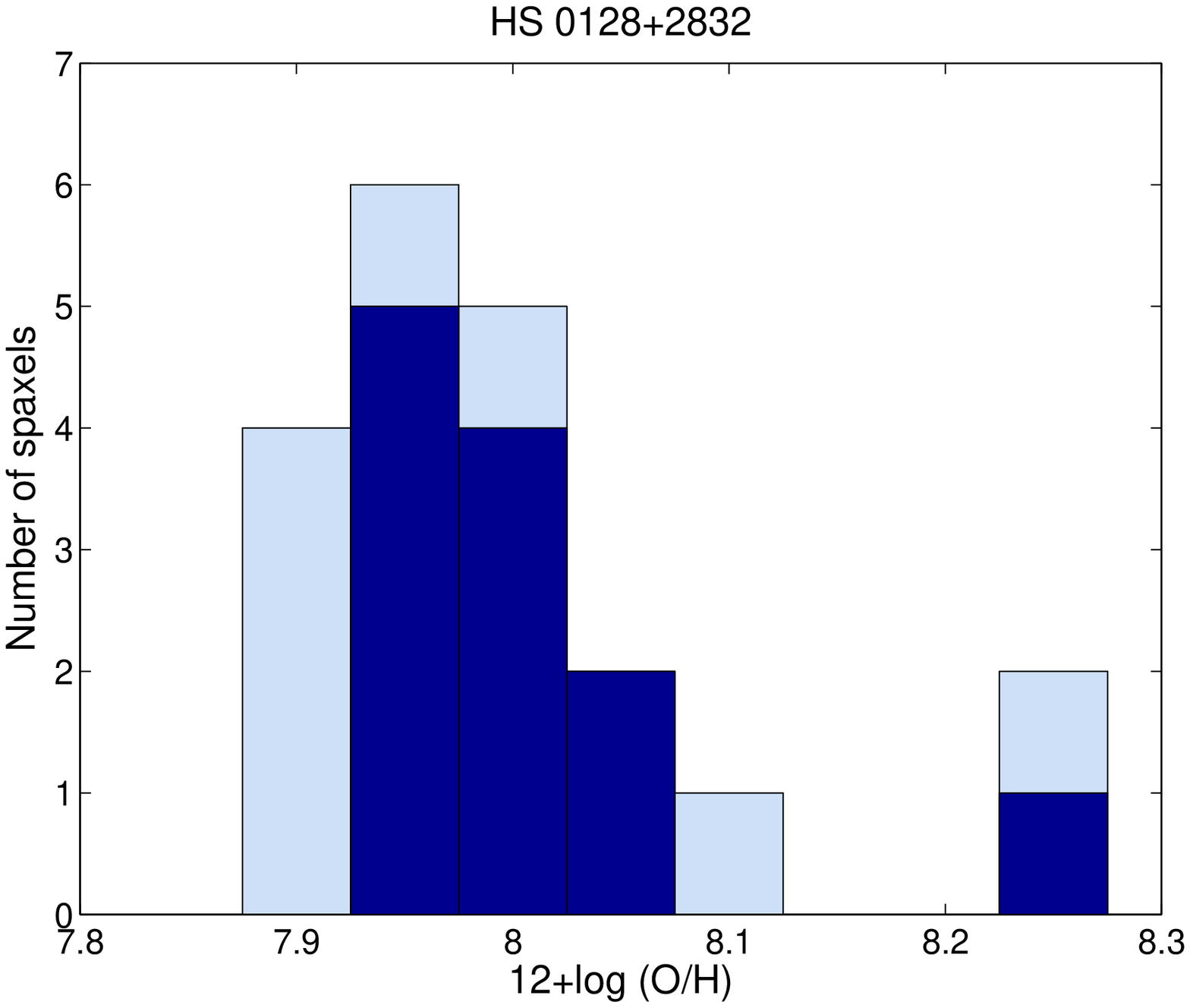}
     \includegraphics[width=6cm,height=5cm,clip=]{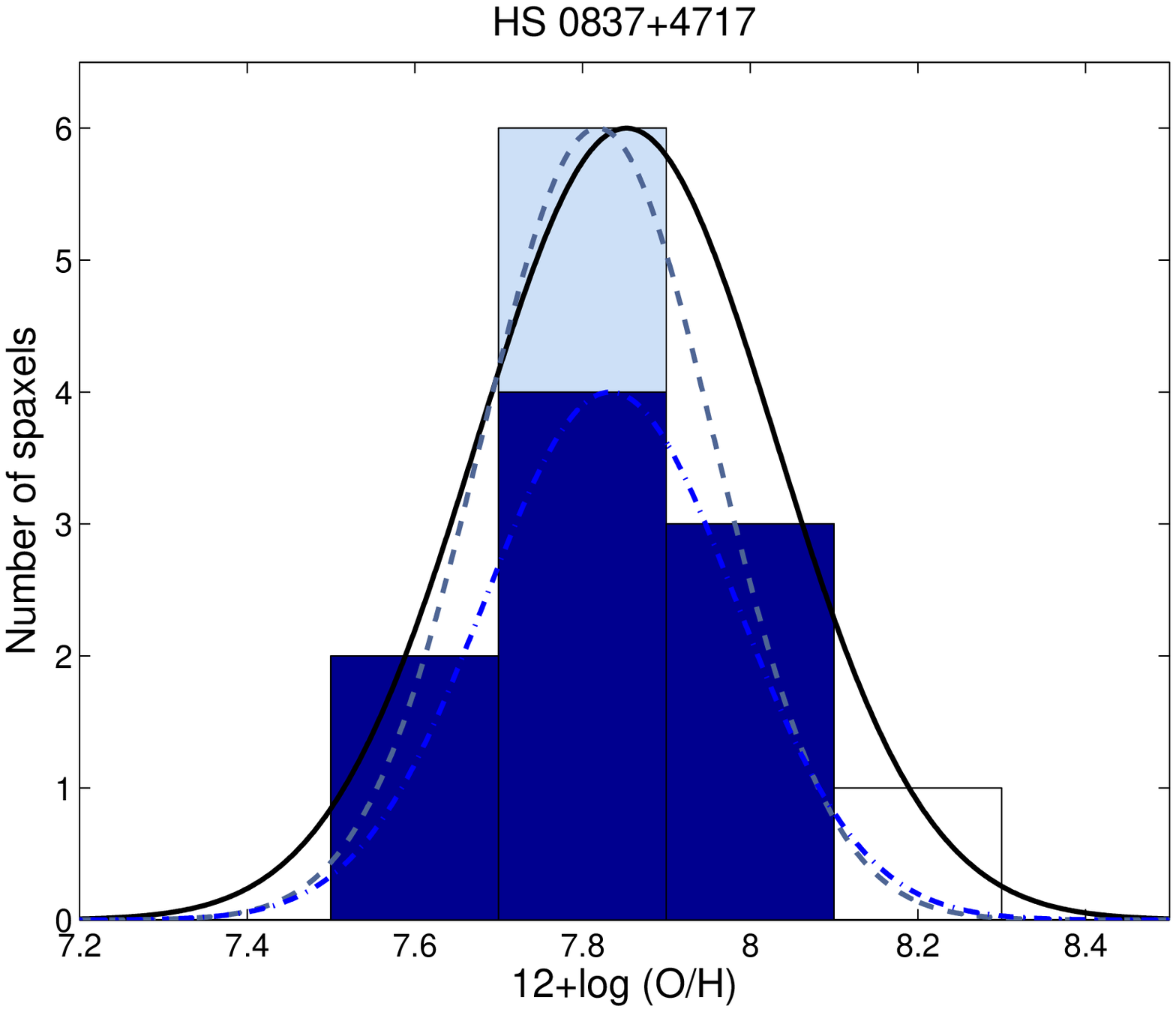}
     \includegraphics[width=6cm,height=5cm,clip=]{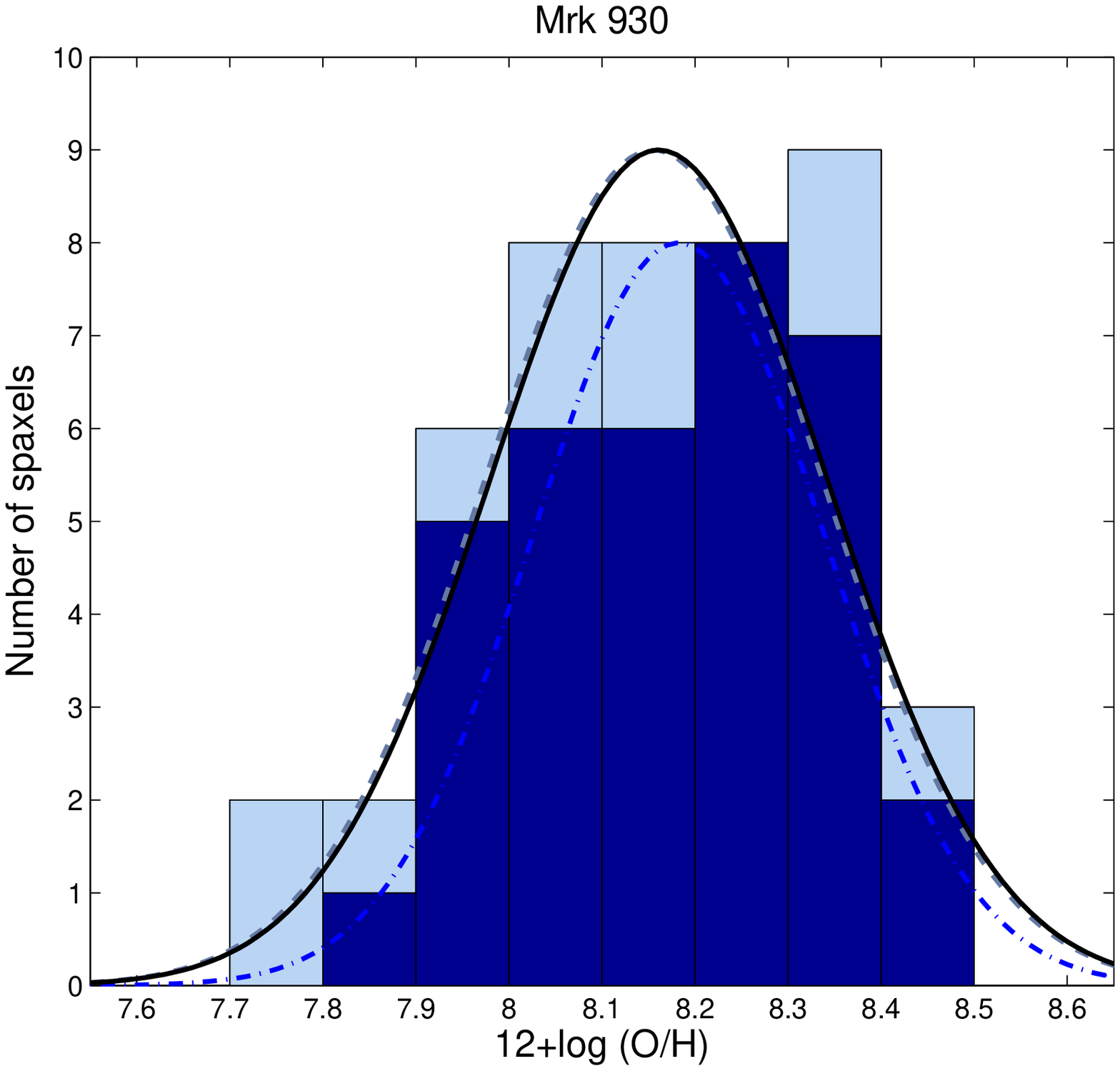}
     \includegraphics[width=6cm,clip=]{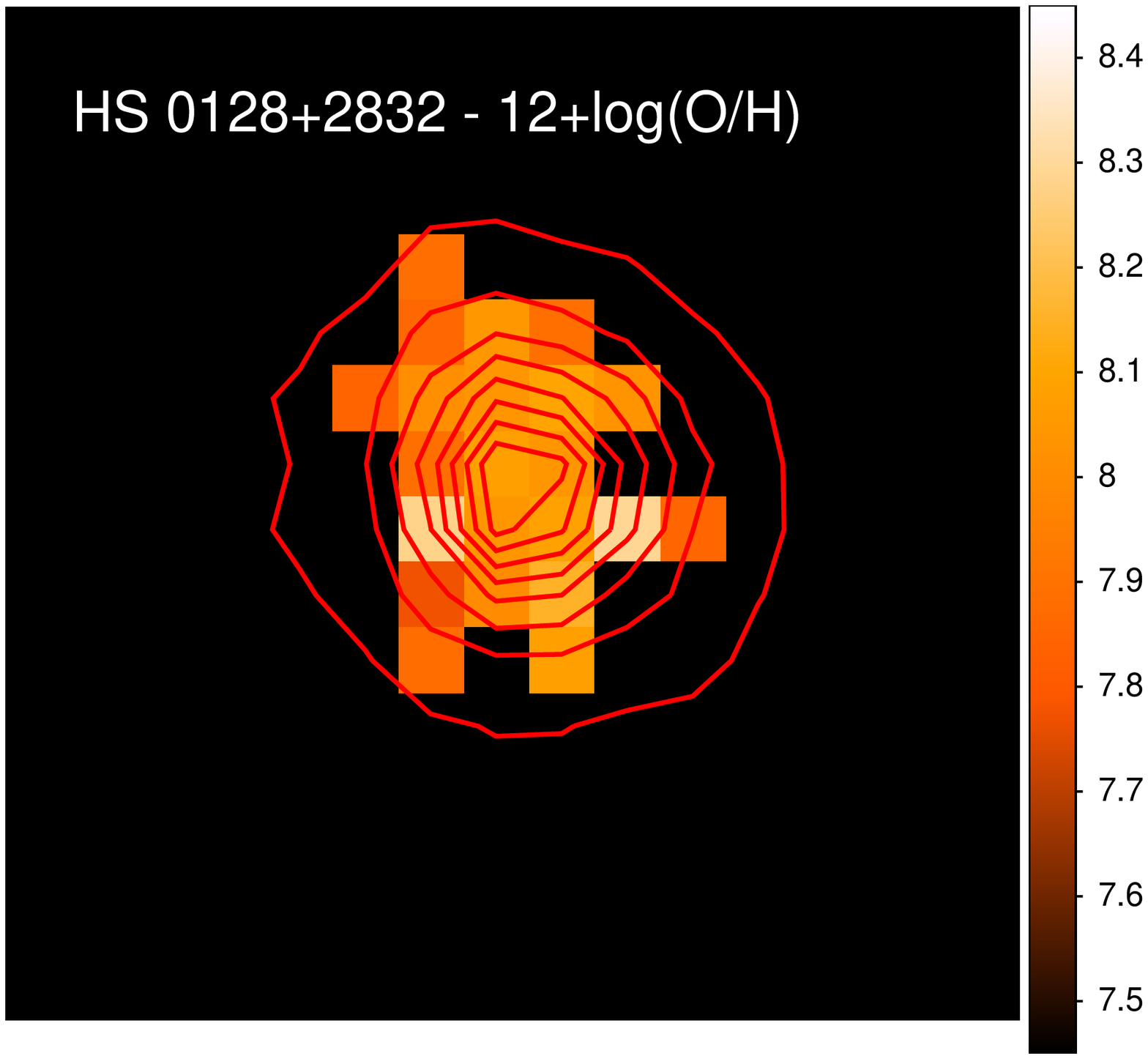}
     \includegraphics[width=6cm,clip=]{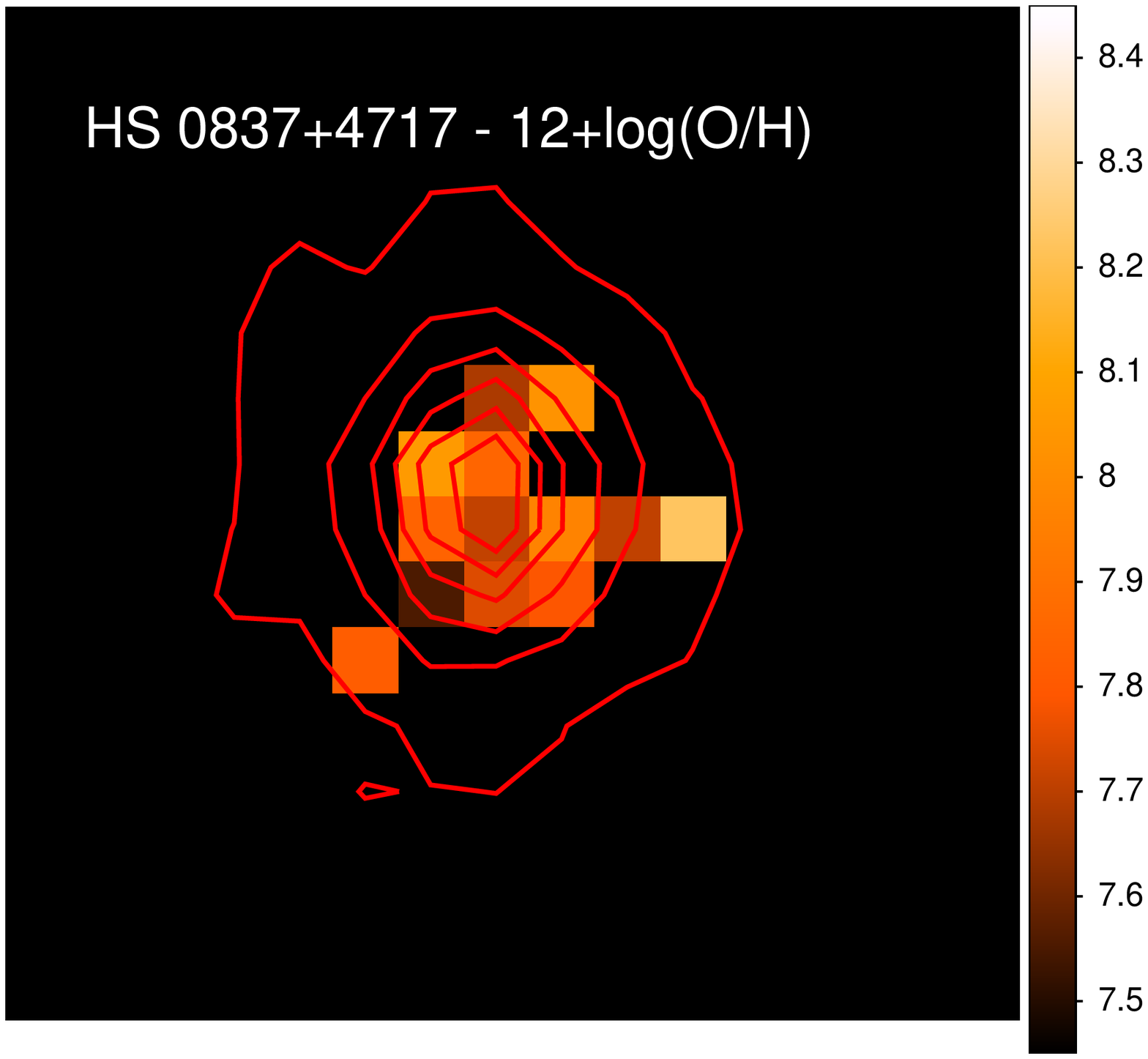}
     \includegraphics[width=6cm,clip=]{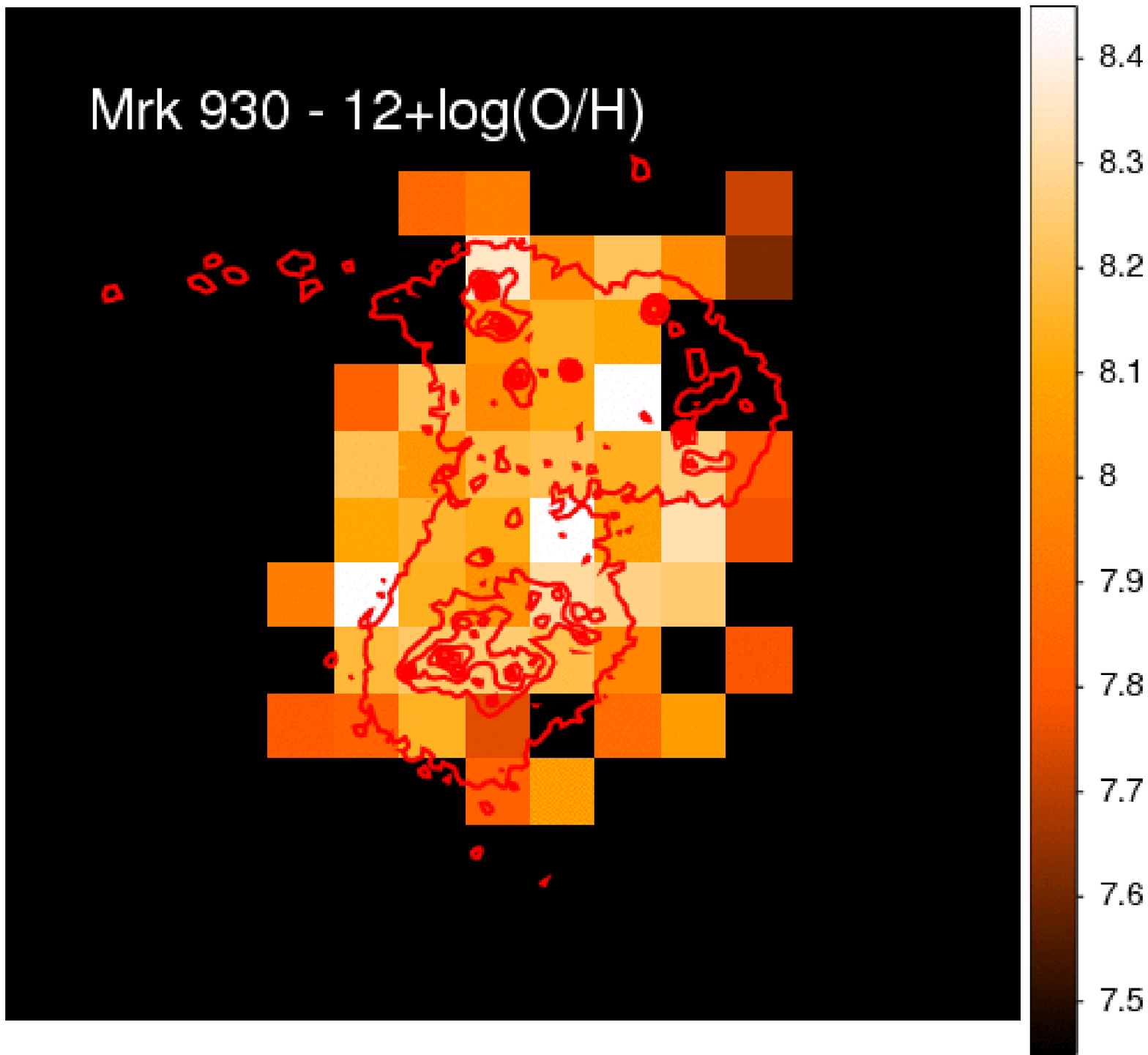}
   
   \caption{Distribution histograms (top) and maps (bottom) of the total oxygen abundance, in terms of 12+log(O/H) calculated using the direct method, in HS 0128+2832, HS 0837+4717, and Mrk 930 from left to right, respectively. The solid lines show the same contours as in Fig. 3. Units are in dex. 
   The distribution regions are the same defined in Fig. \ref{CHb}.}
    \label{oh}
    \end{figure*}



\begin{figure*}
\centering
		    \includegraphics[width=6cm,height=5cm,clip=]{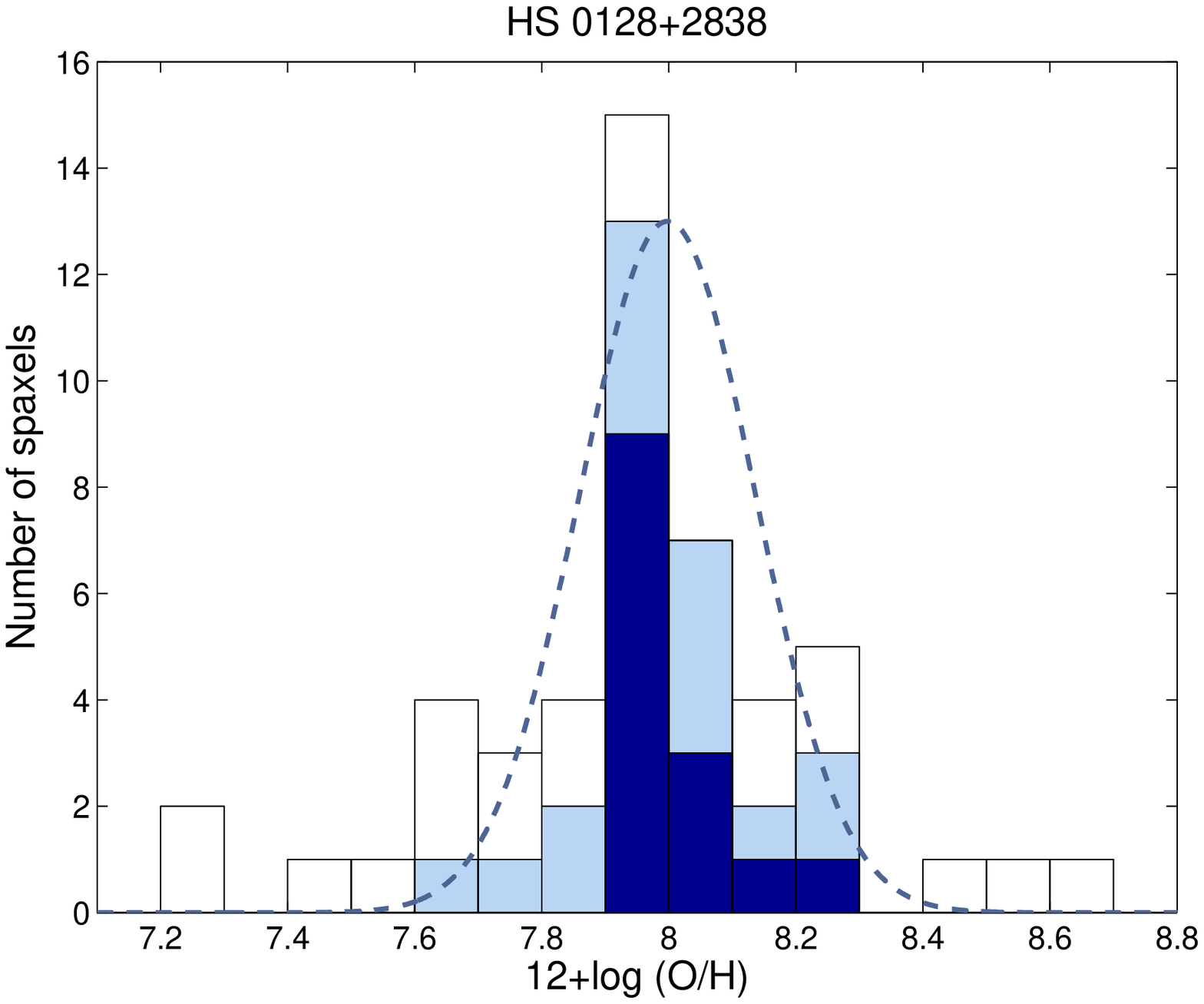}
     \includegraphics[width=6cm,height=5cm,clip=]{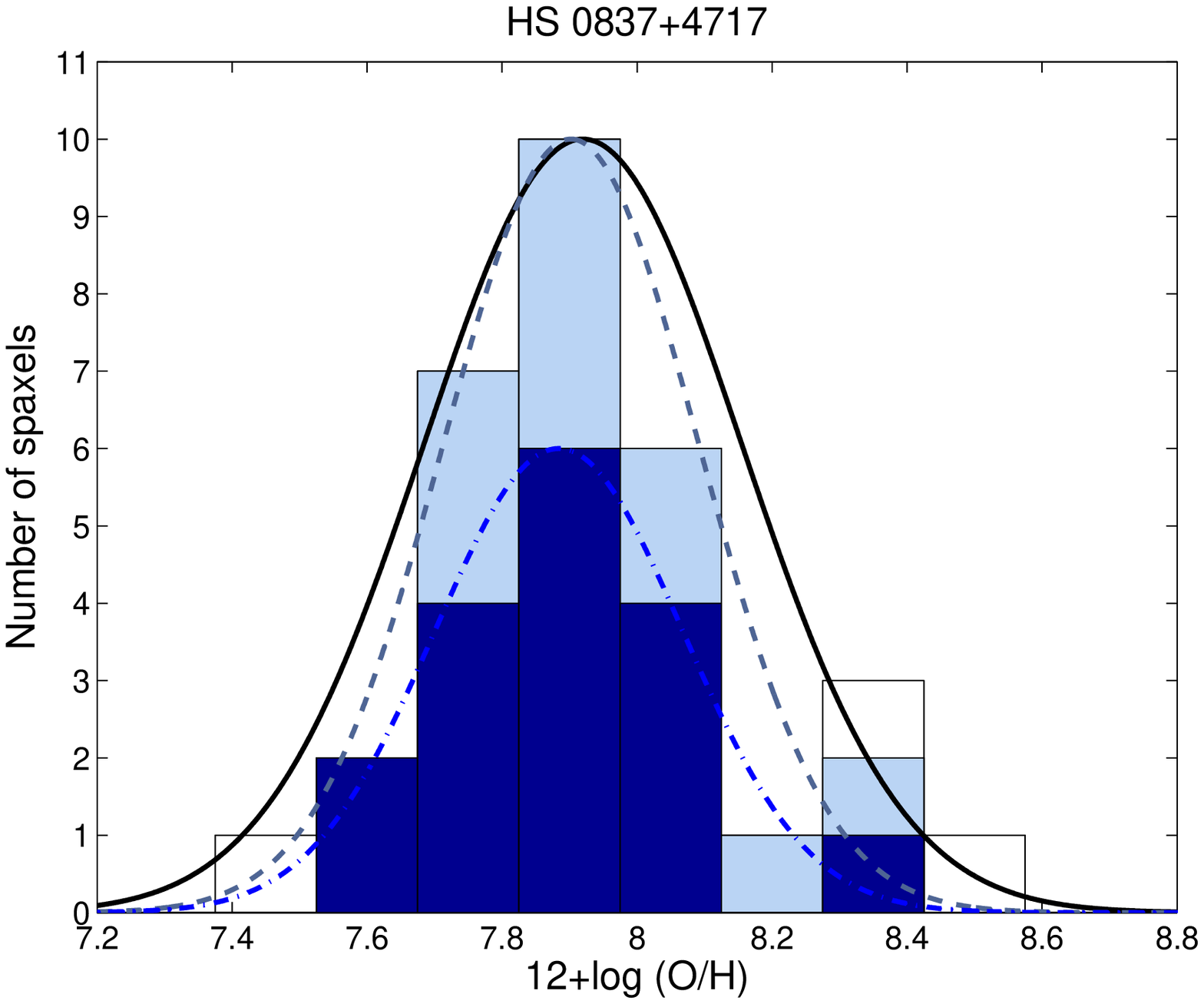}
       \includegraphics[width=6cm,height=5cm,clip=]{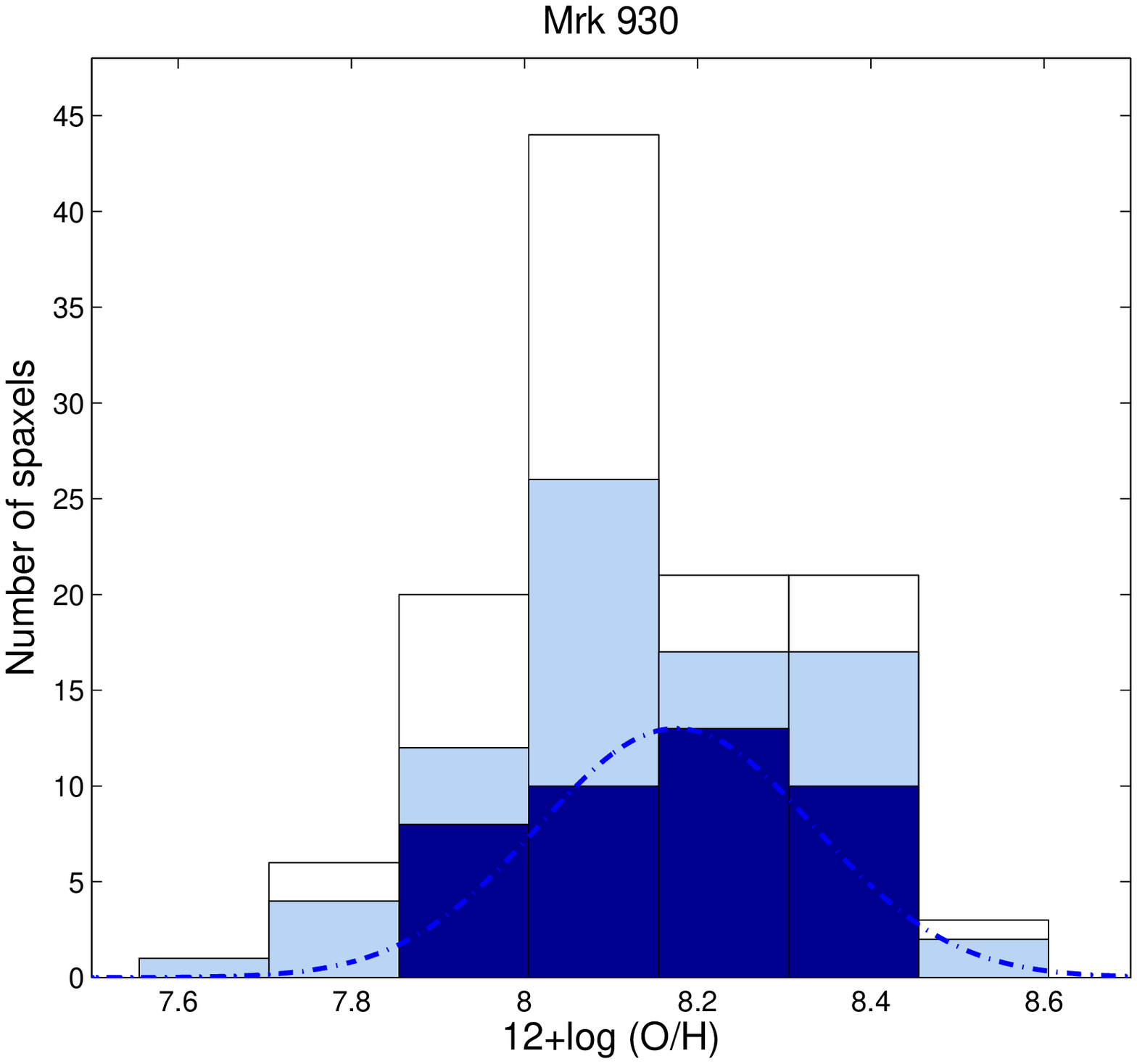}
	    \includegraphics[width=6cm,clip=]{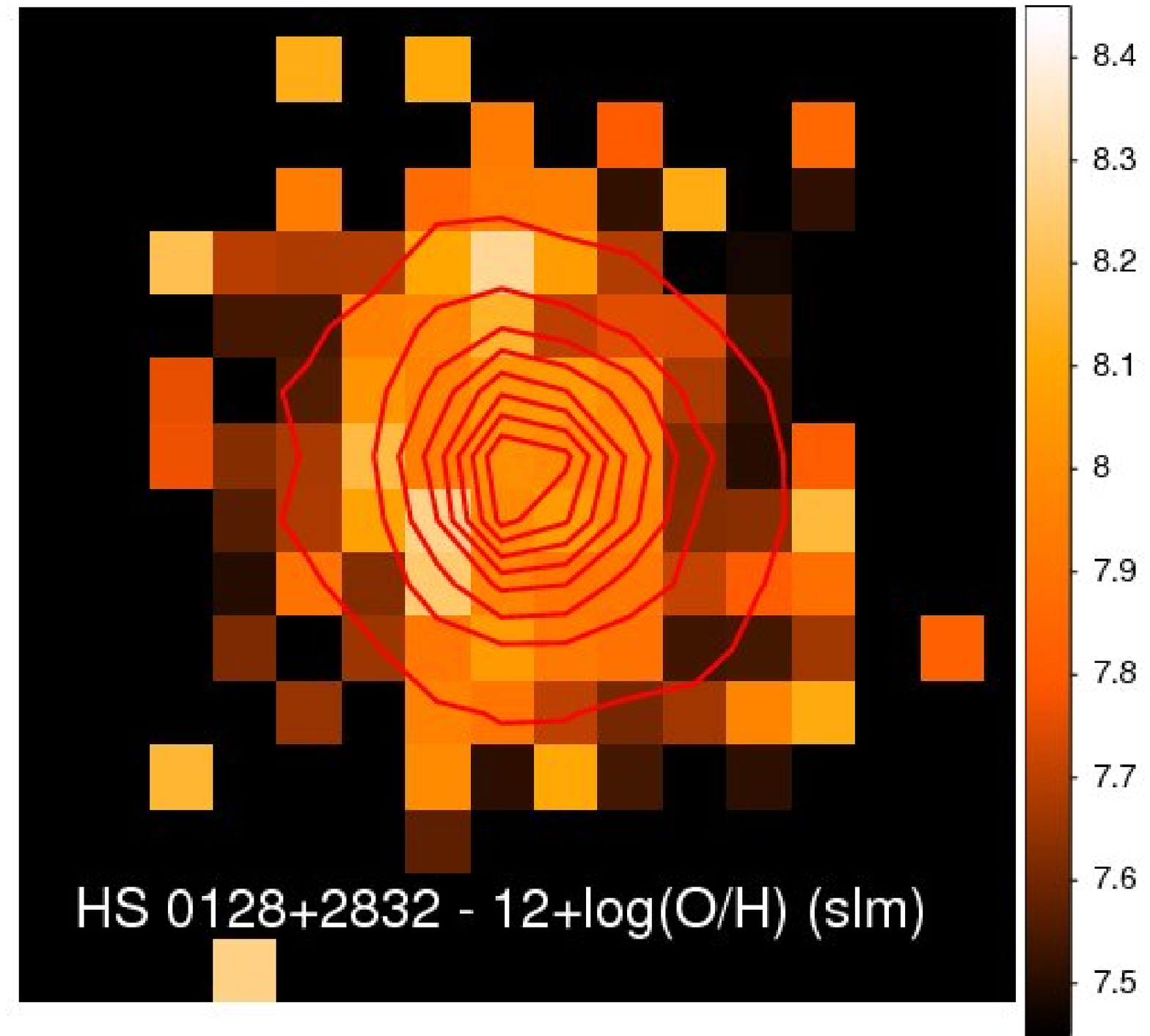}
      \includegraphics[width=6cm,clip=]{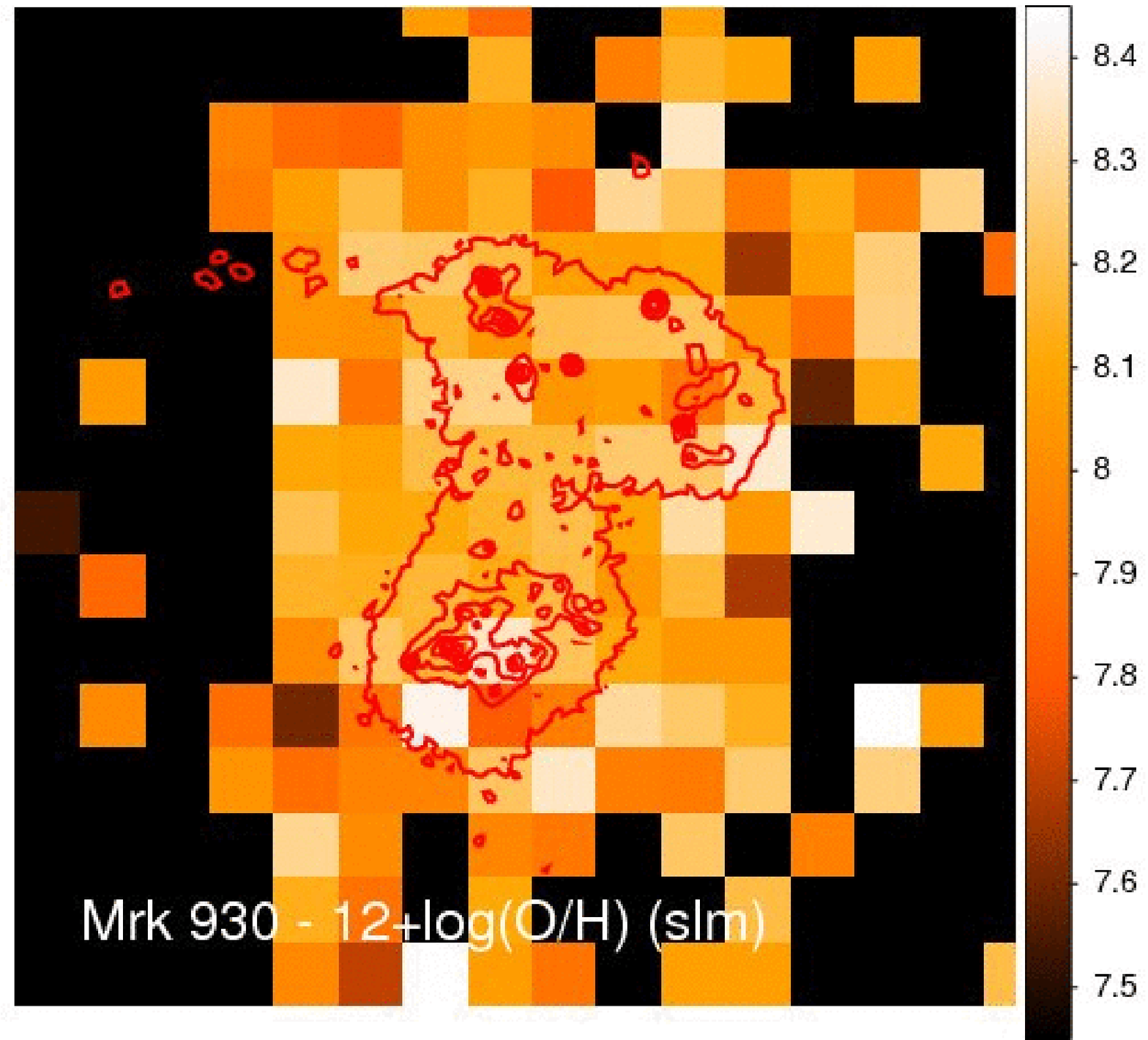}
       \includegraphics[width=6cm,clip=]{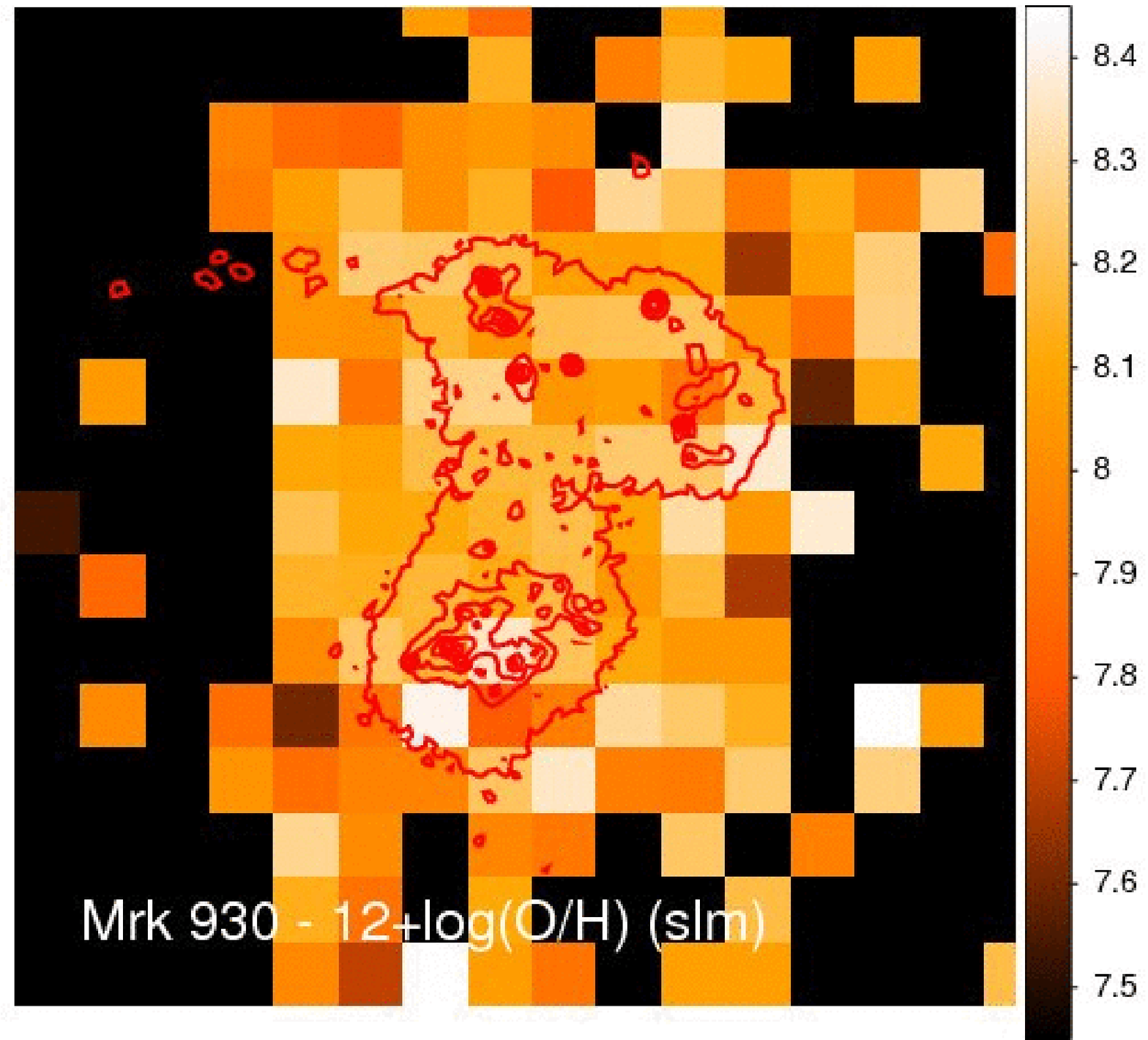}
   
   \caption{Distribution histograms (top) and maps (bottom) of 12+log(O/H) calculated using the strong-line method based on the Pilyugin \& Thuan calibration, 
in HS 0128+2832, HS 0837+4717, and Mrk 930 from left to right, respectively. The solid lines show the same 
contours as in Fig. 3. Units are in dex. The distribution regions are the same defined in Fig. \ref{CHb}.}
    \label{ohe}
    \end{figure*}



\begin{figure*}
\centering
	    \includegraphics[width=6cm,height=5cm,clip=]{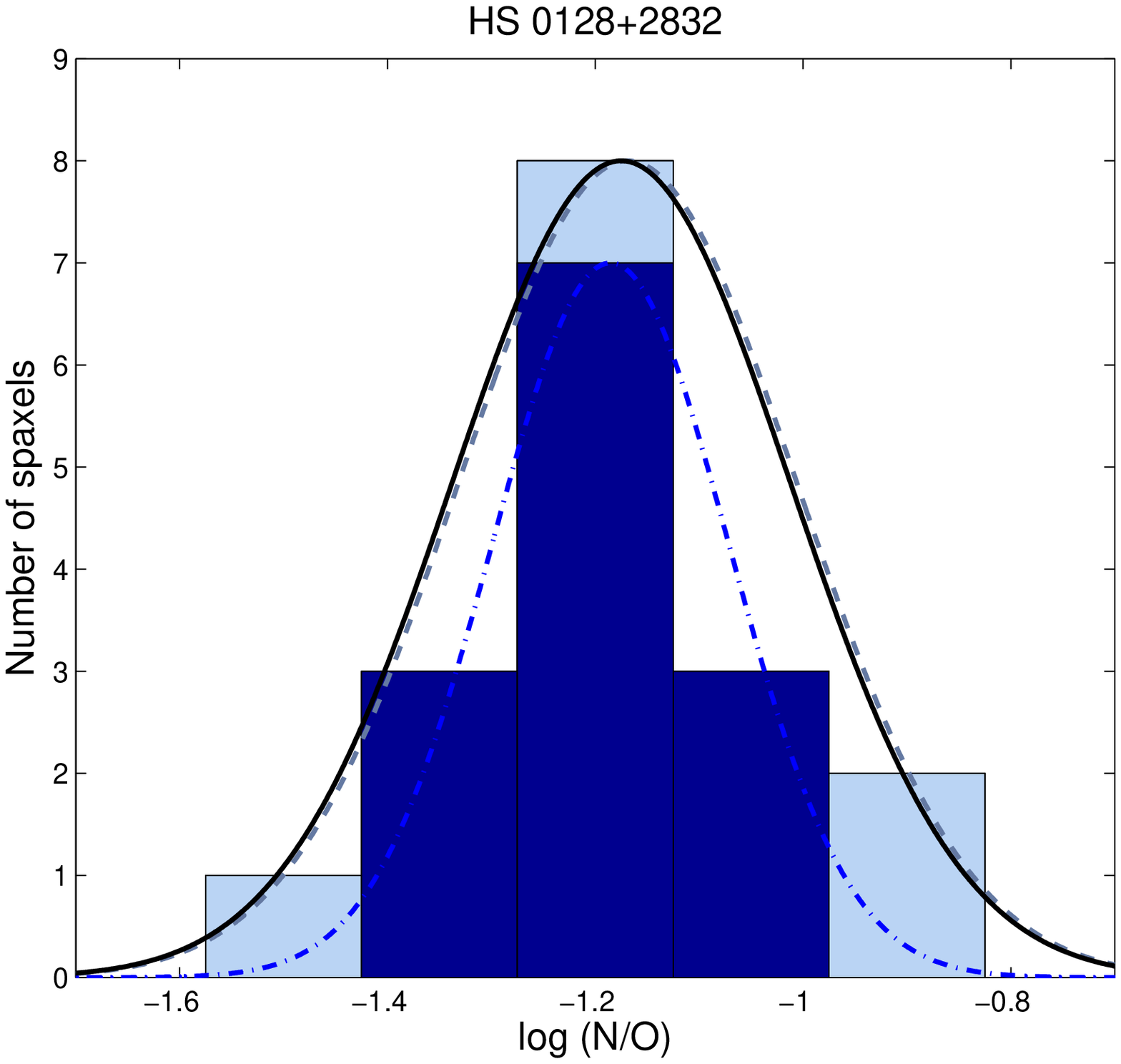}
     \includegraphics[width=6cm,height=5cm,clip=]{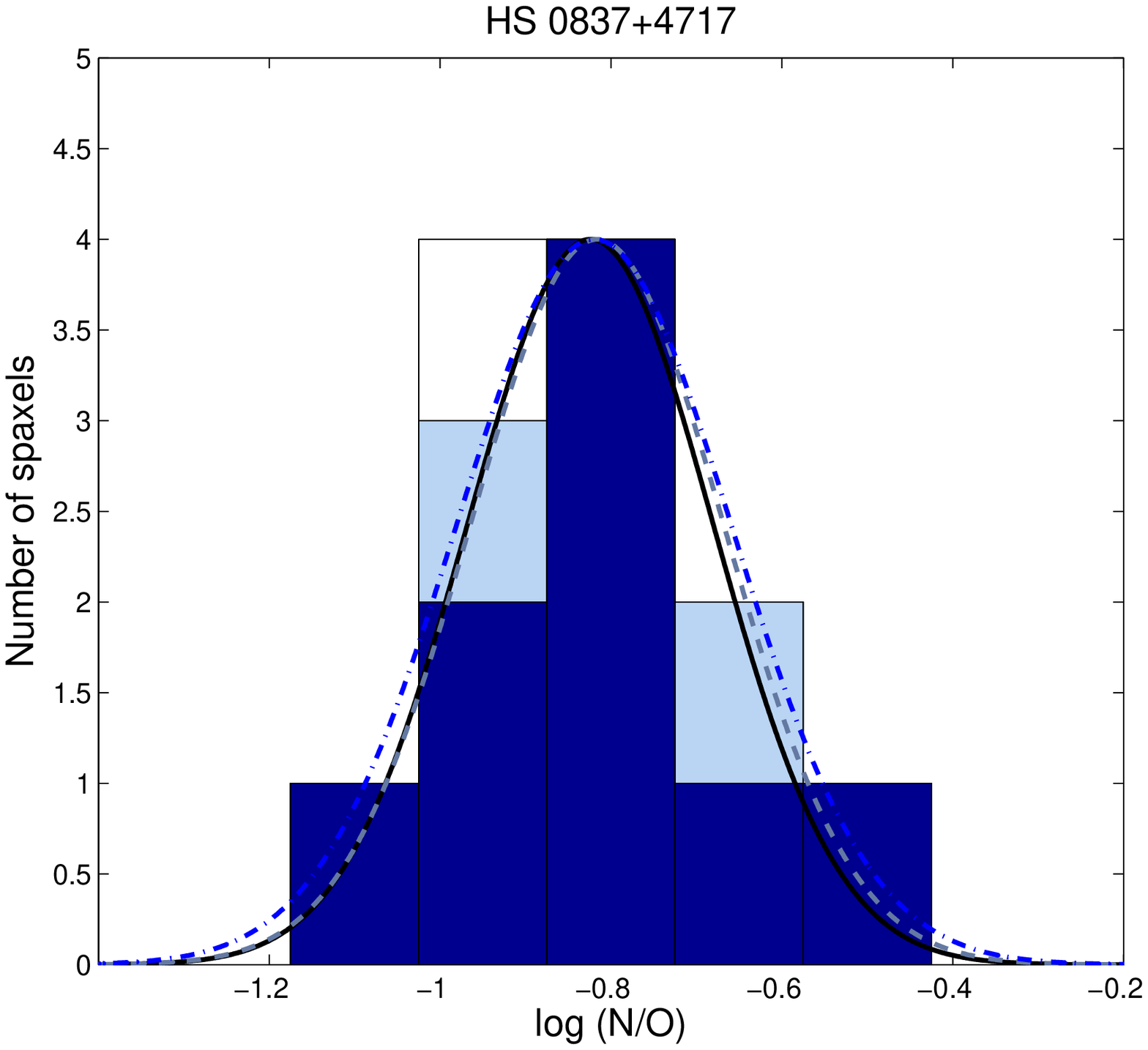}
       \includegraphics[width=6cm,height=5cm,clip=]{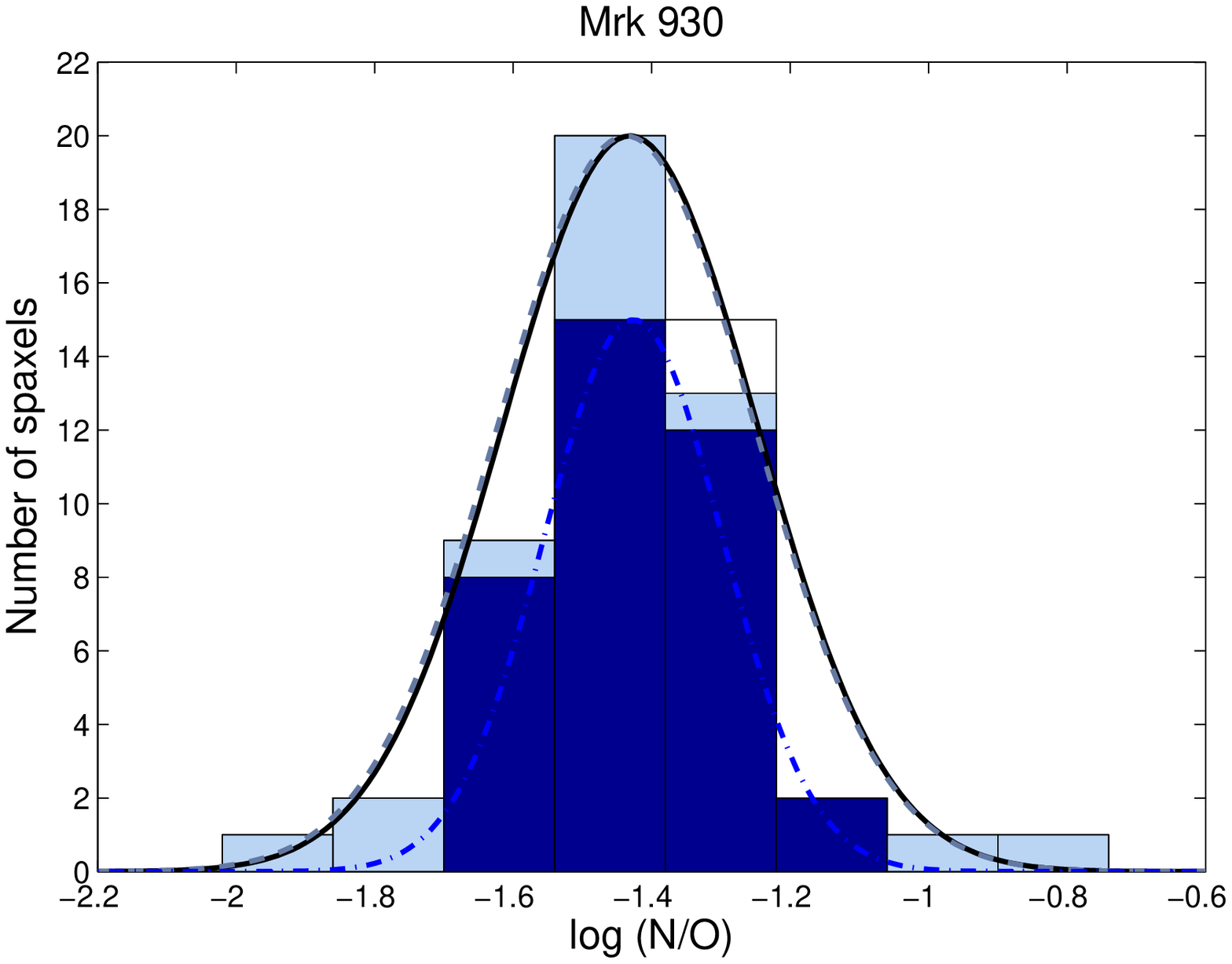}
     \includegraphics[width=6cm,clip=]{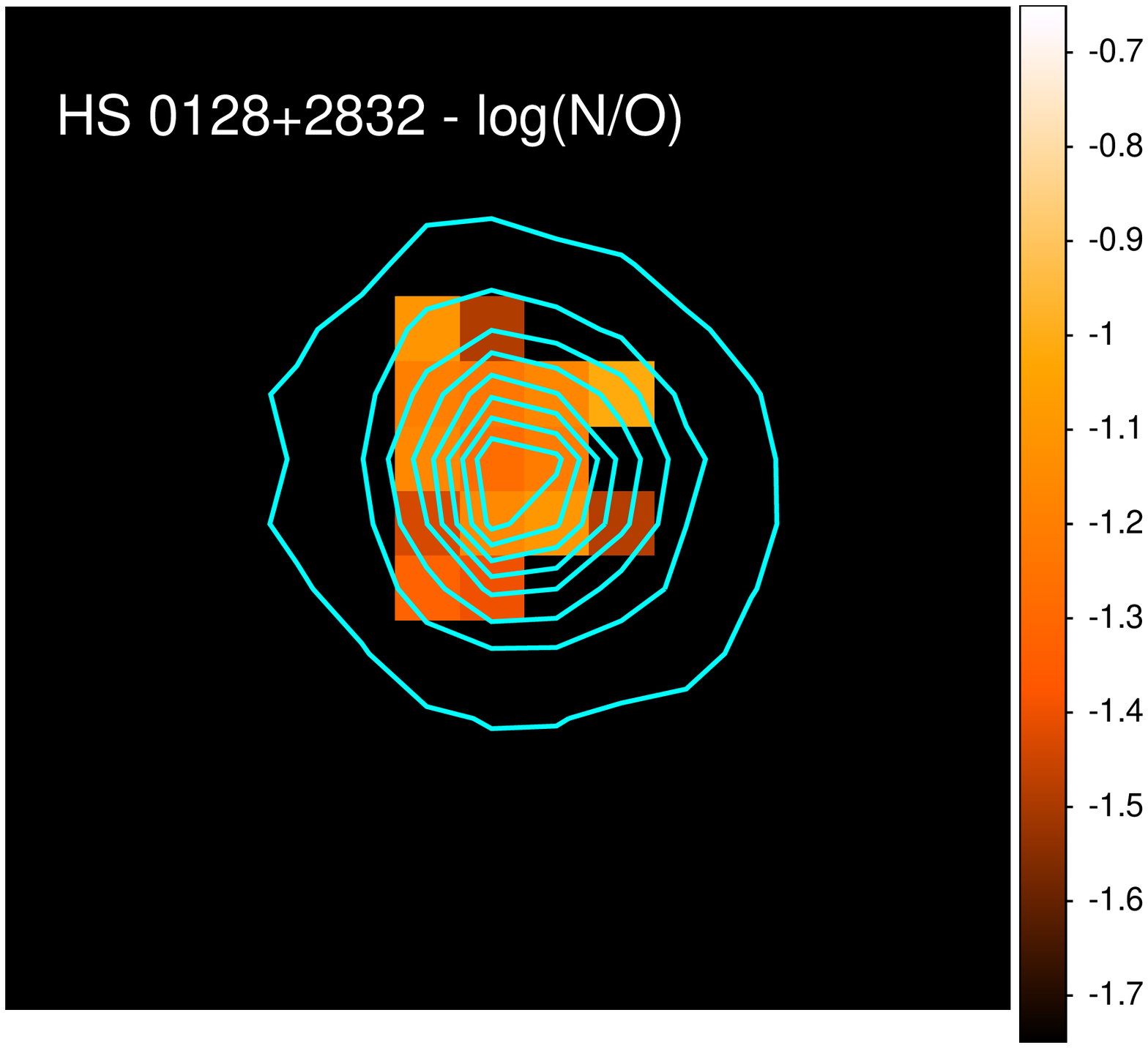}
     \includegraphics[width=6cm,clip=]{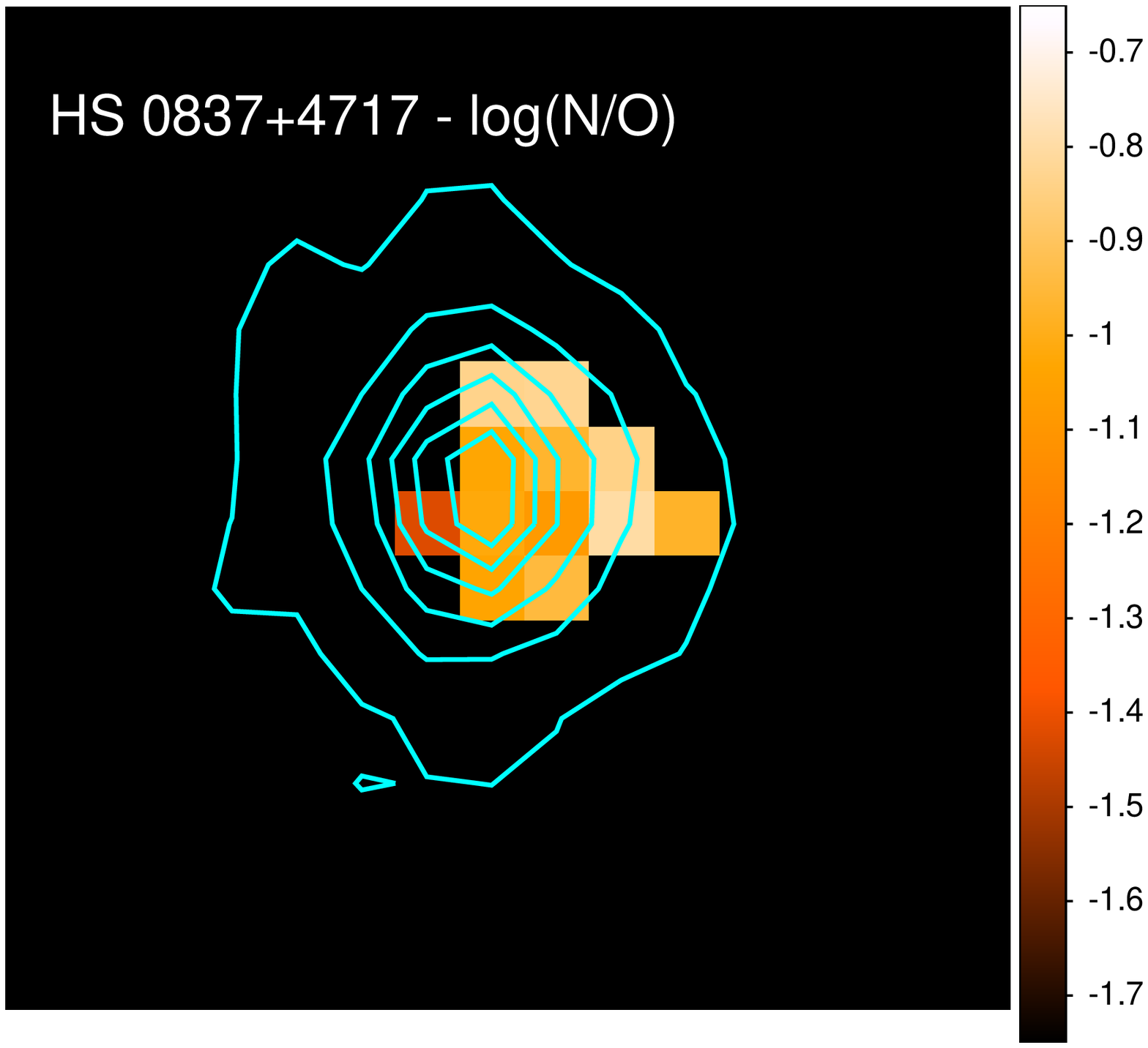}
       \includegraphics[width=6cm,clip=]{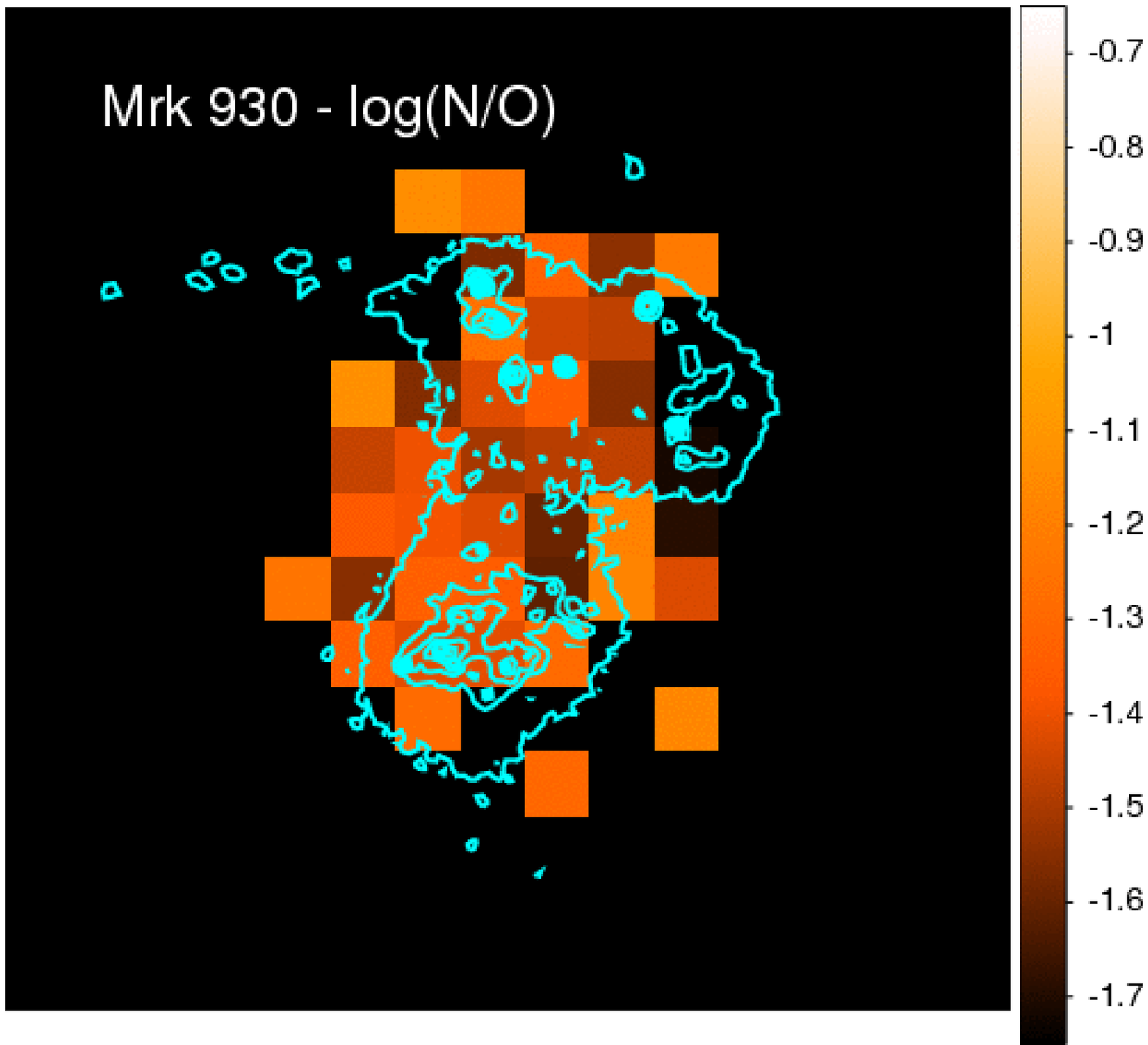}
   
   \caption{Distribution histograms (top) and maps (bottom) of log(N/O) calculated using the direct method, in HS 0128+2832, HS 0837+4717, and Mrk 930 from left to right, respectively. The solid lines show the same contours as in Fig. 3. Units are in dex.
The distribution regions are the same defined in Fig. \ref{CHb}.}
    \label{no}
    \end{figure*}



\begin{figure*}
\centering
	    \includegraphics[width=6cm,height=5cm,clip=]{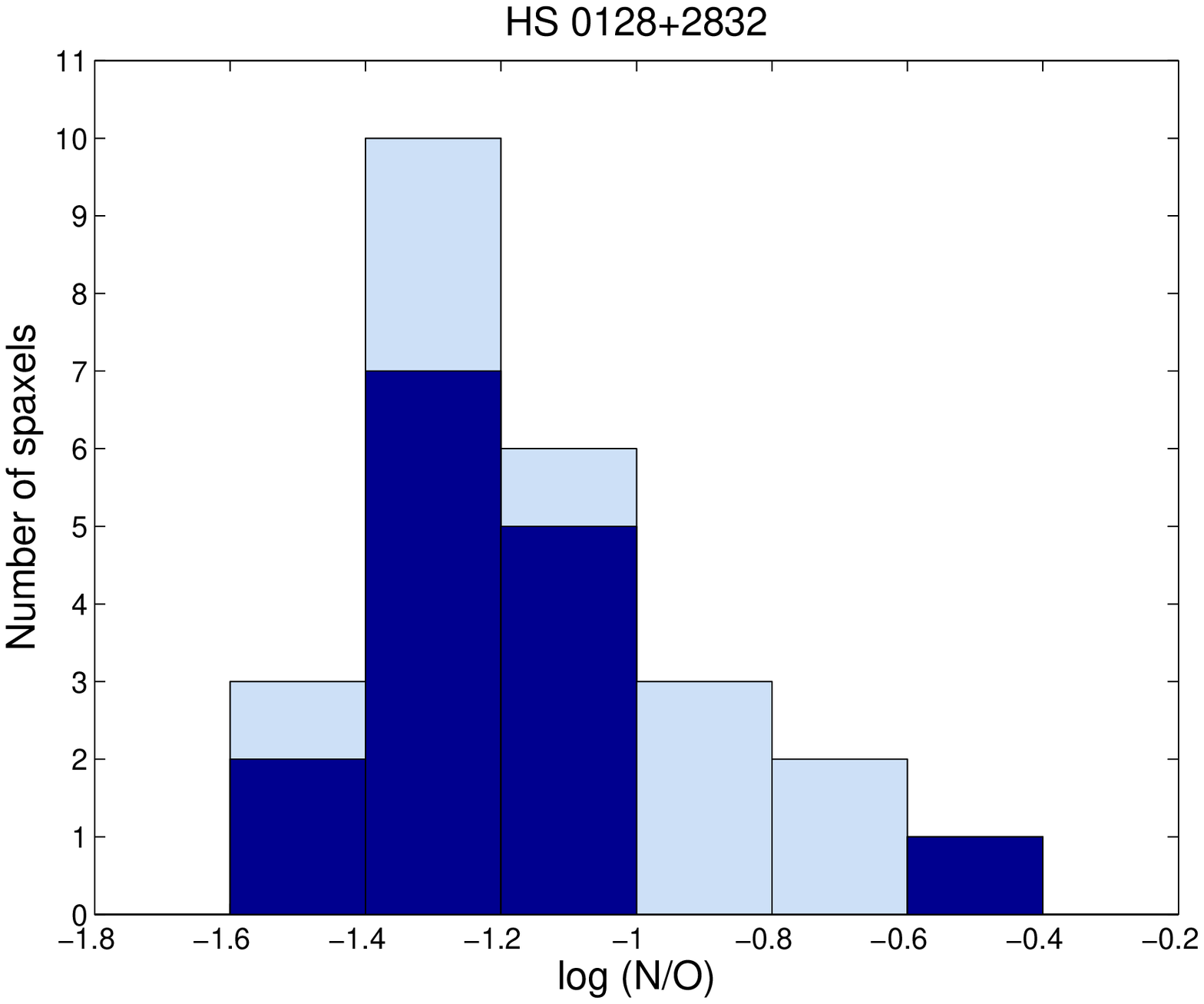}
     \includegraphics[width=6cm,height=5cm,clip=]{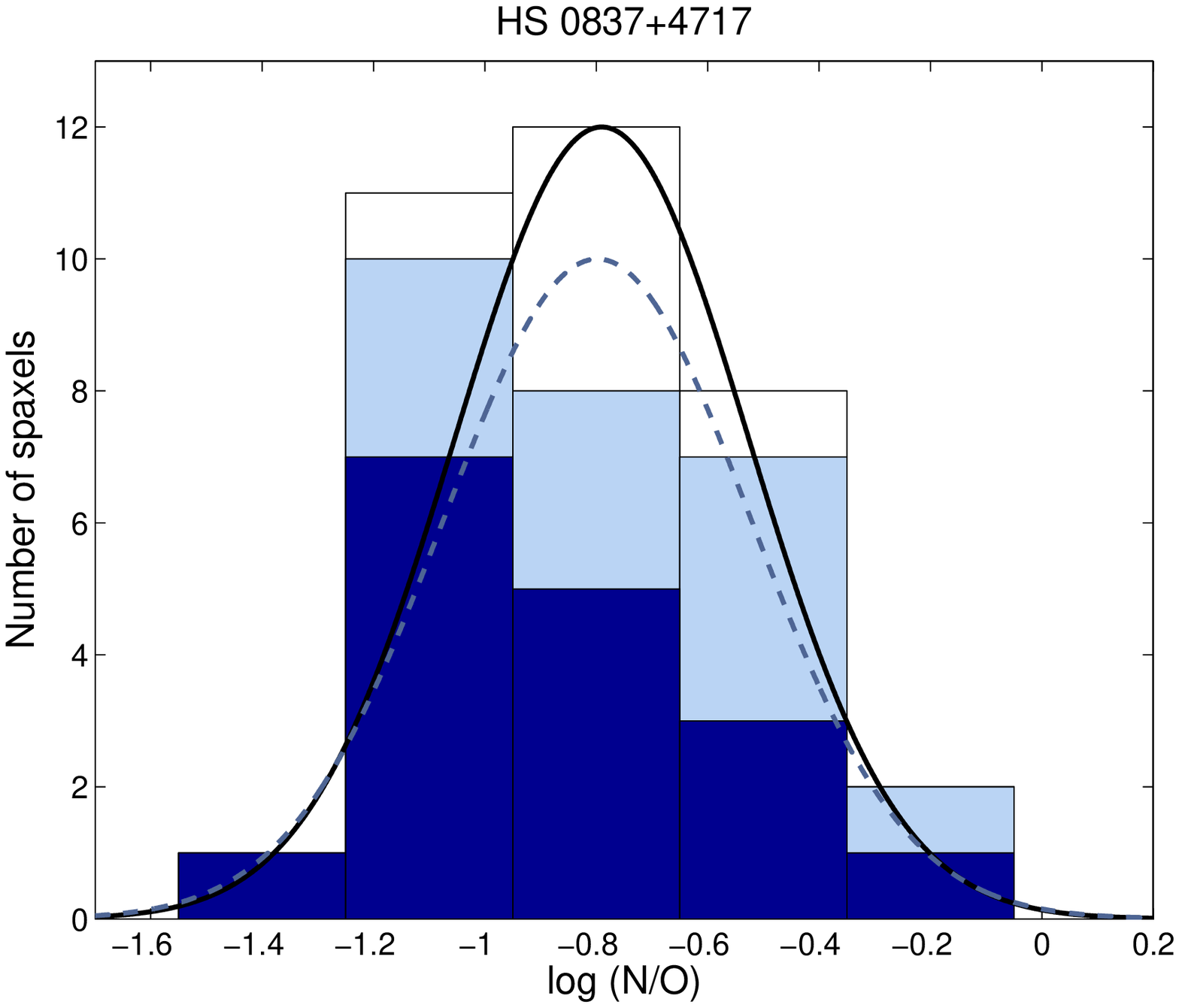}
       \includegraphics[width=6cm,height=5cm,clip=]{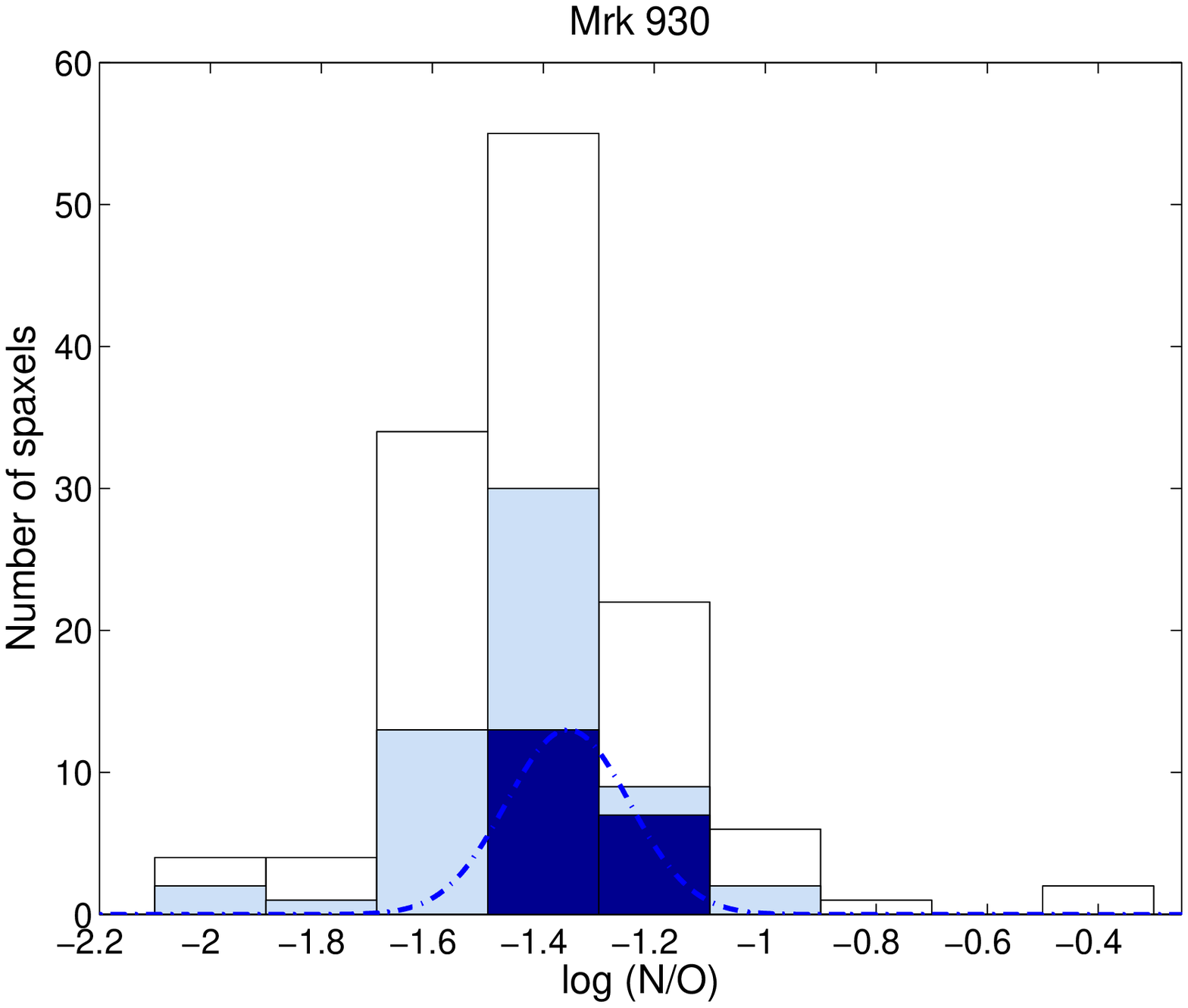}
     \includegraphics[width=6cm,height=5cm,clip=]{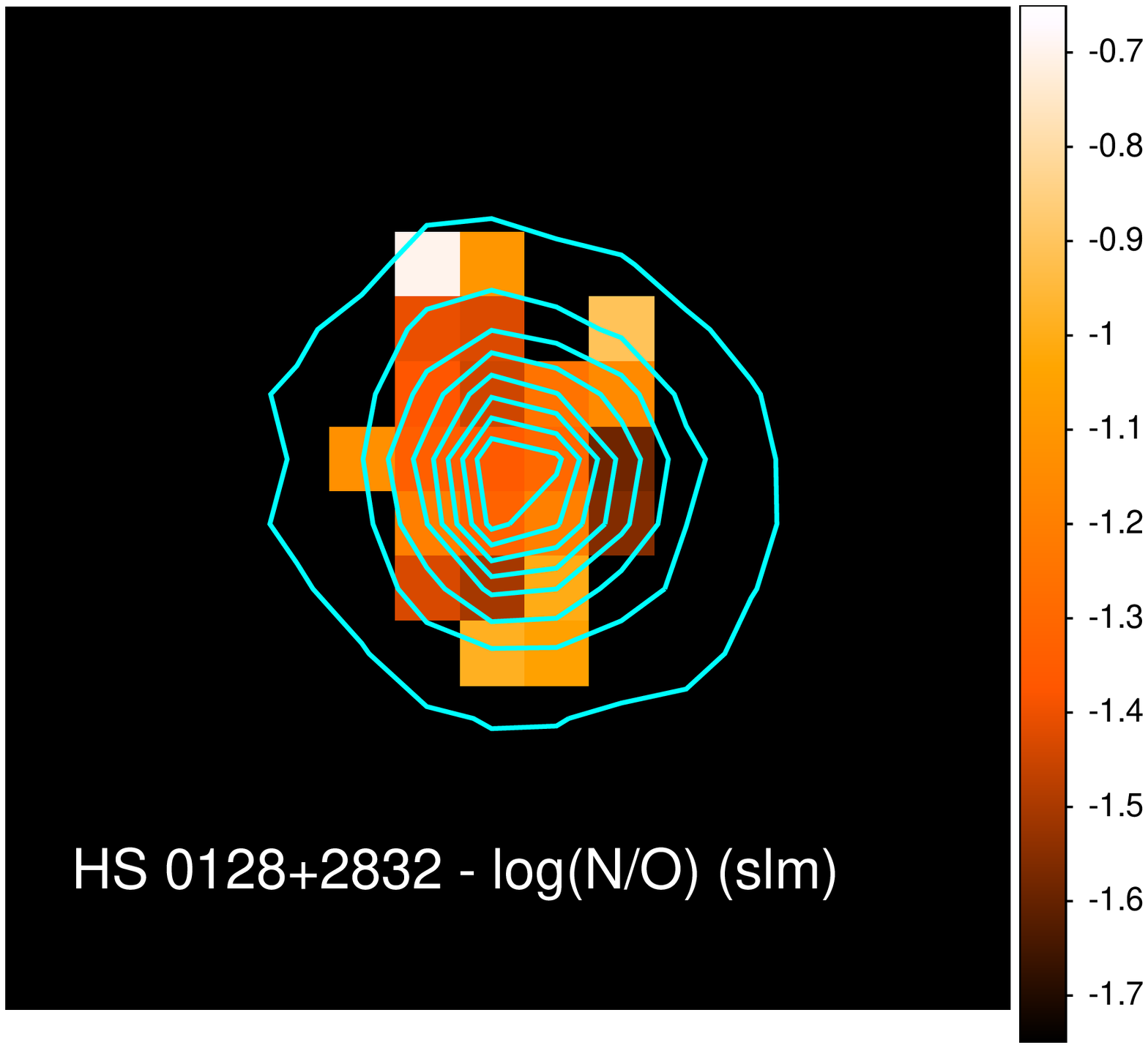}
     \includegraphics[width=6cm,clip=]{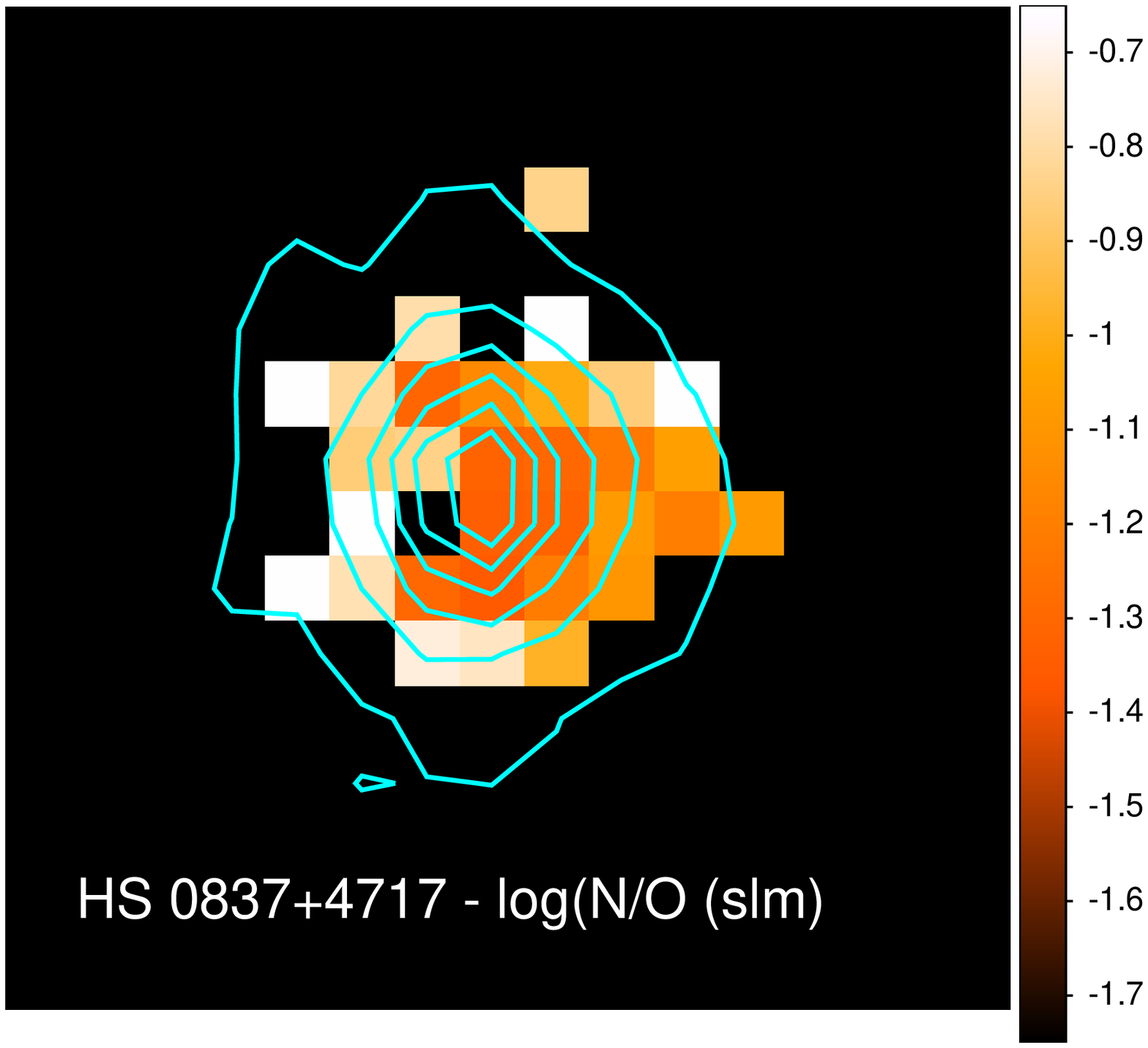}
       \includegraphics[width=6cm,clip=]{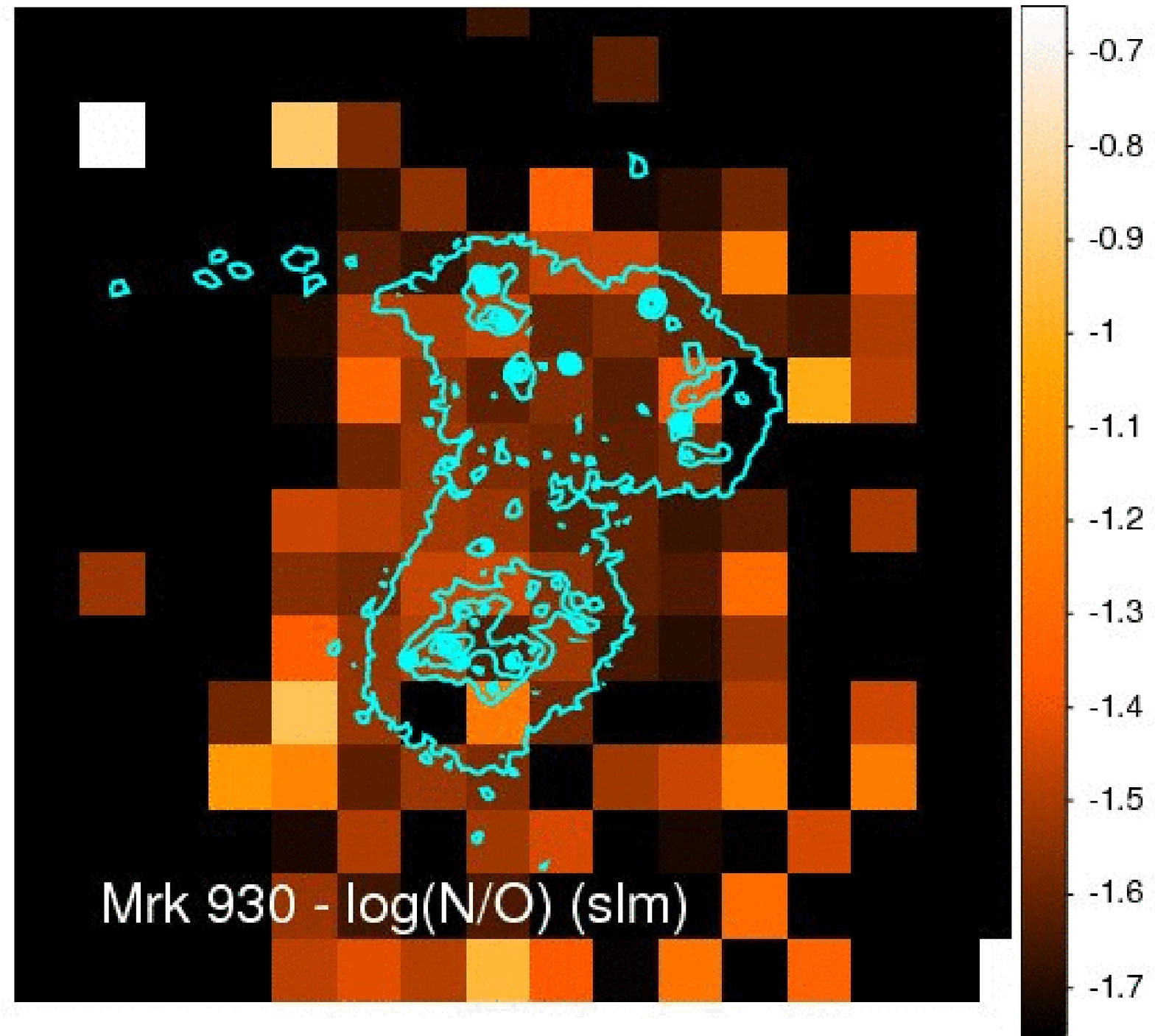}
   
   \caption{Distribution histograms (top) and maps (bottom) of log(N/O) calculated using the strong-line method based on the ratio [NII]/[OII], 
in HS 0128+2832, HS 0837+4717, and Mrk 930 from left to right, respectively. The solid lines show the same 
contours as in Fig. 3. Units are in dex. 
The distribution regions are the same defined in Fig. \ref{CHb}.}
    \label{noe}
    \end{figure*}


\subsection{Oxygen and nitrogen chemical abundances}

In those spaxels with an estimate of the electron temperature it is possible
to derive chemical abundances following the direct method ({\em i.e.} the derivation of abundances
using the relative emission flux of collisional-to-Balmer emission lines and the electron
temperature). O abundance has been derived by calculating the abundances of the main
states of ionization in the nebula: O$^{2+}$ and
O$^+$, the most abundant under the excitation conditions of the gas in these objects.
In both cases, we have used expressions described in H\"agele et al. (2006, 2008) 
based on the task IONIC under the IRAF nebular package.
O$^{2+}$ abundances have been derived using the emission line intensities of 
4959 and 5007 {\AA} relative to H$\beta$ and the electron temperature of [O{\sc iii}], 
which are available for a large number of spaxels.

In the case of O$^+$ abundances, we have used the electron temperature of [O{\sc ii}]
and the intensity of [O{\sc ii}] 3727 {\AA} relative to H$\beta$. Since t([O{\sc ii}]) has
not been measured in any of the spaxels, we have derived it from the available t([O{\sc iii}]) using the
relations based on photoionization models proposed by P\'erez-Montero \& D\'\i az (2003),
which depend on the value of the electron density. In those spaxels without
an estimate of the electron density
({\em i.e.} all in HS 0128+2832 and HS 0837+4717 and some of them in Region 3 of Mrk 930), 
we have assumed a value of the electron density of 100 cm$^{-3}$.
The distribution functions and the maps of the total O abundance as derived using the
direct method are shown in Fig. \ref{oh}. As the main source of uncertainty in
this total abundance comes from the temperature, the average error is higher
in HS 0128+2832 (0.3 dex) than in HS 0837+4717 and Mrk 930 (0.09 and 0.08 dex
respectively).
As in the case of electron temperature a gaussian fit is found in the three regions 
of HS 0837+4717 and in Regions 1 and 2 of Mrk 930, but in none of HS 0128+2832.
The mean value of these distributions agree within the 1$\sigma$ dispersion
range with the values derived from the integrated spectrum, although
in the case of Mrk 930 there is a slight offset between the O abundance of
the southern and northern knots, but it is within the observational errors.

A deeper study of the metallicity distribution of these galaxies has
been performed using strong-line methods. These do not depend on the
previous determination of the electron temperature and they can be
applied in a larger number of spaxels. In the case of O
abundance there are several indicators based on the intensity of the
brightest lines of the spectrum.  In our case, in order to be
consistent with our previous estimation of the O abundance based
on the direct method, we have chosen the calibration proposed by
Pilyugin \& Thuan (2005), which is calibrated using objects whose metallicities
have been derived using electron temperature estimates.This
method relies on the calibration of the parameters R23 and P,
depending on [O{\sc ii}], [O{\sc iii}], and H$\beta$ emission-line
fluxes.  One of the main drawbacks of the strong-line methods based on
R23 is the degeneracy of the parameter with O abundance ({\em
  i.e.} a same value of R23 leads to two different values of the 
  O abundance).  Nevertheless, since our three galaxies are metal-poor objects, the
calibration proposed by Pilyugin \& Thuan for the low metallicity regime has
been used. In Fig. \ref{ohe}, we show the
distribution functions and the maps of the total O abundance in
the three studied galaxies using this method. In all cases, the
average errors in the spaxels of the three galaxies are lower than the
uncertainty associated with the calibration of this parameter, which
is 0.2 dex according to P\'erez-Montero \& D\'\i az (2005).  As in the
case of the O abundance derived using the direct method, in all
three galaxies different gaussian functions are fitted at different
scales.  In the case of the inner-most region of HS 0837+4717, the
level of significance is very large (98\%), but it decreases
substantially for the outer regions. In the case of HS 0128+2832, 
a gaussian fit is found only in Region 2. Finally, in Mrk 930,
contrary to the direct method where gaussians were found in 
Regions 1 and 2, a normal function is only fitted in Region 1.  For the
three objects, a good agreement is found between the mean values of
these distributions and their equivalents using the direct method in a
lower number of spaxels and with the metallicity estimates from the
integrated spectrum using the same parameters.

Nitrogen abundances have been also calculated using both the direct method in those
spaxels with an estimate of the electron temperature and using a strong-line 
method for all the other spaxels where a good measure of the bright emission
lines exist. In the first case, we have used the relative intensity of the
[N{\sc ii}] 6584 {\AA} to H$\beta$ and the electron temperature of [N{\sc ii}], estimated
using the relation with t([O{\sc iii}]) based on models described in P\'erez-Montero
\& Contini (2009). 
The N/O abundance ratio is then assumed to be equal to the N$^+$/O$^+$
ratio. The distribution functions and the maps of this ratio are shown in Fig. \ref{no}.
As in the case of O total abundances derived from the direct method,
the mean error of the N/O ratio in each spaxel is higher in HS 0128+2832 (0.4 dex), 
than in HS 0837+4717 (0.2 dex) and Mrk 930 (0.16 dex).
Contrary to all the other distributions in this work, gaussian functions fit with a good level of confidence 
in all the regions of the three galaxies, in agreement with the values derived in the integrated
spectra.

For the rest of spaxels, with no derivation of the temperature, 
the N/O ratio can be estimated by resorting
to the empirical calibration of the N2O2 parameter, proposed by P\'erez-Montero \& Contini
(2009). This parameter is based on the ratio of [N{\sc ii}] and [O{\sc ii}] emission lines, presenting a linear relation
with the N/O ratio. As previously, we have used this method because it has been calibrated 
using data with a direct estimation of the N/O ratio and it is consequently consistent with
the N/O derivation from the electron temperature.
The distribution functions and the maps of the N/O values using
the strong-line method are shown in Fig. \ref{noe}. The average errors
are in all the spaxels lower than the uncertainty associated with the
calibration of this parameter, which is 0.2 dex, according to P\'erez-Montero \& Contini
(2009).  The mean values of the N/O ratio distributions as derived from this strong-line
method are compatible with the values
found using the direct method. Nevertheless, the gaussian fits appear only
in the inner regions of HS 0837+4717 and no gaussians are found in any region
for Mrk 930 and HS 0128+2818.

\subsection{Kinematics of Mrk 930}

\begin{figure}
\centering
       \includegraphics[width=7cm,clip=]{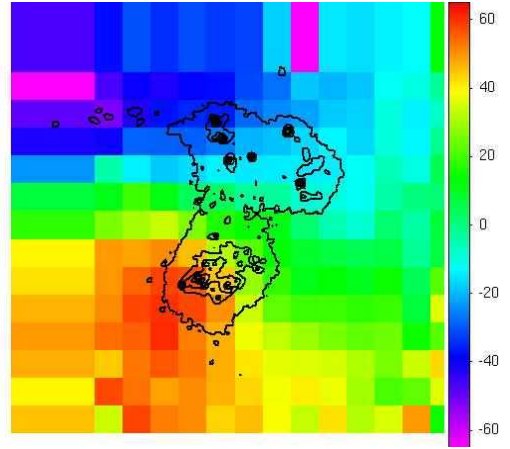}

   \caption{Radial velocity of the ionized gas for the H$\alpha$ emission line in Mrk 930. The solid lines
 show the same contours as in Fig. 3. Units are in km/s. }
    \label{vrad}
    \end{figure}

We have studied the velocity field of the ionized
gas by fitting a gaussian to the H$\alpha$ emission line in the three galaxies,
but only Mrk 930 presents a rotation curve, as shown in the
radial velocity map of Fig. \ref{vrad}.
The spectral resolution of our
data prevents us from deriving reliable velocity
dispersions. The west
side of the galaxy appears to be less disturbed than the east side; a
sharp velocity gradient in the north-south direction is visible in the
eastern half of the map. This velocity field could be the result of an
interaction system.


\section{Discussion}

\subsection{On the spatial uniformity of the derived properties in the sample}

We have studied the distribution of different physical properties and chemical abundances
along the field of view of the PMAS instrument in some BCDs with different degrees
of excess in their N/O ratio in relation to the expected values for their metallicity regime.
IFS allows
a different approach in their study in contrast to integrated long-slit or fiber observations.
In these cases, the integrated spectrum is a weighted mean of the different regions
of the observed objects, whose statistical weights are
the relative luminosities of the different regions. On the contrary, IFS data
can be analyzed considering each observed region independent of
their relative brightness. In this way, in the distributions of the
physical properties and chemical abundances of the three studied BCDs 
in this work, it is remarkable the 
differences between the mean of the distributions in
different regions and the values of the same quantities measured for the
integrated spectrum. These differences are well illustrated by the reddening constant distribution
in the HS 0128+2832 galaxy (see first panel of Fig. \ref{CHb}), whose integrated 
value is c(H$\beta$) = 0.14, while the
mean value of the spaxels distribution is c(H$\beta$) = 0.37 in the inner-most region.
This difference is understood by inspecting the shape of the reddening constant distribution,
where no gaussian fit is found for any of the regions and, on the contrary, an apparent double-peak distribution
appears, with one of the maxima peaking at the same value found for the integrated
spectrum and the other maximum peaking at the same value that the mean of all the data.
Another example of this disagreement is found in the study of the density
distribution of Mrk 930, which is far to be fitted by a gaussian.
This is partially motivated by the presence of a
high density structure well visible in the south part of the galaxy and which
is also seen in the histogram distribution plot (upper panel of Fig. \ref{nS2}).
This structure is also responsible for the difference found in the
electron density between the north (32 cm$^{-3}$) and
south ($<$ 190 cm$^{-3}$) parts of the galaxy. Besides, the mean values found for
different regions of the same galaxy range between 101 and 271 cm$^{-3}$, as
we go from Region 1 ({\em i.e.} the inner-most region, around the peak in H$\alpha$)
and Region 3 ({\em i.e.} all valid spaxels in the IFU), as we can see in Table \ref{gaussian}.
In the most part of the cases, this disagreement between the value found in
the integrated spectrum and the mean value of the spaxels distribution appears
when no gaussian fit is found for this distribution, indicating local
spatial variations which can not be understood only in terms of statistical 
fluctuations.

The histograms of the properties studied in different 
regions have been also used to find out to what extent a certain property can
be considered as homogeneous within the observational and statistical errors in
different regions. This would allow us to find in principle possible 
variations of the excitation as a function of the distance to the ionizing cluster,
to find regions with excesses in electron density or extinction, or to locate
the chemical pollution in specified places of the observed field.
However, the assumption of the hypothesis that a certain
parameter can be considered as uniform if its spaxels distribution can be
fitted by a gaussian have two important limitations in this work which
must be taken into account. Firstly, there is a sampling effect
which makes that, for some very compact objects, the number of spaxels
to be considered as valid can be somehow reduced. This is the case of
the distributions of excitation and N/O ratio, as derived using the N2O2 parameter, in HS 0837+4717 (middle panels of Fig. 
\ref{o2o3} and \ref{noe}, respectively), in which the null hypothesis is rejected
for Region 1. However, a gaussian fit is found when the number of spaxels grow and
Regions 2 and 3 are also considered. This is possibly indicating that the rejection of
the gaussian fit hypothesis does not imply that a certain parameter is not
uniform in a certain region if there is a low number of points to be
considered in the corresponding statistics.
The second limitation comes from the poor seeing measured during the
night of observations, which is worse in all cases than the working spatial resolution.
In this case, serious doubts appear about the no correlation of the 
fluxes between adjacent spaxels. Nevertheless, the typical size of the regions
where the uniformity has been assumed is much larger in all cases than the
typical value of the point spread function for the quoted seeing, and this is
indicative of the validity of the results.

Another important fact which reinforces the assumption of uniformity in
those cases where a gaussian can be fitted, is that most of the times
the normal functions are found in more than a region in each galaxy.
With the exception of the O abundance as derived from the R23-P method
in HS 0128+2832 and the N/O ratio as derived from the N2O2 method
in Mrk 930, the rest of the physical properties and chemical abundances
are fitted in two or even in the three defined regions in the field of view.
This excludes the possibility in those cases of an {\em artificial}
gaussian distribution in a defined region of the space, as the fitting is
found simultaneously in different and independent regions.

The results of these fittings listed in Table 3 indicate that in HS 0128+2832 only the N/O ratio as derived
following the direct method, can be fitted by a normal distribution. On the contrary,
in HS 0837+4717, all the distributions can be fitted by a normal function at
different regions, with the exception of the excitation and N/O ratio 
as derived from the N2O2 method in Region 1, as
explained above. Finally, in Mrk 930 only reddening and density
distributions are not fitted by a gaussian. In this case, it is the disturbed
morphology of this galaxy the most probable cause of the found variations.

It is remarkable the fact that in all the three galaxies the N/O ratio as derived
from the direct method can be considered as uniform at the spatial scales
where it can be measured with enough precision, independently of the
mean value of their distribution or the possible presence of WR stars.
However, this is not the case for the same ratio when it is derived using an
empirical calibrator, N2O2 in this case, which does not appear as uniform
in several regions of HS 0128+2832 and Mrk 930. For total O abundances, 
the same discrepancy between the distributions derived from the direct method and from
the empirical calibration of the R23 parameter given by Pilyugin \& Thuan (2005) is observed.
Hence, in the case of Mrk 930, the O abundance distribution is well fitted
by a gaussian when it is derived using the direct method in Regions 1 and 2, but 
the gaussian is only fitted in the inner-most region when this abundance is derived using R23.
Several limitations related to the calibration of these parameters can be
taken into account to explain this discrepancy, including the sample used to
calibrate them, mostly data coming from integrated observations
of star forming knots, and the dependence of the calibrators on other functional
parameters, such as ionization parameter or effective equivalent 
temperature (P\'erez-Montero \& D\'\i az, 2005), which are not uniform
along a galaxy. Therefore, although the strong-line parameters allow the sampling of
chemical abundances in a larger number of positions, they are
less suitable to study the homogeneity of these abundances in comparison
to other direct method-based techniques.

\subsection{Excitation variations and their implications on diagnostic diagrams}

One of the most interesting results extracted from the analysis of the spatial
distributions of the properties of these three studied objects appears with
the position of several spaxels in Fig. \ref{n2o3},
which represents the [N{\sc ii}]/H$\alpha$ vs. [O{\sc iii}]/H$\beta$ diagnostic diagram.
As we see, the position of the integrated points and the spaxels belonging to
Region 1 lie in the star forming region, although the proximity of these points
to both the theoretical and empirical curves are sensitive to the N/O ratio
({\em i.e.} in HS 0128+2832 and HS 0837+4717, which present the largest
relative N abundance, these points are closer than in Mrk 930).
Nevertheless, a fraction of the outer spaxels, which is larger in HS 0837+4717 and
lower in Mrk 930, spread towards right in the diagrams so they can lie in the so-called composite
region or even in the active galactic nuclei region. This effect is probably due to
the combination of two different effects: i) the high relative N/O ratio which makes
some points to lie in these regions, as shown already by P\'erez-Montero \& Contini
(2009), and ii) the excitation structure of the galaxy, in which the [N{\sc ii}] emission line fluxes
remain bright even at distances where the other emission lines are sensibly less
bright. This makes the [N{\sc ii}]/H$\alpha$ to be higher in these regions, while
[O{\sc iii}]/H$\beta$ remains constant or slightly decreases.
In this case the spectra of some positions of the field of view could be wrongly classified
as AGNs. This effect was already observed by James et al. (2010)
in the BCD Mrk 996 using also IFU observations.


\begin{figure*}[t]
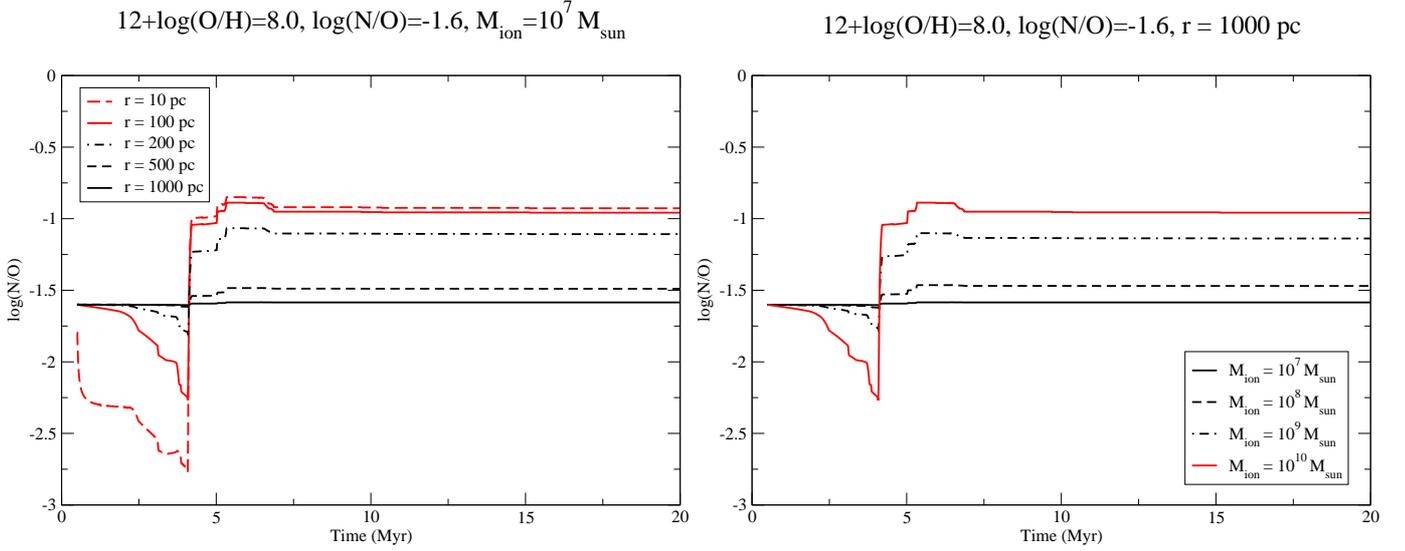

\centering
\includegraphics[width=9cm,clip=]{pol01.eps}
     \includegraphics[width=9cm,clip=]{pol02.eps}
 \caption{Evolution of the N/O ratio for a spherical gaseous distribution of constant density with an initial metallicity
 of 12+log(O/H) = 8.0 and log(N/O) = -1.6, which is polluted by the stellar winds coming from
 a stellar cluster with Z = 0.004. At left, we show the effects of a cluster of mass log(M/M$_{\odot}$) = 7
 for different radii of the distribution. At right, we show the effects of clusters of different stellar masses on 
 a gaseous distribution with a radius of 1 kpc. Stellar wind yields have been taken from Moll\'a \& Terlevich (2011, in prep.)
 }
    \label{polmodels}
    \end{figure*}


\subsection{N/O excess and the presence of WR stars}

Although the WR blue bump has been reported in the integrated spectra of HS 0837+4717
(Pustilnik et al., 2004) and Mrk 930 (Guseva et al., 2000), we have not detected it neither in any
of the individual spaxels of our IFU observations nor in the integrated spectra of the brightest
regions.  The low S/N of the bump combined with poor seeing conditions during the observing night could
be the main reasons of this fact. Nonetheless, it is possible to carry on 
the spatial analysis of the physical properties and the O and N chemical abundances
along the field of view of the IFU. The analysis described in section 3.6 shows that, in most of the
regions of the three galaxies, we can fit a gaussian function to the
O abundance and N/O ratio distributions, obtained from both the direct
method and the strong-line calibration from R23-P for O abundance and N2O2
for N/O ratio.  Besides, the mean values of these gaussian functions match within
the errors the values derived using the integrated spectra from both the
brightest region and the reported long-slit observations in each galaxy.
These values are consistent with a scenario of low metallicity with an excess
of the N/O ratio in relation to the expected value at this metallicity regime. 
Adopting $\rm log(N/O) = -1.6$ as the most typical value of the plateau in the 12+log/O/H) vs.
log(N/O) relation, as can be seen in Fig. 1, this excess is $\Delta$log(N/O) 
= +0.43 dex in HS 0128+2832, +0.78 dex in HS 0837+4717 and +0.17 dex in Mrk 930, 
taking as a reference the mean value of the gaussians
fitted to the distributions of log(N/O) as derived using the direct method
in the respective Region 3 of the galaxies.
At same time, we can take into account the mean radii of the spatial distributions
of log(N/O), considering the spatial size of the spaxels at the adopted distances for each galaxy,
which are 330 pc in HS 0128+2832, 840 pc in HS 0837+4717, and 380 pc in Mrk 930.
By inspecting the corresponding maps, the sizes of the log(N/O) distributions 
as derived from the direct method are approximately
1.2 x 1.7 kpc$^{2}$ for HS 0128+2832, 2.5 x 3.4 kpc$^{2}$ for HS 0837+4717, and
1.9 x 3.0 kpc$^{2}$ for Mrk 930, assuming that these are upper limits due to
the seeing of the observing night.

These scale lengths lead us to question if the reported WR stars can be
the actual source responsible for the global N pollution detected in our
sample of BCDs. To check the validity
of this possibility we have taken the chemical yields of different
elemental species ejected to the ISM by massive stellar clusters
proposed by Moll\'a \& Terlevich (in prep.)
for a metallicity of the cluster Z = 0.004 ($\approx$ 1/5·Z$_{\odot}$), which
is close to the metallicity of our 3 observed BCDs.
We have studied the effects of these winds on a spherical homogeneous distribution of gas with
constant density of 100 cm$^{ -3}$ and an initial O
abundance of 12+log(O/H) = 8.0 and $\rm log(N/O) = -1.6$. 
We also assume that the mixing of the ejected elements with the 
material of the spherical distribution is instantaneous, which is
not reallistic at these galactic scales, but it is
the most favourable hypothesis for the assumption of the 
pollution of the ISM by stellar winds.

The evolution of the log(N/O) ratio with time is shown in Fig. \ref{polmodels} under
different assumptions. In the left panel, we show 
the effect of the ejecta from a massive cluster with log(M/M$_{\odot}$) = 7
to a spherical gaseous distribution of different radii. As we can see, for the 
typical radii of the N/O distributions in our galaxies, which are of the
order of 1 kpc, the effect is negligible.
For lower radii, it begins to be apparent how the ejection of this material changes the
relative chemical composition of the ISM. At first, the ejection of O by very massive stars
makes the N/O ratio to decrease and the later appearance of WR star winds
makes the N/O ratio to increase again, reaching a maximum and an asymptotic value.
The value of this maximum matches only the measured mean values of
our three galaxies for radii much lower than those measured in the IFU observations.
In the case of HS 0837+4717, the measured N/O is even higher than
the maximum value of log(N/O) reached at any radius.

In the right panel of Fig. \ref{polmodels} we have fixed the radius of the spherical gas distribution to 1 kpc, and
we have varied the stellar mass of the ionizing cluster. As can be seen,
the effects are only noticeable at this scale when the cluster is more
massive than log(M/M$_{\odot}$) = 8.   
We can estimate the masses of the ionizing stellar clusters in the three BCDs,
taking the following expression proposed by D\'\i az (1998)

\[\log(M/M_{\odot}) = \log L(H\alpha) - 0.86 \cdot EW(H\beta) - 32.61
\]                                                                                                                                                                                     

\noindent with the H$\alpha$ luminosities measured from the area covered by the IFU and the 
EW(H$\beta$) of the integrated spectrum of the corresponding
brightest regions. Therefore, the ionizing masses are of
10$^{6.47}$ M$_{\odot}$ for HS 0128+2832, 10$^{7.45}$ M$_{\odot}$ for HS 0837+4717, 
and 10$^{7.72}$ M$_{\odot}$ for Mrk  930.
Although these estimates are only lower limits, because escaping photons
and dust absorption are not taken into account, they are much lower than the mass of the ionizing cluster
required to pollute the ISM to the spatial scales measured in our IFU observations.

Therefore, although a number of WR stars have been detected in two of these galaxies
({\em i.e.} of the order of 10$^3$ both in HS 0837+4717 by Pustilnik et al., 2004 and
in Mrk 930 by Guseva et al., 2000), and according to stellar evolutionary models they are also
expected to be present in HS 0128+2832, 
the global high N/O ratio observed in these objects is not likely produced by the 
pollution from stellar wind ejecta coming from WR stars, even
in the most favourable conditions of mixing and dillution
 and, on the contrary, could be more related to other
global processes affecting the metal content of the whole galaxy, as was
already suggested for the {\em green pea} galaxies analyzed by Amor\'\i n et al. (2010). 
They propose a combination of outflows of enriched gas and inflows of metal-poor
gas, capable to trigger the star formation processes in these objects, decreasing the
total content of metals without noticeably changing the N/O ratio
(K\"oppen \& Hensler, 2005).

\section{Conclusions}

We have studied, by means of PMAS optical integral field spectroscopy, 
the spatial distribution of ISM physical-chemical properties for a sample of three BCDs 
(HS 0128+2832, HS 0837+4717, Mrk 930) with
low metallicity and an excess of N/O, as measured with previous
integrated long-slit spectrophotometry. The spectral range covering from 3700 to
6900 {\AA} and the spatial resolution of 1''x1'' in a field of view of
16''x16'' ensure the characterization of the physical properties and the chemical
abundances across the objects, although the spatial resolution is somehow 
reduced due to the seeing during the observing night ($\approx$ 2'').

H$\alpha$ intensity maps show that both HS 0128+2832 and HS 0837+4717
have compact aspect, with their luminosities dominated by a single central massive stellar cluster.
Mrk 930, whose morphology has been also compared with available UV
HST-ACS images, shows a more complex structure, with knots in the northern and southern
part of the galaxy. Besides, it shows a clear rotation curve along the north-south axis,
which could be consequence of an interaction system.

All properties derived from the emission-line intensities with enough
S/N have been analyzed in different spaxels of the
field of view. Their corresponding distribution in different regions have been also compared to
the properties of the objects as derived from the integrated spectrum
collected in the brightest region of each galaxy.  For instance, while
all integrated spectra lie in the star-forming region of the [N{\sc ii}]/H$\alpha$ vs. 
[O{\sc iii}]/H$\beta$ diagram, a sample of the spectra in the outer regions of each
galaxy extends towards right in this diagram, even lying in the so-called {\em composite}
and AGN regions. This is due to the high N/O ratio of these spaxels, combined with
the excitation structure of the gas, causing the [N{\sc ii}] emission lines to remain bright even at
very large distances to the ionizing stellar source.

We have assumed
that a certain property of the ISM can be considered as uniform within the galaxy or a 
determined region if the null hypothese of the Lilliefors test ({\em i.e.} the
fitting of a gaussian function to the corresponding distribution) cannot be rejected.
We find that in those cases where a gaussian does not fit a given
distribution, the disagreement between the mean value of the
distribution and the value derived from the integrated spectrum is larger.

We find that extinction, as derived from the decrement of Balmer, is only
uniform in HS 0837+4717.  Mrk 930 shows a
substructure of higher extinction in the outer southern part of the galaxy which breaks the uniformity.
Excitation, as estimated using the [O{\sc ii}]/[O{\sc iii}] emission-line ratio is
only uniform in the inner-most regions of HS 0837+4717 and Mrk 930, due possibly
to the lower excitation in the outer parts of these galaxies, as the distance to
the corresponding ionizing sources increases.  Electron density has
been only estimated with enough confidence in Mrk 930, where 
a structure of high density appears in the south part of the galaxy, so the distribution
of this property cannot be considered as uniform.

Electron temperature has been derived in a sample of spaxels
of the three galaxies from the [O{\sc iii}] 4363 {\AA} to 
[O{\sc iii}] 4959,5007 {\AA}  emission line-ratio. Its 
distribution appears as uniform in HS 0837+4717 and Mrk 930. This is also the
case for O chemical abundances and N/O ratios as derived
using these electron temperatures through the direct method. In the
case of HS 0128+2832, the N/O derived using the direct method is the only ISM
property in this galaxy which is uniform after our analysis.  The distribution of O/H and N/O
derived using strong-line methods are uniform in a lower number of
regions, possibly due to the dependence of the corresponding 
strong-line parameters on the
ionization structure of the gas.

Finally, we have not detected neither the broad WR blue bump ($\sim$ 4650 \AA)
nor the red one ($\sim$ 5808 \AA) suggested by previous works in two of the
observed galaxies. Even though we have investigated if these stars 
can be considered as the main cause of the measured excess of N/O.
We have demonstrated, using the stellar yields of massive star winds at the corresponding
metallicity of these objects (Z=0.004, Moll\'a \& Terlevich, in prep.) and taking into account
the stellar masses of the ionizing clusters as estimated from the integrated H$\alpha$ flux,
that at the scale lengths where we detect the uniform N/O high values ({\em i.e.}
$\sim$ 1 kpc),
the WR stars cannot be responsible for the enhancement in the N abundance.
It would be necesary between 2 and 3 orders of magnitude higher
in the ionizing cluster mass to reach such a degree of pollution, assuming an 
instantaneous mixing of the ejected material with the surrounding gas. 
Thus, for the case of these three studied galaxies, another  chemical evolution
scenario ({\em e.g.} metal-rich outflows, infall of metal-poor gas or interacting scenario as the morphology
and dynamics of Mrk 930 suggest) is required.

\begin{acknowledgements}
Based on observations collected at the Centro Astron\'omico Hispano Alem\'an (CAHA) at Calar Alto, operated jointly by the Max-Planck Institut f\"ur Astronomie and the Instituto de Astrof\'\i sica de Andaluc\'\i a (CSIC).

This work has been partially supported by projects AYA2007-67965-C03-02 and AYA2007-67965-C03-03 
 of the Spanish National Plan for Astronomy and Astrophysics, by the project TIC114  {\em Galaxias y Cosmolog\'\i a} of the
 Junta de Andaluc\'\i a and project CSD2006 00070 {\em 1st Science with GTC} of the Spanish Ministry of Science and Innovation
(MICINN).

Some of the data presented in this paper were obtained from the Multimission Archive at the Space Telescope Science Institute (MAST). STScI is operated by the Association of Universities for Research in Astronomy, Inc., under NASA contract NAS5-26555. Support for MAST for non-HST data is provided by the NASA Office of Space Science via grant NAG5-7584 and by other grants and contracts

CK, as a Humboldt Fellow, acknowledges support from the Alexander von Humboldt Foundation, Germany
RGB acknowledges additional funding by the China National Postdoc Fund Grant No. 20100480144

\end{acknowledgements}
RGB acknowledges additional funding by the China National Postdoc Fund Grant No. 20100480144

\end{document}